\documentclass [12pt,letterpaper]{report}

\usepackage{cite}
\usepackage{setspace}       
\usepackage{url}

\usepackage{graphicx}
\usepackage{subfigure}
\usepackage{float}
\usepackage{longtable}  
\usepackage{multirow} 

\usepackage[fancyhdr]{McECEThesis}  
\usepackage{McGillLogo}     
\usepackage{color}
\definecolor{red}{rgb}{1,0,0}
\def\headrulehook{\color{red}}      

\usepackage{anysize}
\marginsize{1in}{1in}{1.1in}{0.75in}

\usepackage{amsmath}        
\usepackage{amssymb}
\usepackage{bm}
\usepackage{amsthm}
\usepackage[linesnumbered, algoruled, longend, boxed]{algorithm2e}
\usepackage{url}

\graphicspath{{images/}}

\begin{document}

\title{Content Distribution Strategies in Opportunistic Networks \\ \vspace*{20pt} \Large Technical Report \vspace*{-28pt}}
\author{S.H. Masood, S.A. Raza, M.J. Coates}
\organization{%
  \\[0.2in]
  \\[0.1in]
  Computer Networks Laboratory\\
  Department of Electrical \& Computer Engineering\\
  McGill University\\
  Montreal, Canada}
 
\maketitle
\raggedbottom
\onehalfspacing
\pagenumbering{roman}

\section*{\centering Abstract}

The use of mobile data services is in high demand ever since the advent of smartphones and is expected to increase further with the evolution of various services and applications for future mobile devices. This excessive use of data has caused severe bottlenecks within the mobile networks due to overloading. The route towards network upgrades is an expensive one especially due to the high licensing fees attached to spectrum acquisition. Traffic offloading through opportunistic communication is a new paradigm exploiting the meeting opportunities within emerging Mobile Social Networks for the dissemination of delay tolerant contents without any additional investment. The idea is to save the network bandwidth by pushing the content optimally to as few users as possible from base station and then exploit the social interactions of social group members for content distribution among themselves by using point to point communications through WiFi or Bluetooth. Several studies suggests that there exists strong community architecture within contact graph of Mobile Social Networks i.e. the edges are not randomly distributed over nodes but in the form of clusters. We study the problem of dissemination of a large file in community based mobile social network where base station seeds the contents of the file initially to some users within each community. Furthermore, we determine whether it is more efficient in terms of latency and number of transmissions to transfer encoded or uncoded contents during opportunistic meetings by comparing different coded (Network Coding, Erasure Coding) and uncoded (Flooding, Epidemic Routing) data dissemination strategies in mobile social network.

 \newpage
 \tableofcontents{}
 \listoffigures
 \listoftables
\newpage
 \cleardoublepage
\pagenumbering{arabic}

 \typeout{}

\chapter{Introduction}\label{Ch1}

\section{Motivation}

Global mobile data usage has been rapidly increasing for the past few years. In 2010,  global mobile data traffic grew 2.6-fold~\cite{54} and is expected to continue to grow at a fast pace for the next five years. A significant portion of this traffic includes video. Mobile video traffic will exceed 50$\%$ of the total mobile data usage for the first time in 2011\cite{54} . 

Every day, thousands of mobile devices (phones, tablets, cars, etc) use the wireless infrastructure to retrieve content from Internet-based sources. The content also has become increasingly larger in size (video, software upgrades on smartphones). This has started to put an immense burden on the limited spectrum of infrastructure networks. The exceedingly expensive license fees placed on spectrum acquisition makes bandwidth expansion an expensive route for the service providers. Efforts are being made  to look for alternatives which can be used to offload some of this traffic~\cite{50,67,72,47}. In some areas Wi-Fi base stations and hotspots have been deployed, with some success, to shift some of the demands from the mobile infrastructure. Interoperability issues with the cellular network and interference concerns between adjacent Wi-Fi access points need to be addressed before it can be adopted on a larger scale. 

Recently, there has been a growing interest in opportunistic networking which is a cost-effective way of offloading some of the traffic from the mobile infrastructure. The scheme is based on peer-to-peer data sharing among mobile wireless devices and hence no infrastructure is required to support it. Communication is usually done using short-range Wi-Fi or Bluetooth connectivity. It has been shown that short-range communication usually consumes much less energy than long-range \cite{106}. The hand-to-hand distribution of content makes it ideal to target applications which have high spatial significance such as local and general news, sports, schedules. If subsets of users can be identified which subscribe to the same service and have high spatial locality, some of these users can be seeded with information to be spread to the entire population through opportunistic contacts. Our study involves devising a practical scheme to disseminate such content to  subscribers. 

\section{Background}

Opportunistic networks have gained significant attention in the literature recently. Several potential applications of opportunistic networks have attracted researchers to study the problems of data routing and forwarding in these networks.  There are various application scenarios for opportunistic networks. These include providing connectivity to nomadic and rural communities (DakNet\cite{Pentland2004}, SNC\cite{Doria2002}), interplanetary communication networks, wild life monitoring (ZebraNet\cite{Juang2002}), underwater species monitoring (SWIM\cite{Small2003}) and content distribution in urban settings like the Cambridge Pocket Switched Network (PSN) \cite{Hui2005}. We are considering the last application in our study.

Mobile social networks are an evolving class of opportunistic network and a new communication paradigm where mobile users communicate with each other through occasional communication opportunities. Such opportunities can be either scheduled or completely random. The widespread use of advanced small portable mobile devices like smartphones and tablets, with abundant local resources including storage space, local connectivity (WiFi, Bluetooth) and computing power enable their use for content sharing whenever the devices come in contact with each other through inter-device contacts due to human mobility. The communication opportunities arise where peoples belonging to different walks of life interact with each other socially in offices, class rooms, conferences, community centres etc.

\subsection{Contact Graph and Community Architecture} \label{CGCA}
In mobile social networks contacts happen due to mobility of individuals carrying mobile devices. Such contacts reflect the complex structure in the movement of people in the form of chance meetings with strangers, intentional meetings with colleagues, friends, family or familiar strangers due to similarity in mobility patterns. The structure of such mobility scenarios can be represented by aggregating the entire sequence of contacts of a trace to a static, weighted contact graph \textit{G(N,\textbf{W})} with nodes \textit{N} and weight matrix $\textbf{W}={w_{ij}}$ \cite{Hossmann2011}. In this graph each device or a person carrying a device is represented as node and the link weight $w_{ij}$ represents the relationship strength between nodes \textit{i} and \textit{j}. The weights that have been chosen in the literature are frequency of contacts \cite{Hui2008}, age of last contact \cite{Dubois2003}, aggregate contact duration \cite{Hui2008} and the combination of contact frequency and duration \cite{Hossmann2011}. For experimental studies of mobile social networks, user mobility that governs pair-wise meeting of users in social networks has been based on real mobility measurements or synthetic mobility models. Real mobility datasets include Dartmouth (DART)\cite{Henderson2004}, ETH Campus (ETH) \cite{Tuduce2005}, Gowalla (GOW)\cite{Hossmann2011}, MIT Reality Mining (MIT)\cite{Eagle2009}. The synthetic mobility models that use methodologies to map contacts to a conceptual social graph include Time Variant Community Model(TVCM), Home-cell Community-based Mobility(HCMM) and SLAW \cite{Weijen2007,Boldrini2010,Kyunghan2009}.    
 
 
The mobility scenarios from the contact graphs reveal the presence of underlying communities, bridges, hubs and other social structures common in social networks. Communities are modular structures or subsets of nodes with stronger connections between them than other nodes. They generally represent social groups formed among friends, co-workers etc. In \cite{Hossmann2011} Hossmann et al. apply the Louvian algorithm \cite{Boldrini2008} to identify communities in the aforementioned real and synthetic mobility models as depicted in Table 2.1. Their results provide evidence for the presence of communities in mobile social networks. 

\begin{table}[]
	\centering
    	\begin{tabular}{|c|c|c|}
	\hline
					   	\multicolumn{3}{|c|}{Identified Communities}    \\    \cline{1-3} 
	Trace/Model 	&	No. of Nodes  &	No. of Communities   \\ \hline

	DART	&  1044    &	24	       \\
	ETH		&  285     &	30		  \\
	GOW		&  473     &	29				 \\ 
	MIT		&   92     &	6				 \\ \hline
	TVCM	&	505    &	10				 \\
	HCMM	&	100    &	10		 \\
	SLAW	&	100    &	2			 \\  \hline
	
    \end{tabular}
    \caption[Identified Communities in Different Mobility traces and Synthetic Models]{Number of Communities identified in \cite{Hossmann2011} using Louvian algorithm \cite{Boldrini2008} on different mobility traces and models.}
\end{table}

To capture the structural and pair-wise statistics in people's movements, we use the LFR benchmark software \cite{Lancichinetti2008,Lancichinetti2009} to generate weighted contact graphs \textit{G(V;E;w)}, where individuals are represented by vertices V , relationships between individuals by the edges \textit{E} and the weights \textit{w} represent the expected contact rates between pairs of users connected by an edge. There are many tools for creating these social graphs; but the LFR benchmark software is effective in providing a model which captures real-world characteristics present in human-mobility traces (power-law degree distribution, community structure, power-law based community size distribution, adjustable ratio between inter-community and intra-community weights).


LFR benchmark is a special case of the \textit{planted l}-partition model \cite{Liu2010} in which communities are of different sizes and nodes have different degrees. With the LFR benchmark, graphs having different communities and groups can be generated. A mixing parameter $\mu_t$ expresses the ratio between the external degree (number of edges outside community) and the total degree of the node. This ratio can be controlled to vary the connectivity between communities. Each node shares a fraction $1 - \mu_t$ of its links with nodes in its community and a fraction $\mu_t$ with nodes outside its community. Other parameters that can be customized include the number of nodes \textit{N}, the average degree of a node \textit{k}, the minimum community size \textit{minc} and the maximum community size \textit{maxc}. Each node is given a degree drawn from a power law with exponent $\gamma=2$. The community sizes also follow power law distribution with exponent ranging from 1 to 2. The values of power law exponents are typical of those found in real networks \cite{Lancichinetti2008,Arenas2004}. Assignment of weights on edges lying within the community as well as outside the community is handled by two parameters $\mu_w$ and $\beta$. The parameter $\beta$ is used to assign a strength $s_i$ to each node, $s_i=k_i^{\beta}$, where $k_i$ is the degree of node \textit{i}. Such a power law relationship between the strength and the degree of a node is frequently observed in real weighted networks \cite{Barrat2004}. A high value for strength $s_i$ means the user \textit{i} meets its neighbours more frequently. The parameter $\mu_w$ is used to assign strength \textit{s} to each node. There are two type of strengths associated with each node. An internal strength $s_i^{(in)}=(1-\mu_w)s_i$ indicates how strongly the node is connected to other nodes within its community and the external strength $s_i^{(out)}=\mu_ws_i$ shows the same relationship outside the node's community. The weights on individual edges can be assigned ensuring that the intra-community edge weights sum to $s_i^{(in)}$ while the inter-community weights aggregate to $s_i^{(out)}$. This can be accomplished by minimizing the variance of squared error \cite{Lancichinetti2009}, which is given by:\\

\hspace{1 in} $Var(w_{ij}) = \Sigma_{i} (s_i-\rho_i)^2 + (s_i^{(in)}-\rho_i^{(in)})^2 + (s_i^{(out)}-\rho_i^{(out)})^2$.\\

Here, $\rho_i = \Sigma_jw_{ij}$ , $\rho_i^{(in)} = \Sigma_jw_{ij}. k(i,j)$, $\rho_i^{(out)} = \Sigma_jw_{ij}. (1-k(i,j))$ , where the function $k(i,j)=1$ if nodes \textit{i} and \textit{j} lie in the same community, and $k(i,j)=0$ otherwise. The mean inter-contact interval $(c_{ij})$ between nodes \textit{i} and \textit{j} is $(c_{ij})= \frac{1}{w_{ij}}$.\\ 
 
The output of the LFR benchmark is a weighted undirected graph. The membership of each node is also provided. Membership in our context indicate the community with which a node is associated. We consider the membership of nodes to be \textit{hard}; each node can only be associated with a single community. The hard membership assumption is also considered by the work done in \cite{Xiangwei2009,Zonghua2005} for epidemic dissemination in scale free and social networks.

\subsection{Content Distribution in Opportunistic Networks}

Content distribution on the Internet refers to the delivery of digital multimedia for example multimedia files, streaming audio and video and software to a large number of users over internet. P2P networks like BitTorrent, KaZaA etc. are the famous content distribution systems over the Internet. Today a large percentage of Internet traffic is related to content distribution. The collaboration of devices in opportunistic networks can increase the chances of content dissemination but it is not an easy task to carry out. The inherent properties of opportunistic networks make it more difficult and challenging. The pairwise contact times and contact durations are important parameters in opportunistic networks. The contact durations are usually short so the type and the amount of content to share should be simple and compact in formats. This pairwise content exchange paradigm has motivated researchers to study the development of algorithms that can fully exploit the contacts to opportunistically disseminate data among all devices in the network.   


Papadopouli et al.\ proposed 7DS \cite{Papadopoul2001} a peer-to-peer dissemination and sharing system for mobile devices having intermittent connectivity. In 7DS nodes which are experiencing intermittent connectivity can query the peers in their proximity, to determine if they either have data cached or have Internet access and thus can forward or relay the data.\ In \cite{Lindemann2005} Lindemann et al.\ proposed a variant of 7DS using epidemic routing for information dissemination in MANETS. In \cite{Vukadinovic2010} mobility-assisted wireless podcasting is proposed to offload traffic from the cellular operator's network. The approach reduces spectrum usage in cellular networks by distributing content to some percentage of users that have the strongest propagation channels.  

Pocket Switched Networks (PSNs) strive to convey messages in networks where users are mobile by exploiting the local and global connectivity \cite{hui2005_1,Hui2005}. They are based on a set of assumptions such as a user carry mobile device with significant storage space, the users are willing to carry messages for other users in network, the devices have local connectivity in them along with global connectivity. Haggle \cite{hui2005haggle} is a real implementation of a PSN to explore the contact and inter-contact time values in order to design appropriate forwarding policies for opportunistic exchange of information between devices. Haggle keeps track of mobility traces for human movements in real experiments by extracting the mobility information from mobile device carried by individuals. 
        
In \cite{Leguay2006} Leguay et al.\ have done an experimental study for a period of two months in an urban setting by collecting device contacts with each other and have analysed their properties for data forwarding. This connectivity information was then utilized to study the feasibility of city-wide content distribution networking. By classifying users with different behavioural patterns they analysed the effectiveness of this user population in distributing contents. 

In \cite{LeBrun2006} LeBrun et al.\ explored the feasibility of spreading content of interest using transit buses in the University of California, Davis campus. Each bus has a Bluetooth Content Distribution cache called BlueSpot installed, which holds the content of interest, and any device that has a bluetooth connectivity switched on can connect to these caches and download the content of interest during the idle time while the bus is en route. The authors predict the provision of services like on demand paid iTunes music file as a future potential application.

Karlsson et al.\ propose a receiver-driven broadcasting system in \cite{Karlsson2006}. The system enhances the infrastructure based broadcast system that exploit the pair-wise contacts of mobile nodes to spread the content. The content distribution problem has been studied through simulation and a practical test bed. The test bed based system uses Bluetooth as a wireless communication mechanism to spread out the content among mobile nodes when they interact each other through their pair-wise meetings. Instead of flooding to everyone who is met, the system smartly allows nodes to decide on what to download and from whom to download during encounters.


In \cite{Ioannidis2009,Niyato2010}, optimal and scalable content distribution policies are presented in publish subscribe mobile social networks where users subscribed a news-feed, a blog, or a service that monitors stock prices or traffic congestion, assist each other in retrieval of common subscribed services. Such transmission policies strive to increase the freshness of the content such that at any point in time a large percentage of users within network have the most up-to-date information.

\subsection{Routing In Opportunistic Networks} \label{RouteOppNet}
Routing in opportunistic networks is a challenging task. The inherent properties of the networks make it difficult because the network topology evolves frequently and this information in most cases is not communicated to nodes. In \cite{Pelusi2006} a detailed hierarchy of the routing protocols for opportunistic networks is presented. For the applications of content distribution, dissemination based routing has been considered in various studies. Dissemination based routing performs delivery of message by simply diffusing it all over the network. The dissemination routing protocols for opportunistic networks can broadly be classified as replication based and coding based \cite{Chen2006}.\ In replication based routing multiple identical copies of data are injected into the network and distribution of data to all nodes of network relies on node movement. In coding based routing the data is initially transformed into coded blocks. These coded blocks are then disseminated within network and upon reception of sufficient encoded data, the original data is recovered through decoding. To clarify the difference, replication based schemes require the successful delivery of each individual data block whereas for coded schemes the data is recovered from subset of any sufficiently large encoded blocks. We are considering flooding and epidemic routings for replication based routing and network coding and erasure coding based routing for coding based routing.


\subsubsection{Uncoded Content Dissemination}


\subsubsection*{Flooding}
In flooding, every node forwards non-duplicated packets (those which it has not forwarded to the same node earlier) to the nodes it meet. In \cite{Wang2005}, flooding has been considered in opportunistic networks. A node does not relay back a packet to the node from which it has received this packet. In this approach, the sending node does not have any information regarding the packets that the receiving node has already downloaded. Let the two meeting nodes be \textit{A} and \textit{B}. During a specific meeting node \textit{A} selects a packet randomly which it has not yet forwarded to node \textit{B}. If node \textit{A} does not find such a packet, the transmission opportunity is missed during that meeting. Node \textit{B} follows the same procedure during the meeting. In this case there could be redundant transmissions because the receiving node may already have received the packet being sent from some other node.
  
\subsubsection*{Epidemic Routing} 
With Epidemic Routing \cite{Vahdat2000} content diffuses in the network similar to a disease or virus \cite{Pelusi2006}. We are implementing epidemic routing as described in \cite{Lin2007}. When two nodes \textit{A} and \textit{B} meet, through the exchange of packet identifiers we assume that both the nodes know the packets in each others' buffers. Let $S_A$ and $S_B$ represent the set of packets stored by node \textit{A} and \textit{B} respectively. Node \textit{A} chooses a packet from the set $S_A - S_B$ to transmit to node \textit{B} such that the packet transmitted is always new to node \textit{B}. If the set is empty, node \textit{A} misses the transmission opportunity. Node B repeats the same procedure.

In \cite{Lin2007,Gkantsidis2005} three policies were discussed as to which packet to exchange from set $S_A - S_B $. First in the \textit{random} policy node \textit{A} may choose any packet from the set $S_A - S_B$ with equal probability. Second, in the \textit{local rarest} policy, a node downloads the packet which is rarest in its neighbourhood. The rarity of the packet is determined by the number of copies of a packet present in its neighbourhood. BitTorrent also utilizes a local rarest policy for peer to peer content dissemination. Third, in the \textit{global rarest} policy, the packet is downloaded based on the rarity of packet within the complete network which is very hard to track, hence \textit{global rarest} is unrealistic and is not considered in our implementation. We implement random and local rarest policy based epidemic routing for comparison among different content dissemination schemes.        
  
\subsubsection{Encoded Content Dissemination}  

\subsubsection*{Network Coding}
The concept of network coding was introduced in the pioneering work of Ahlswede et al. that established the value of coding in routers and provided theoretical bounds on the capacity of networks that employ coding \cite{Ahlswede2000}. As proved in \cite{Li2003,Koetter2003,Jaggi2005} for multicast traffic, linear codes achieve the maximum capacity bounds and coding and decoding can be done in polynomial time, Ho et al. proved that it is still true in case even if the routers choose random coefficients \cite{Ho2003}. 

Network coding was initially proposed for throughput improvements but it has also been applied for content distribution. In \cite{Gkantsidis2005} Gkantsidis et al. developed a content distribution system (Avalanche) based on network coding. In \cite{Ma2007} Ma et al. developed a content dissemination system based on sparse network coding using the Chord protocol that generates independent network encoded packets with high probability. In \cite{Widmer2005} network coding based on probabilistic routing is presented for efficient communication on extreme networks. The study of \cite{Lin2007} revealed through simulations and analysis that incorporating network coding in epidemic routing is beneficial than replication based epidemic routing in Delay Tolerant Networks (DTNs).         

With network coding, the file that has to be distributed is divided into \textit{k} packets of the same size. The network coded packets that are distributed among the users are the linear combination of all such packets. The network coded packet also contains the vector of linear coefficients used to create it.  For content distribution, the base station provides such network coded packets to the nodes. Upon meeting, the two nodes share with each other the linear combination of the network coded packets present in their respective buffers by generating a coefficient vector within the finite field. The generated coefficient vector is then updated within the resulting packet by multiplying with the already existed coefficient vector within the network coded packets used to generate this new packet. Once the receiver accumulates \textit{k} independent packets it can recover the complete file by inverting the coefficient matrix.\\

\hspace*{2.2 in} $ \textbf{P}$ = $\textbf{C}^{-1} \textbf{P}' $\\
where \textbf{P} is the vector of \textit{k} packets belonging to original file that have to be recovered, \textbf{C} is the $ k \times k$ matrix of coefficient vectors and $\textbf{P}'$ is the vector of \textit{k} independent network coded packets.  

\hspace*{1.8 in}$\begin{bmatrix} p_1 \\ \vdots \\ p_k \end{bmatrix}$ = $\begin{bmatrix}  c_{11} &\hdots &c_{1k} \\ \vdots &\ddots &\vdots \\ c_{k1} &\hdots &c_{kk} \end{bmatrix} ^{-1} $  $\begin{bmatrix} p_1' \\ \vdots \\ p_k' \end{bmatrix}$ \\                
  
\subsubsection*{Erasure Coding}
Erasure coding is also a scheme used for data dissemination. Spoto et al. presented a BitTorrent system based on LT fountain codes \cite{Spoto2010}. In \cite{Magnetto2010} ToroVerde -- a push based P2P content distribution system employing fountain codes is presented. In \cite{Wang2005} Wang et al.\ study erasure coding based routing in an opportunistic network formed by the real mobility dataset of ZebraNet \cite{Juang2002}. Their result showed that erasure coding can achieve the best worst case delay performance from source to destination routing with fixed overhead. In \cite{Jain2005} erasure coding has been employed to exploit redundancy to cope with failures in Delay Tolerant Networks (DTNs). 

Unlike network coding where each node encodes packets before forwarding, in erasure coding only the source or server encodes the file.\ The idea of erasure coding is to incorporate redundancy so that in case of a lost packet a new encoded packet can be used to recover the lost one.\ In this way sequential transmission is not required. In the case of opportunistic networks, erasure coding provides the means to have successful meeting (in which packet transfer occurs). Because nodes can download the missing packets with more probability than no encoding. The reason is erasure coded packet represents the exclusive-OR (XOR-ed) version of original packets instead of single packet and hence the chances of receiving the missing packets increase.   
 
Depending on the count of a number of packets erasure coding is classified as \textit{k/n} rate (where \textit{n} encoded packets are generated by the source from \textit{k} original file packets) or as rateless coding. The idea of rateless coding is associated with codes also termed as fountain codes \cite{Mitzenmacher2004} where a theoretically infinite number of encoded packets can be generated by the source. With ideal source coding only \textit{k} encoded packets are needed to recover or decode a file; practically, slightly more than \textit{k} packets are needed. Based on the encoding/decoding complexity and time, different types of erasure codes have been proposed. They are classified as \textbf{Rate based} which includes Reed Solomon , Tornado Codes and \textbf{Rateless} which includes Luby Transform (LT) and Raptor Codes. We implement LT codes based erasure coding for content dissemination because of their simple encoding and decoding process. Below is the brief description of LT codes.   

\subsubsection*{LT (Luby Transform) Codes}
LT codes \cite{Luby2002} are the first practical realization of the rateless or fountain codes i.e., the number of encoding symbols that can be generated from the data is potentially limitless. The process of encoding is very simple.

\begin{enumerate}
\item Randomly choose the degree \textit{d} of the encoding symbol from the degree distribution.\ The degree distribution is very important for successful operation of LT Codes and we will discuss it later.
\item Randomly choose \textit{d} input symbols as neighbours of the encoding symbol from the uniform distribution. 
\item The value of the encoding symbol is the exclusive-or (XOR) of the \textit{d} neighbours.  
\end{enumerate}

The decoder can decode the encoded symbols to recover the input data if it knows the degree and set of neighbours of each encoding symbol. The decoding process works as follows. \textit{``If there is at least one encoding symbol that has exactly one neighbour then the neighbour can be recovered immediately since it is a copy of the encoding symbol. The value of the recovered input symbol is exclusive-ored (XORed) into any remaining encoding symbols that also have that input symbol as a neighbour, the recovered input symbol is removed as a neighbour from each of these encoding symbols and the degree of each such encoded symbol is decreased by one to reflect this removal''} \cite{Luby2002}.\\

Luby discussed that the design criteria for degree distribution is based on two goals.
\begin{enumerate}

\item The number of encoding symbols required to ensure decoding success should be as few as possible. 

\item The average degree of the encoding symbols should be as low as possible.

\end{enumerate}

He identified a degree distribution called the Robust Soliton distribution. If $\delta$ is the decoding failure probability of the decoder to recover the data from a given number \textit{K} encoding symbols then \textit{k} input symbols can be recovered from a set of $K = k + O(\ln^2(k/\delta ) \sqrt{k}) $ fully randomly generated encoding symbols. For all \textit{d}, $\mu(d)$ is the probability that an encoding symbol has degree \textit{d}. The degree distribution is given by,\\

\hspace*{1.4 in}$\mu(i)=(\rho(i)+\tau(i))/\beta $ For all $ i=1,....,k$\\

\begin{equation*}
 \tau{(i)} =  \left\{ \begin{array}{cl}
   R/ik &for\;i=1,...,k/R-1 \\
   R\ln(R/\delta)/k &for\;i=k/R\\
   0 &for\;i=k/R+1,...,k
  \end{array}\right.
\end{equation*}\\    
where $R=c\ln(k/\delta)\sqrt{k}$ for some suitable constant $c > 0$, which as per \cite{Cataldi2006} is bounded by\\

 
\begin{equation*}
\frac{1}{k-1}\frac{\sqrt{k}}{\ln(k/\delta)} \leq c \leq \frac{1}{2}\frac{\sqrt{k}}{\ln(k/\delta)}
\end{equation*} 
 
\begin{equation*}
 \rho{(i)} =  \left\{ \begin{array}{cl}
   1/k &for\;i=1 \\
   1/i(i-1) &for\;i=2,...,k
   \end{array}\right.
\end{equation*}\\    

\hspace*{2 in}$\beta=\Sigma_{i=1}^k \rho(i)+\tau(i) $\\

In order to decode, a receiver needs to receive slightly more packets than the number of packets in the original file. These excess packets constitute the overhead of erasure coding and determine its efficiency.

\section{Problem Statement}
We are interested in determining the importance of detecting communities in opportunistic networks with mobile carriers. We analyze the deviation in performance, in terms of the rate at which mobile users obtain a file, between seeding each community independently with the complete file versus randomly seeding file packets to the network for different degrees of community structure in the network. Network coding has been shown to be efficient in large scale file distribution \cite{37}. In our analysis, we compare dominant encoded and unencoded strategies for file dissemination. 

\typeout{}
\chapter{Content Dissemination in Community-based Networks} \label{Ch3}

In this chapter, we investigate the performance of encoded and unencoded file dissemination techniques discussed in Section \ref{RouteOppNet} for networks exhibiting varying degrees of community structure. For every dissemination scheme whether coded or uncoded, the base station initially seeds some of the nodes within each community or network using some predefined seeding strategy. For every dissemination scheme whether coded or uncoded, the base station initially seeds some of the nodes within each community or network using some predefined seeding strategy.\ The different seeding strategies are discussed in Section \ref{ExpDet}.\ After this initial seeding, all the nodes gather missing packets via the employed dissemination strategy upon meetings with other nodes opportunistically. The complete dissemination process is shown in Figure \ref{ExpDet}. 

Our analysis is based on networks generated using the LFR benchmark graph generation software. The software allows us to customize the depth of community structure in the synthetically generated network. The following sections discuss experimental details for the comparison like the file (content) and packet size, the seeding strategies that the base station employs for initial seeding and the underlying graph that governs the social interactions among users. 

\begin{figure}[H]
    \centering
	\includegraphics[angle=0,scale=0.4,  clip,type=pdf,ext=.pdf,read=.pdf]{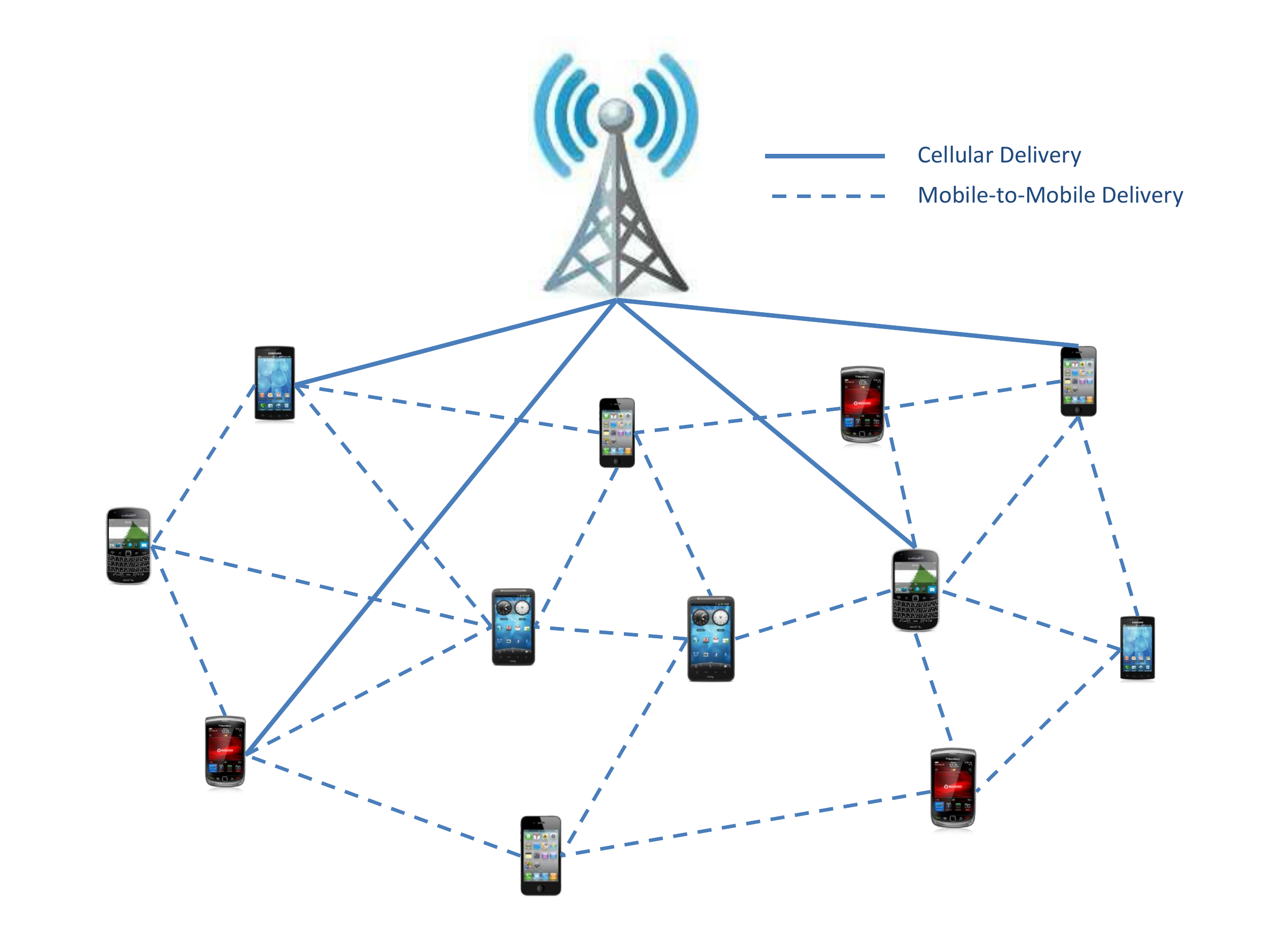}
	\label{fig:Ch1_Exp1_Graph}
	\caption[Content Dissemination Process]{Content Dissemination Process showing initial seeding from Base Station and Opportunistic content sharing among nodes themselves after initial seeding.}
	\end{figure}

\section{Experimental Details} \label{ExpDet}

\subsection{File and Packet Size}
We assume the content size i.e. the file size to be 10 MBytes which is equal to 10240 KBytes. We divide this file into 128 KBytes equally sized packets or blocks. Total number of packets is now n = 10240/128 = 80. Each of such packet will comprises of 131072 symbols of 1 byte thus forming a packet of size 128 KBytes similar to the block size definition of \cite{Ma2007} where the block size of 256 KBytes was considered.
 
We assume that during a contact; flooding, epidemic routing, network coding and erasure coding can exchange one packet of size 128 Kbytes which is realistic if we assume 802.11n MAC layer. Assuming the communication takes place over 802.11n enabled devices, a close range point-to-point communication can support a data-rate of up to 150 Mbit/s or 19200 Kbytes/s on a bi-directional link. For such a setup, transmitting a couple of packets of length 128 Kbyte each would take $(2 \times 128)/19200 = $ 0.01secs or 10ms. If we consider data transmission in 802.11n without frame aggregation, the maximum data payload that can be sent in one 802.11n frame is 2304 bytes. So during a single contact between two users, each user sends 128/2.25 = 57 frames to the other to complete the transfer of a single packet (coded or uncoded) of size 128Kbytes each way. 

\subsection{Seeding Strategies}
Next to compare the performance of coded and uncoded dissemination schemes in different scenarios of base station's available bandwidth, we will consider the following seeding strategies;

\begin{enumerate}

\item Seeding 50\% more packets than the minimum number of packets required to download the complete file in each community (150\% Seeding). 

\item Seeding the minimum number of packets required to download the complete file in each community (100\% Seeding). 

\item Seeding 10\% less than the number of packets required to download the complete file in each community (90\% Seeding).

\item Seeding 20\% less than the number of packets required to download the complete file in each community (80\% Seeding).  

\item Randomly seeding the minimum number of packets i.e. seeding the same number of packets as injected in 100\% seeding strategy without consideration of community and randomly choosing the nodes within the whole network. 
  
\end{enumerate}            

For all cases of seeding within communities i.e. 80\%, 90\%, 100\% and 150\%, the base station selects the nodes within each cluster uniformly at random and seed them with packets until the given percentage of packets is seeded. After such seeding it may be possible that there will be some nodes  which do not posses any packet however it is possible that some nodes may have more than one packet. For the random seeding case the total number of packets seeded is the same as of 100\% seeding however nodes are selected uniformly at random from the whole network rather than within each cluster. 

After the initial seeding from the server, nodes collect the missing packets during meetings with other nodes of the same or different clusters.\ The packets seeded during Network coding are linear combinations of all file packets.\ For erasure coding each seeded packet represents the XOR of a subset of the original file's packets and for flooding and epidemic routing the packets are the same as the file's original packets. For the network coding case, each node further encodes the received packets before forwarding to another node during an opportunistic contact. For erasure coding (coding only at server), epidemic routing and flooding (no coding) nodes do not alter the packets and share the packets in the original form as provided to them by the base station. For every seeding case we perform 50 trials and calculate the average performance metrics.

\subsection{Social Graph Generation}\label{SGG}

Social networks are often modelled as graphs where individuals are represented as nodes and the edge between two nodes shows the existence of a relationship between them. There are many ways of creating these social graphs, some mainstream approaches are listed in Section \ref{CGCA}. We use the LFR benchmark \cite{Lancichinetti2009,Lancichinetti2008} to generate social graphs for our experiments. The LFR benchmark as discussed in the literature review is a special case of the planted partition model (with different partition sizes) and gives more control over the parameters that can be tuned to generate graphs having different properties. We are considering a social network of 200 nodes where each node is assigned a degree taken from a power law. The community sizes also obey a power law distribution. For LFR graphs we discussed that the parameter $\mu_t$ controls the fraction of links within and outside community. We have selected this parameter equal to 0.1 which means 10\% of the count of total links of nodes within community are inter-community.\ Recall that for LFR benchmarks $\mu_w$ is used to control the frequencies of the meetings of nodes within the same community and between nodes of different communities. For our experiment we have chosen this parameter to be $\mu_w=0.001$, which means nodes meet other nodes from their own communities more frequently than with nodes outside their respective communities. More precisely this means that the average inter-meeting interval for nodes belonging to different communities is 10 times higher than average inter-meeting interval for nodes belonging to same communities.\ In real networks we expect average inter-contact time between pairs of users in different communities to be closer to 10-15 times larger than the meetings between users in the same community \cite{Hossmann2011}.\ Figure \ref{fig:CH1_Exp1_individual_nodes_150percent0001_a} shows the generated graph using the discussed parameters for a graph of 200 nodes that comprised of 14 communities. 

The pairwise meeting time is considered to be exponential in much of the analytical work done for opportunistic networks \cite{Groenevelt2005,Neglia2006,Zhang2007}. In other words a Markov mobility model and the measured meeting rates from simulations are used to set parameters in the mobility model \cite{Small2007,Haas2006}. In \cite{Groenevelt2005} Groenvelt et al.\ show that the inter meeting time for any pair of node is exponentially distributed if (i) nodes move according to common mobility models like random direction or random way point; (ii) the node transmission range is small compared to its area of movement (iii) and each node's speed is sufficiently high. In our simulation we model the intervals between the meetings of node pairs as independent, exponentially distributed random variables. The graph weights determine the mean meeting period for the exponentially distributed inter meeting time per node pair. In other words, we are modelling the pair wise meetings among nodes as a Poisson process.

\begin{figure}[H]
	\centering
	\includegraphics[scale=0.6, angle=0, type=pdf,ext=.pdf,read=.pdf]{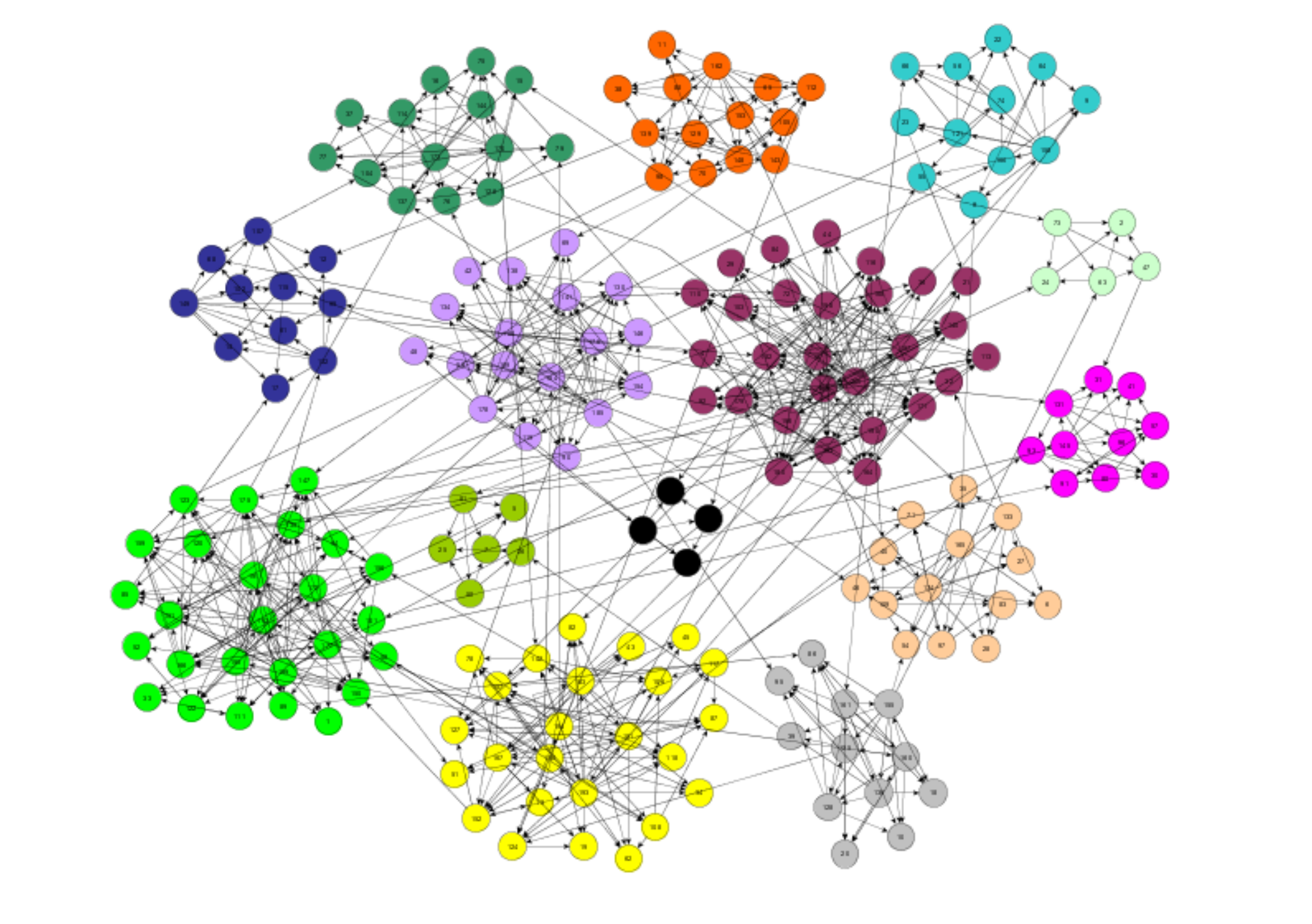}
\caption[Network of 200 Subscribers with 14 Communities]{Graph for $\mu_w = 0.001$, $\mu_t = 0.1$, $\gamma = 2$, $\beta = 1$  with users belonging to same community represented by the same color. A user can only meet other users with which it is connected. 14 communities are identified in this 200 node social graph.}
	\label{fig:CH1_Exp1_individual_nodes_150percent0001_a}
\end{figure}

\section{Performance Metrics} \label{PM}

The considered forwarding strategies namely Flooding, Epidemic routing, Erasure Coding and Network Coding are compared using the following performance parameters for the above experimental settings.

\begin{enumerate}

\item \textbf{Latency:} Latency represents the time required by the users to completely receive the file. The plot for this performance parameter depicts the percentage of nodes within the network that have received the complete file at any time \textit{t}. It is important to note that we have assumed that the time needed for the base station for initial seeding is negligible and initial seeding is done at \textit{t=0}. 

\item \textbf{Users Finish Time:} This represents the median finish time over all the experiment runs that each user takes to completely receive the file in the network. The users are sorted cluster wise to check if there exists correlation among completion times for the users belonging to the same cluster for different seeding strategies. Within each cluster the users are sorted by their degrees to have a better understanding of the impact of how connected a node is on the content retrieval rate.

\item \textbf{Transmissions and Innovativeness:}
This performance metric comprises three sub metrics.

\begin{enumerate}

\item \textbf{Packets Transmitted:} This metric represents the total number of data packets transmitted for the complete content reception at all nodes of the network. This is an important parameter as it inversely affects the life time of the mobile devices.   

\item \textbf{Total Contacts:} This parameter represents the total number of meetings that have taken place until all the nodes receive the complete content.

\item \textbf{Non Innovative Transmissions:} In case of network coding not all the transmitted packets are useful. These are the transmissions that do not increase the rank of the linear coefficient matrix \textit{C}, which is required to have rank at least equivalent to the number of packets in the file for decoding.  
 
\end{enumerate}

\end{enumerate}

\section{Results} \label{Ch4}

This chapter discusses the results derived for flooding, epidemic routing, erasure coding and network coding for different seeding strategies and experimental settings described in the previous chapter. All the techniques are compared using the performance parameters detailed in section 3.2. Matlab is used as the simulation platform for the implementation of all the techniques.	

\subsection{Dissemination Strategies Comparison} \label{DSCsec}

In this section we compare the performance of different dissemination schemes for different seeding strategies in terms of latency, i.e., the percentage of users that have received the file at any time \textit{t}. 

It is evident from Figure~\ref{DSC}(a),(b) that for 80\% and 90\% seeding strategies network coding (NC) outperformed other strategies. The reason for this is that the graph we are considering is tightly knotted i.e. the meetings within communities are much more frequent as compared to inter community meetings. In network coding, intermediate nodes can also encode packets and the importance of each encoded packet is the same because it represents the linear combination of all the file packets. Hence the meetings within a cluster could lead to the generation of enough packets to yield the rank of coefficient (\textit{C}) matrix necessary for file decoding. In other words the useful packets which we referred to as innovative packets can be generated within the communities themselves. For the rest of the strategies the missing packets within the communities must be extracted from neighbouring communities which causes their degraded performance in comparison to network coding. In the case of Epidemic Routing using Random (EP-R) and Local Rarest (EP-LR) packet selection, both the strategies are observed to perform almost the same. This effect is again due to the presence of the strong community architecture because by the time inter community meetings take place almost all nodes within the community have acquired all of the packets already possessed by the community. Any new packet from neighbouring community will be rare in the node's neighbourhood and within the whole community even if it is just selected randomly form the set $S_A - S_B$. This makes local rarest and random packet selection  behave almost the same. With Erasure Coding (ER), the source encoded packets are shared based on the local rarest policy. The reason for the performance degradation in comparison to epidemic routing is that there is some overhead associated with erasure coding. Unlike network coding in erasure coding not all of the packets are of the same importance, so some of the packets shared during precious inter-community meetings represent the encoded version of packets already decoded by the node. Secondly, it may be possible that the encoded packet is the combination of more than one missing packets so other packets are needed for decoding which again introduces delay. With flooding (FD) packets being transmitted during inter-community meetings may already be available in community so the precious transmission is wasted causing further delays.   

For 100\% and 150\% seeding strategies Figure~\ref{DSC}(c),(d) the performance of network coding and epidemic routing are almost the same. A sufficient number of packets whether coded in case of network coding or uncoded in case of epidemic routing or flooding, to recover a complete file are seeded to each community so the content dissemination is achieved based on local meetings. Flooding still performs poorly compared to epidemic routing and network coding because of wasted transmissions of already existing packets.\ For erasure coding with 100\% seeding the file can not be decoded using packets available within each community because some additional packets (overhead) are needed so the decoding process has to rely on packets extraction from neighbouring community which causes delay. With 150\% seeding some additional packets are already seeded hence the performance of erasure coding in 150\% is better in comparison to 100\% seeding.  

In the Random seeding strategy (see Figure~\ref{DSC}(e)) the latency is worst for all dissemination schemes however the relative performance of dissemination schemes is same i.e. Network coding performs the best, followed by epidemic routing, erasure coding and the worst is flooding. Erasure coding performs similar to network coding and better than epidemic routing during the start of dissemination process. Packets are seeded by selecting nodes from the whole network uniformly at random, so the large communities are assigned more packets and hence these communities are provided initially with sufficient packets for erasure decoding. The small communities must rely more on their neighbouring communities and hence with time erasure coding starts performing poorly as compared to epidemic dissemination. It still performs better than flooding as local rarest encoded packets are being exchanged resulting in more useful transmission than blind flooding.

\begin{figure}[H]
    \centering
	\subfigure[]{
	\includegraphics[angle=270,scale=0.45,  clip,type=pdf,ext=.pdf,read=.pdf]{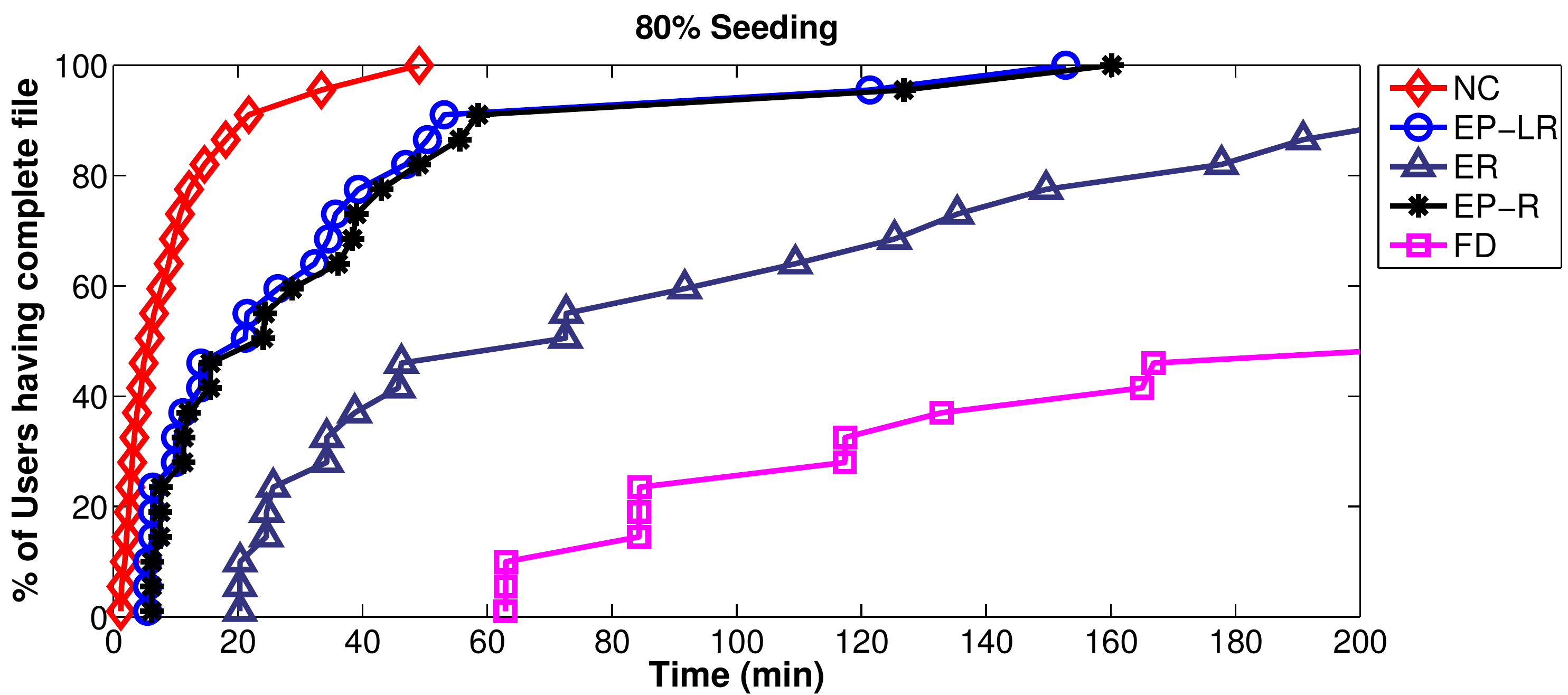}}
	\label{DSCa}
	
	\subfigure[]{	
	\includegraphics[angle=270,scale=0.45,  clip, type=pdf,ext=.pdf,read=.pdf]{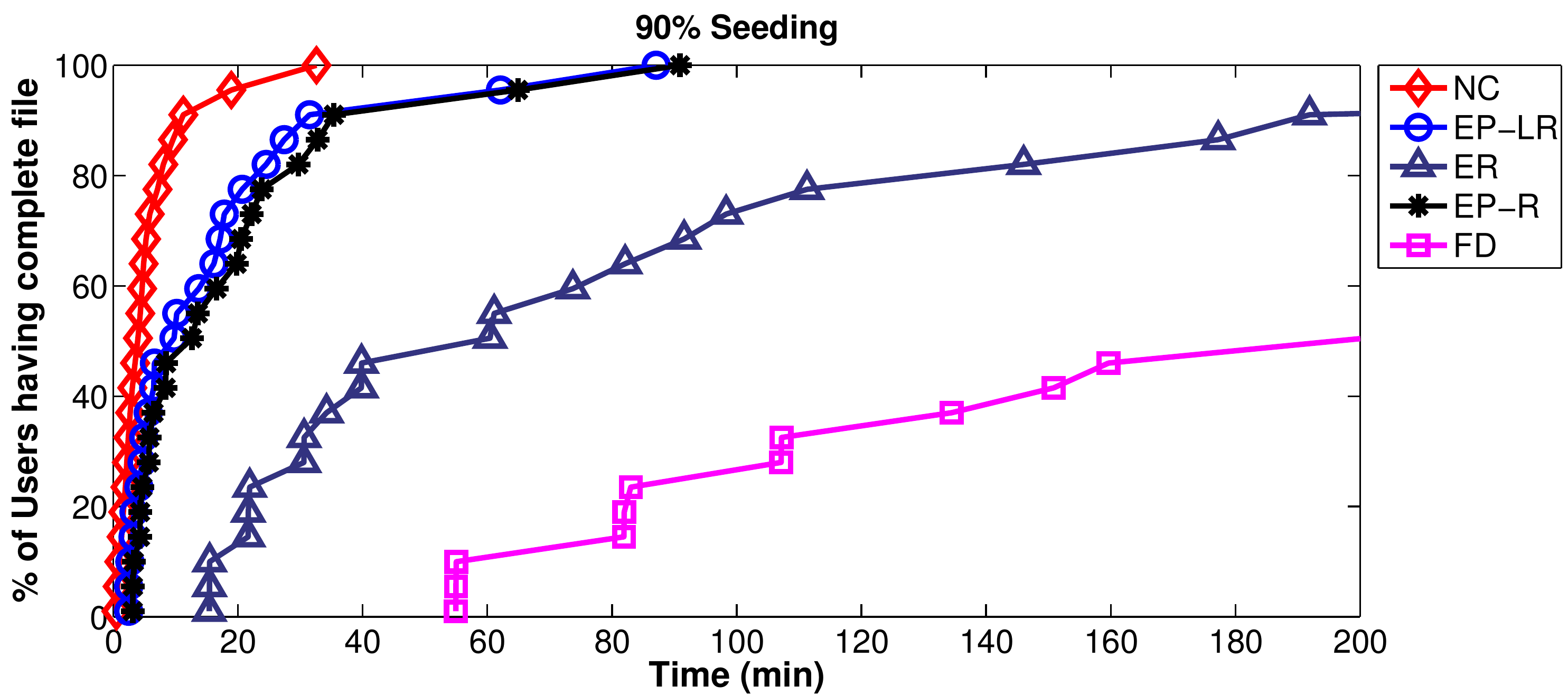}}
	\label{DSCb}
	\caption[Dissemination Schemes Comparison : 80\%,90\%,100\% Seeding]{cont.}
	\addtocounter{figure}{-1}
\end{figure}

\begin{figure}
    \centering

\subfigure[]{
	\includegraphics[angle=270,scale=0.45,  clip, type=pdf,ext=.pdf,read=.pdf]{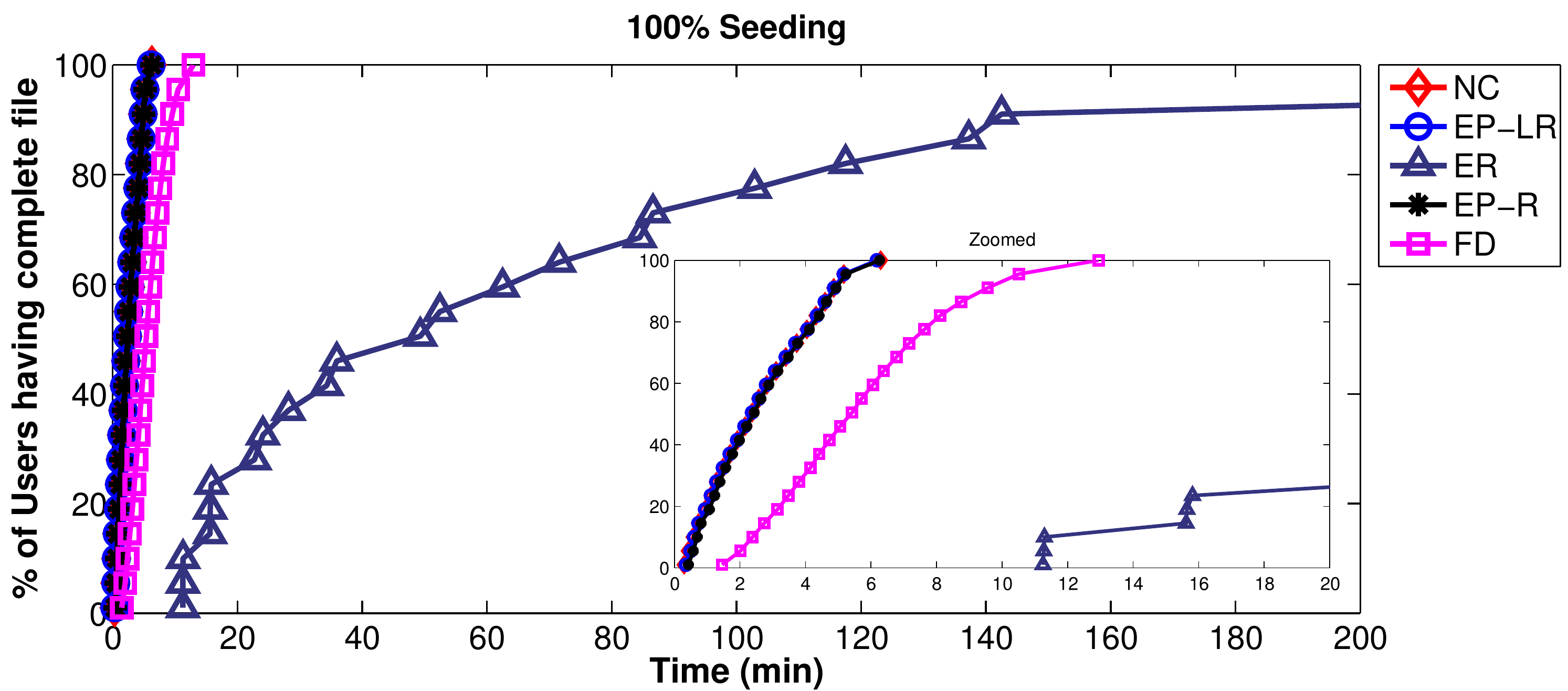}}
	\label{DSCc}

	\subfigure[]{
	\includegraphics[angle=270,scale=0.45,  clip, type=pdf,ext=.pdf,read=.pdf]{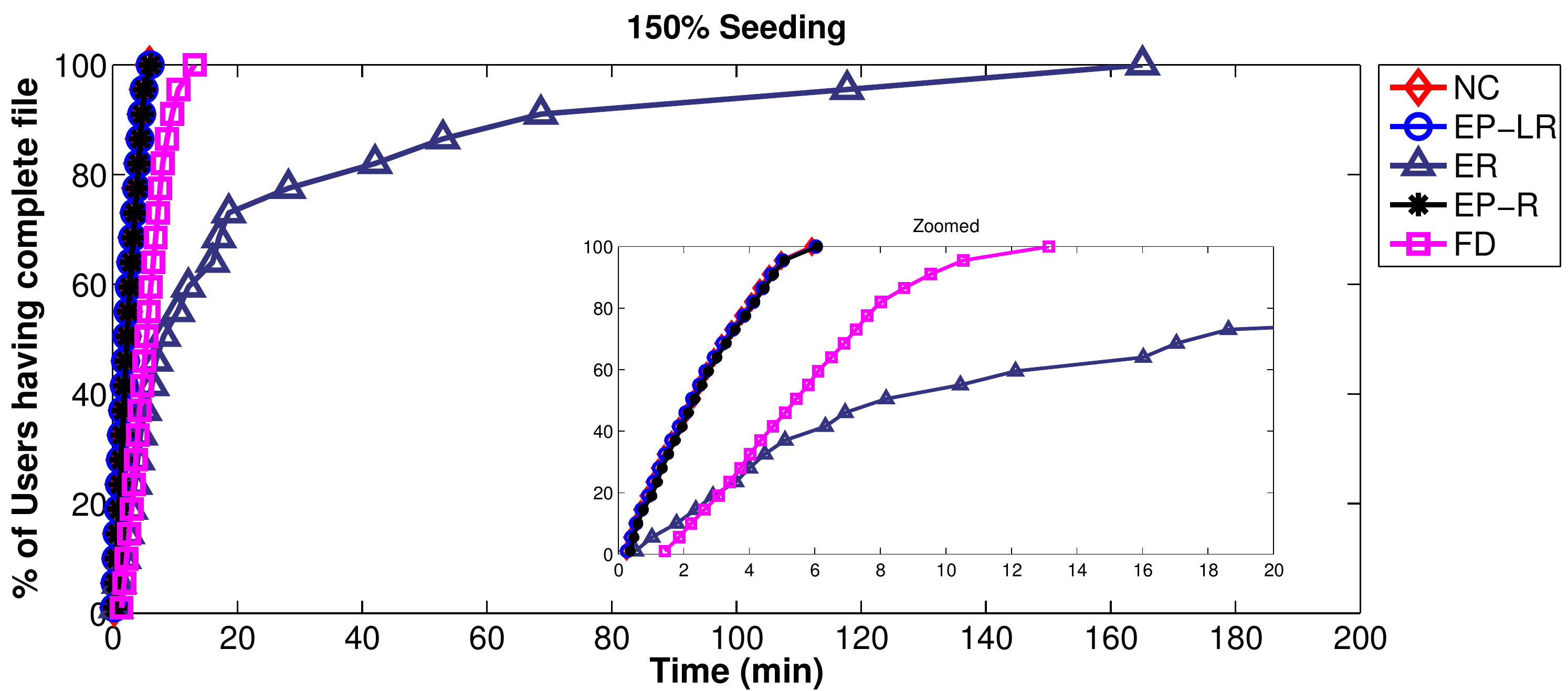}}
	\label{DSCd}
		
	\subfigure[]{
	\includegraphics[angle=270,scale=0.45,  clip, type=pdf,ext=.pdf,read=.pdf]{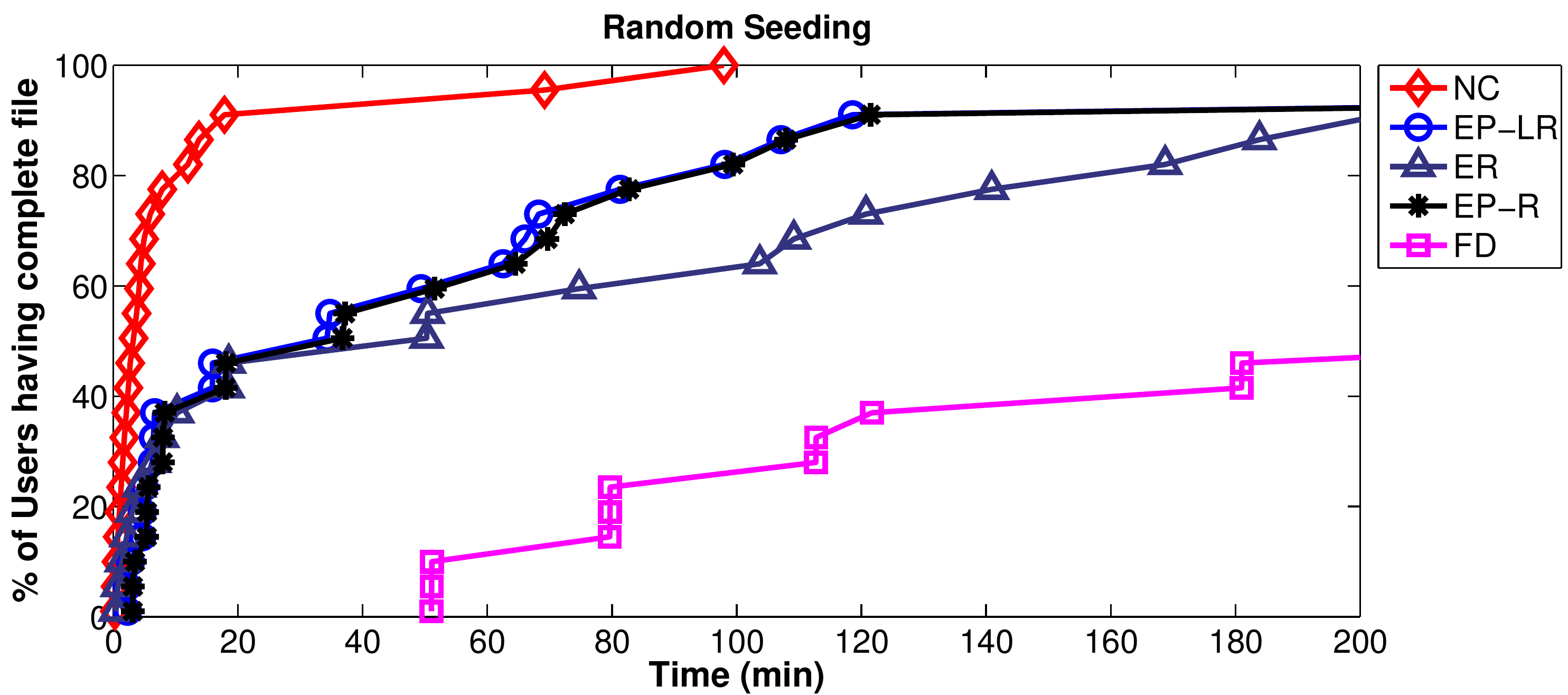}}
	\label{DSCe}
	
	\caption[Dissemination Schemes Comparison : 150\%, Random Seeding]{Comparison of the expected percentage of users that obtain the file over time for different Dissemination strategies. The comparison is shown for the following Seeding: (a) 80\% (b) 90\% (c) 100\% (d) 150\% and (e) Random.}
\label{DSC}

\end{figure}
\newpage
\subsection{Seeding Strategies Comparison}

In this section we compare the performance of different seeding schemes for each dissemination approach in terms of percentage of users that have received the file at any time \textit{t}. 

In all the dissemination approaches, seeding 100\% packets needed to recover the file has less delay as compared to other schemes as shown in Figure~\ref{SSC}. As the percentage of seeding reduced to 80\% and 90\% the performance degrades proportionally in terms of latency. Except for Erasure Coding, seeding more packets than required $(> 100\%)$ for file retrieval has no improvement on latency and the results are identical to the 100\% seeding strategy. With erasure coding as shown in Figure~\ref{SSC}(d), this difference in behaviour is because seeding more packets is helpful as the extra packets allow decoding to be performed within the community itself; the decoding process does not rely on inter-community packet transfer. For each dissemination approach, random seeding has by far the worst latency characteristics.  

\begin{figure}[H]
    \centering
	\subfigure[]{
	\includegraphics[angle=270,scale=0.45,  clip,type=pdf,ext=.pdf,read=.pdf]{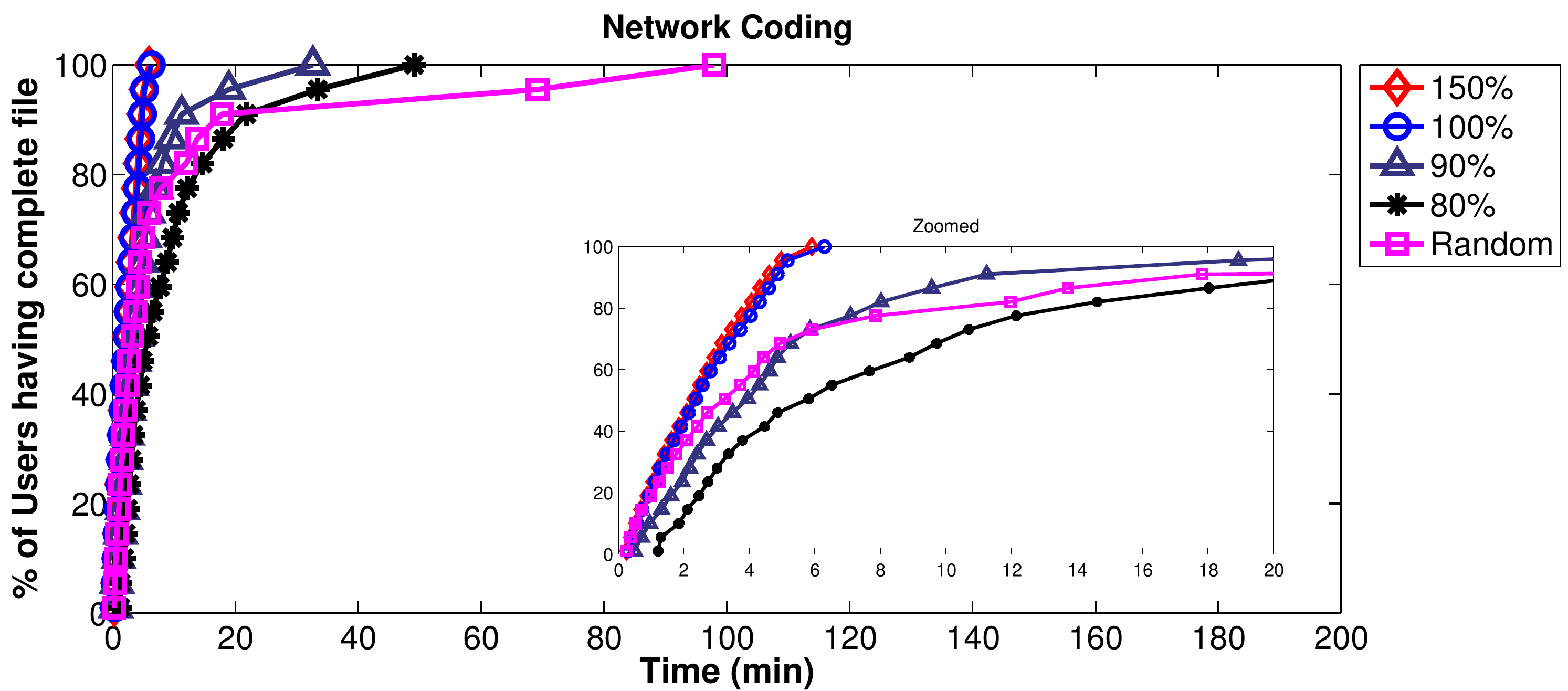}}
	\label{fig:Ch1_Exp1_Graph}
	

	\caption[Seeding Strategies Comparison for Network Coding]{cont.}	
		
\end{figure}	
	
\begin{figure}[H]
    \centering

\subfigure[]{
	\includegraphics[angle=270,scale=0.45,  clip, type=pdf,ext=.pdf,read=.pdf]{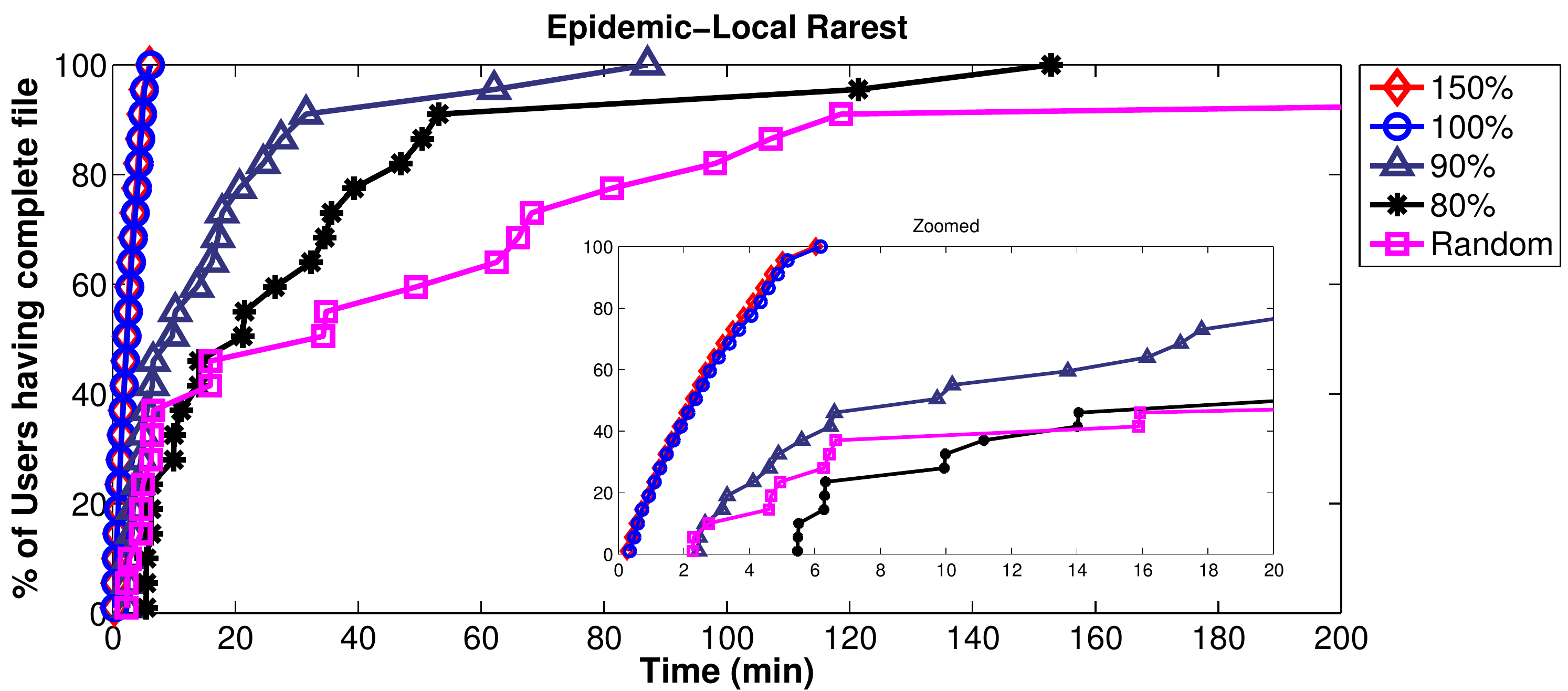}}
	\label{fig:Ch1_Exp1_Graphclus}

\subfigure[]{
	\includegraphics[angle=270,scale=0.45,  clip, type=pdf,ext=.pdf,read=.pdf]{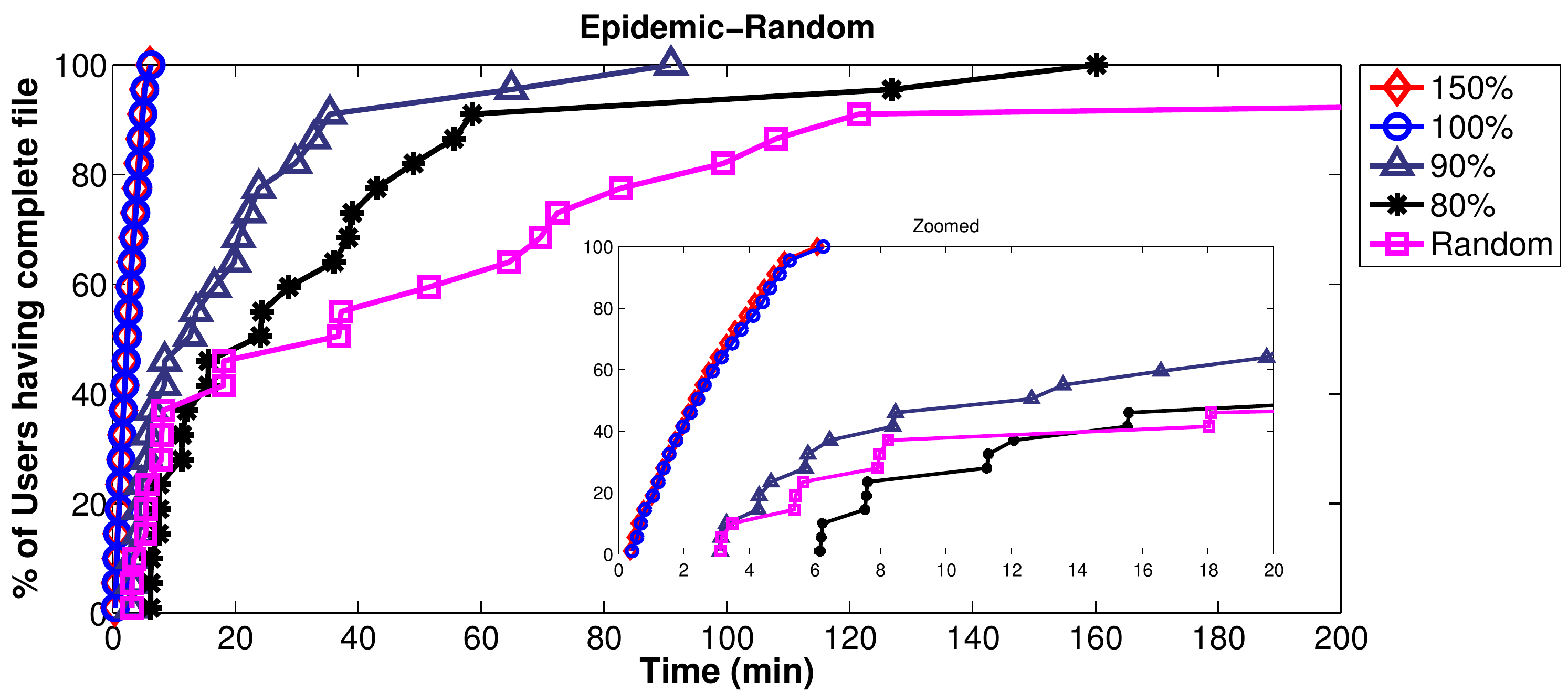}}

	\label{fig:Ch1_Exp1_Graphclus}

	\subfigure[]{
	\includegraphics[angle=270,scale=0.45,  clip, type=pdf,ext=.pdf,read=.pdf]{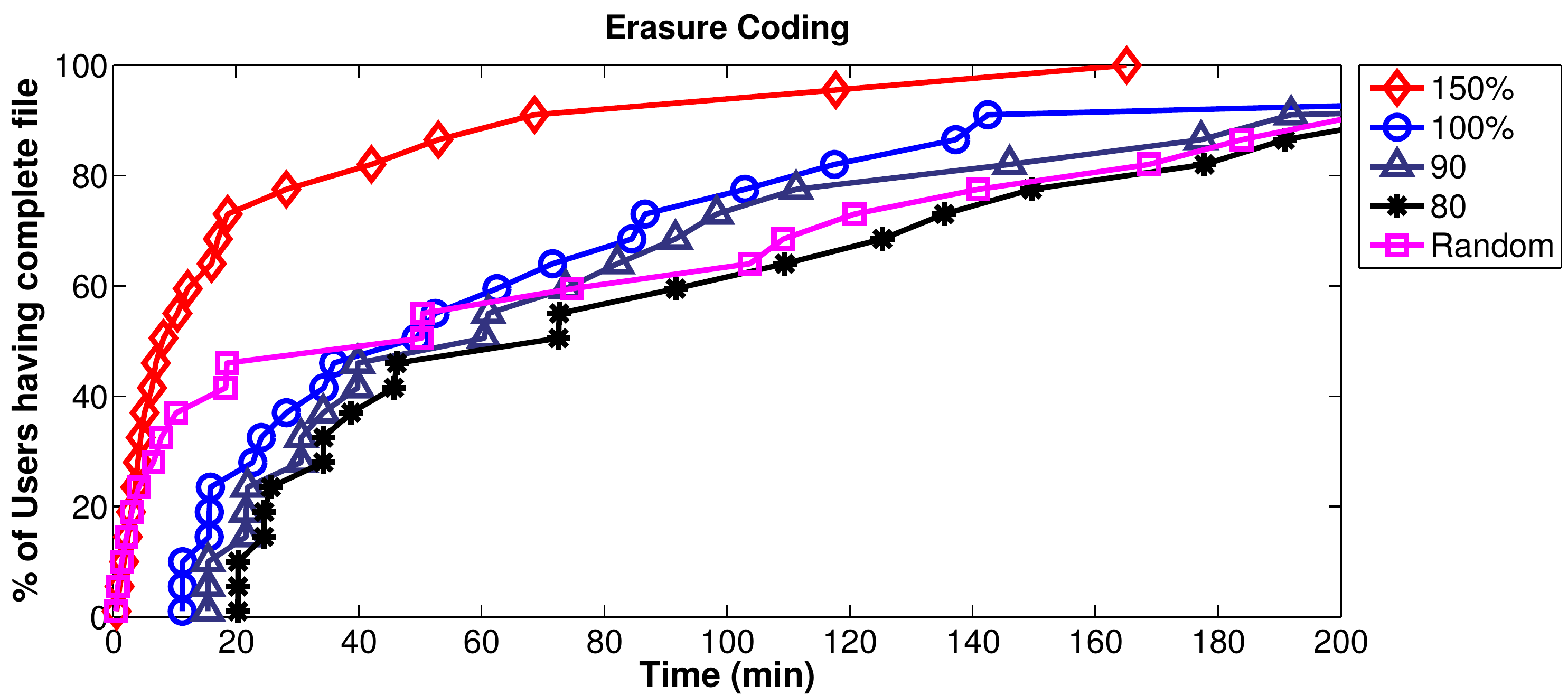}}
	\label{fig:Ch1_Exp1_Graphclus}
	\caption[Seeding Strategies Comparison for Epidemic-Local rarest, Epidemic-Random, Erasure Coding]{cont.}	
\end{figure}	
		
	\begin{figure}[H]
	\subfigure[]{
	\includegraphics[angle=270,scale=0.45,  clip, type=pdf,ext=.pdf,read=.pdf]{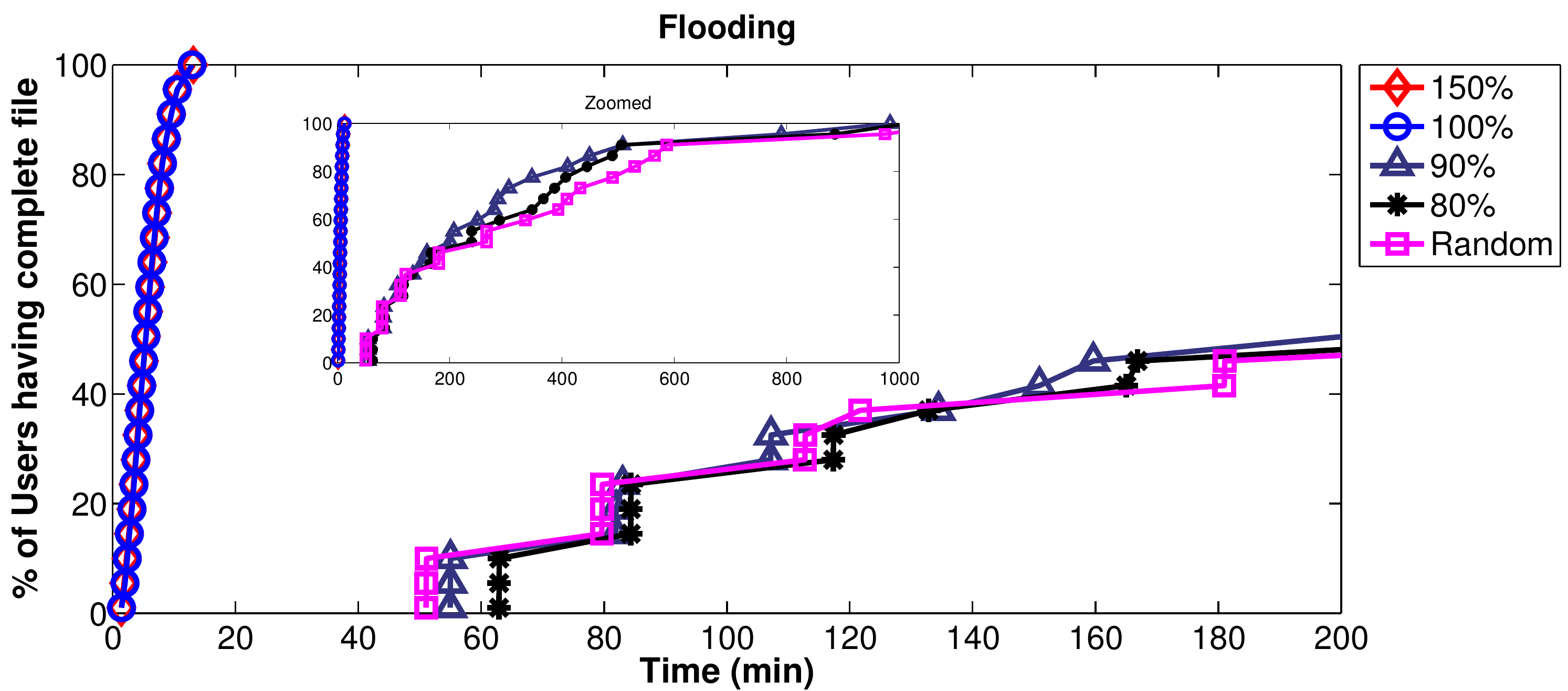}}
	\label{fig:Ch1_Exp1_Graphclus}
		
	\caption[Seeding Strategies Comparison for Flooding]{Comparison of the expected percentage of users that obtain the file over time for different community seeding strategies. The comparison is shown for the following Dissemination Strategies: (a) Network Coding (b) Epidemic-Local rarest (c) Epidemic-Random (d) Erasure Coding and (e) Flooding.}
\label{SSC}
\end{figure}

\subsection{Median and Standard Deviation of Finish Times}

Table~\ref{Table:MFT} summarizes the median finish time for different dissemination schemes with mentioned seeding strategies along with standard deviation.

\begin{table}[H]
	\centering
    	\begin{tabular}{|c|c|c|c|c|c|}
	\hline
					   	\multicolumn{6}{|c|}{Finish Times}    \\    \cline{1-6} 
	Seeding 	&	Network &	Epidemic	&	Epidemic     &	Erasure  & 	Flooding   \\ 
		
	\%           &  Coding 	&	(Local Rarest)	& (Random)	&	Coding 
		&  	 	   \\ \hline
	150\%	    &	5.84(0.44)	    &	6.00(0.4326)		&	6.07(0.41)		&	149.28(65.75)	    & 	12.85(0.89)   \\
	100\%		&	6.19(0.39)		&	6.05(0.38)		&	6.18(0.37)		&	494.13(118.00)		&	12.83(0.85)  \\
	90\%		&	26.92(20.98)		&	76.55(28.55)		&	83.50(21.77)		&	463.33(130.82)		&	988.03(164.68)  \\ 
	80\%		&	42(31.02)		&	149.5(30.24)		&	161.9(20.12)		&	508.4(90.85)		&	1001.30(115.74)  \\
	Random		&	84(36.03)		&	502.6(70.50)		&	483.4(85.73)		&	782.7(173.61)		&	1061.1(86.45)  	\\ \hline
    \end{tabular}
    \caption[Median and Standard Deviation of Finish Times for all users]{Table shows the comparison of median finish times (time for all users in the network to obtain the complete file) for all seeding across networks with different routing strategies . The standard deviation is shown in brackets.}
    \label{Table:MFT}
\end{table}


\subsection{Average Total Meetings and Transmissions}

This section compares the performance of the discussed content dissemination schemes with respect to the average total meetings and average total transmissions needed to disseminate the complete file in whole network. Table~\ref{Table:MFT2} presents the comparison for meetings in case of different seeding strategies averaged out over different runs of experiment for each dissemination scheme. The total meetings count is directly linked to the delay in the complete reception of file at all nodes.  It is evident that as the seeding percentage within the community is reduced, the number of meetings increase because the nodes have to rely on inter-community meetings for missing content and the meetings within communities are wasted. However this effect is not as prominent in network coding where more intra community meetings are still useful as already discussed.    

\begin{table}[H]
	\centering
    	\begin{tabular}{|c|c|c|c|c|c|}
	\hline
					   	\multicolumn{6}{|c|}{Average Total Meetings}    \\    \cline{1-6} 
	Seeding 	&	Network &	Epidemic	&	Epidemic     &	Erasure  & 	Flooding   \\ 
		
	\%           &  Coding 	&	(Local Rarest)	& (Random)	&	Coding 
		&  	 	   \\ \hline
	150\%	    &	32759	    &	33347		&	33623		&	915330	    & 	72864   \\
	100\%		&	34802		&	34169		&	34622		&	2752100		&	71771  \\
	90\%		&	180650		&	482940		&	504180		&	2619600		&	5458100  \\ 
	80\%		&	247650		&	847400		&	888290		&	2955200		&	5631100  \\
	Random		&	449620		&	2713250		&	2860839		&	4693000		&	6066800  	\\ \hline
    \end{tabular}
    \caption[Average Total Meetings]{Comparison Table for Average Total Meetings in Different Forwarding Schemes with Different Seeding Strategies.}
    \label{Table:MFT2}
\end{table}

Table~\ref{Table:MFT3} shows the comparison of schemes in terms of the average number of transmissions for complete retrieval of file at all nodes. This is the important performance indicator as the more transmissions the less is the lifetime of nodes and hence the life time of the opportunistic network. It is desirable to have a dissemination scheme that requires fewer transmissions for content distribution. It is important to note that in case of Epidemic routing with both local rarest or random selection there are no wasted transmissions. Content is only shared if the set $S_A-S_B$ is not empty otherwise the transmission opportunity is missed. As discussed previously for Erasure Coding (ER) the source encoded packets are shared based on the local rarest policy too. But since the unique encoded packets from server are more (different encoded packets can be generated by selecting different file packets) than the original file packets so with erasure coding the number of transmissions are more compared to epidemic dissemination because the probability for $S_A-S_B$ being empty is lesser in comparison to epidemic routing. With network coding every meeting yields unique packet because it represents the linear combination of already acquired packets so transmission is involved until the node retrieve the complete file as a result as depicted in Table 4.3 for 80\% and 90\% seeding strategies network coding has much more transmissions as compared to epidemic routing and erasure coding. In fact it it due to these increased transmissions that network coding performs better in terms of latency as discussed in Section 4.1. For each transmission there is a chance of innovative packet generation. 

\begin{table}[H]
	\centering
    	\begin{tabular}{|c|c|c|c|c|c|}
	\hline
					   	\multicolumn{6}{|c|}{Average Total Transmissions}    \\    \cline{1-6} 
	Seeding 	&	Network &	Epidemic	&	Epidemic     &	Erasure  & 	Flooding   \\ 
		
	\%           &  Coding 	&	(Local Rarest)	& (Random)	&	Coding 
		&  	 	   \\ \hline
	150\%	    &	14333	    &	14320		&	14320		&	24861	    & 	40102   \\
	100\%		&	15098		&	14880		&	14880		&	23842		&	40701  \\
	90\%		&	35263		&	14992		&	14992		&	24342		&	60680  \\ 
	80\%		&	63592		&	15104		&	15104		&	24573		&	60883  \\
	Random		&	46605		&	14880		&	14880		&	23947		&	61051  	\\ \hline
    \end{tabular}
    \caption[Average Total Transmissions]{Comparison Table for Average Total Transmissions in Different Forwarding Schemes with Different Seeding Strategies.}
    \label{Table:MFT3}
\end{table}

\subsubsection{Innovative and Non-Innovative Transmissions for Network Coding}

Table~\ref{Table:MFT4} indicates the number of non-innovative transmissions involved during data dissemination using network coding. Clearly non-innovative transmissions increase as seeding percentage reduces because with fewer seeded server encoded packets the probability of non-innovative packet generation increases.      

\begin{table}[H]
	\centering
    	\begin{tabular}{|c|c|c|}
	\hline
					   	\multicolumn{3}{|c|}{Network Coding}    \\    \cline{1-3} 
	Seeding 	&	Innovative Transmissions &	Non-Innovative Transmissions	   \\ \hline

	150\%	    &	14320	    &	13		   \\
	100\%		&	14880		&	218		  \\
	90\%		&	14992		&	20271		 \\ 
	80\%		&	15104		&	48488		 \\
	Random		&	14880		&	31725		 	\\ \hline
    \end{tabular}
    \caption[Innovative and Non-Innovative Transmissions for Network Coding]{Table shows Innovative and Non-Innovative Transmissions in Network Coding.} 
        \label{Table:MFT4}
\end{table}

\newpage
\subsection{Effect of strength of connections between communities}

\begin{figure}[H]
    \centering
	\subfigure[]{
	\includegraphics[angle=270,scale=0.45,  clip,type=pdf,ext=.pdf,read=.pdf]{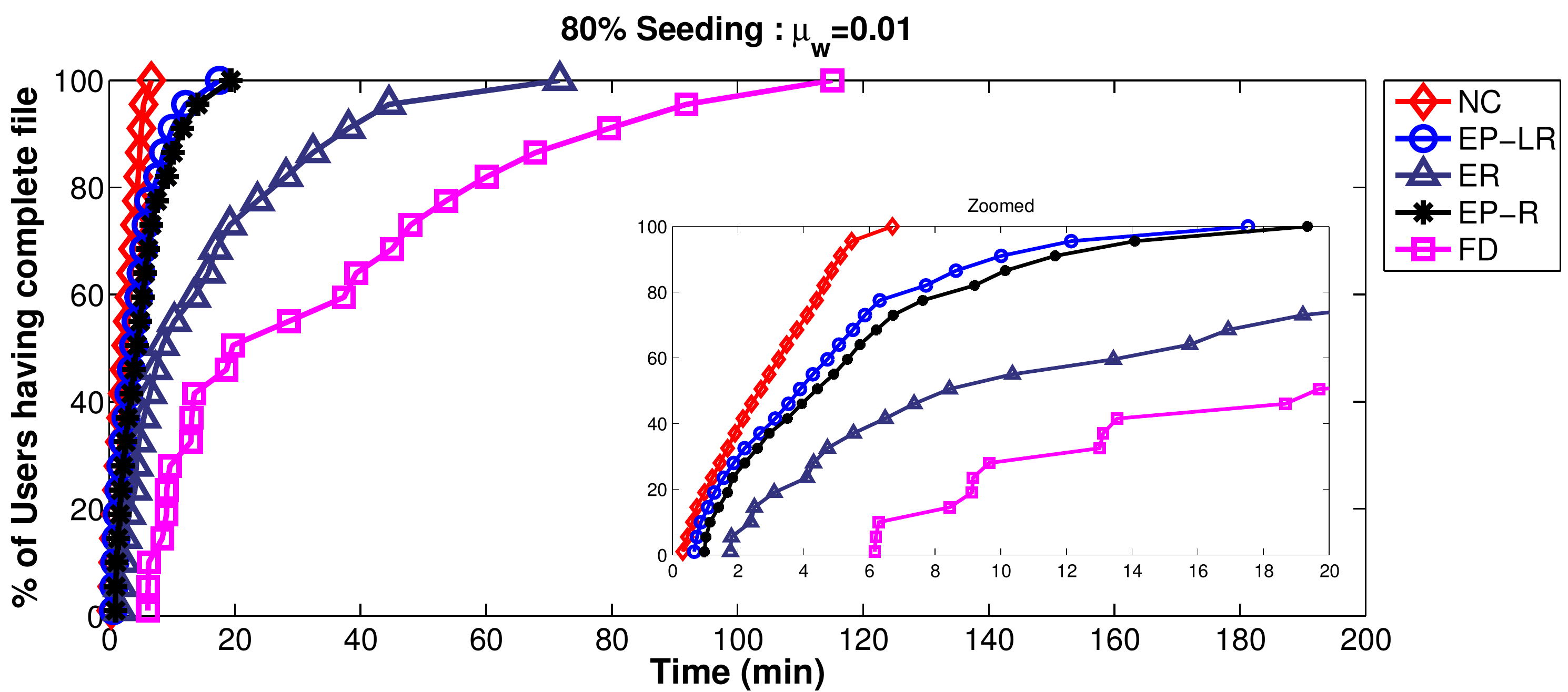}}
	\label{fig:Ch1_Exp1_Graph}
	

	
\subfigure[]{	
	\includegraphics[angle=270,scale=0.45,  clip, type=pdf,ext=.pdf,read=.pdf]{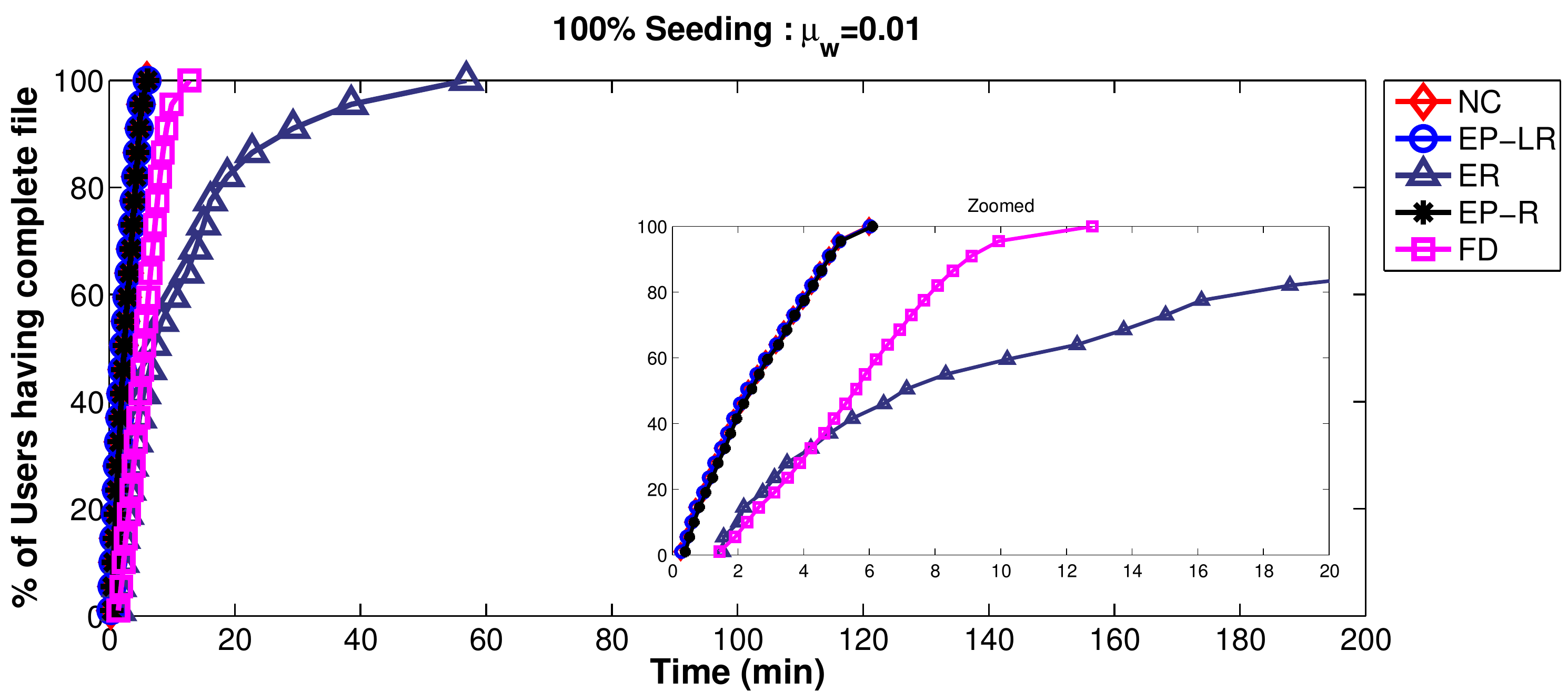}}
	\label{fig:Ch1_Exp1_Graphclus}

\subfigure[]{
	\includegraphics[angle=270,scale=0.45,  clip, type=pdf,ext=.pdf,read=.pdf]{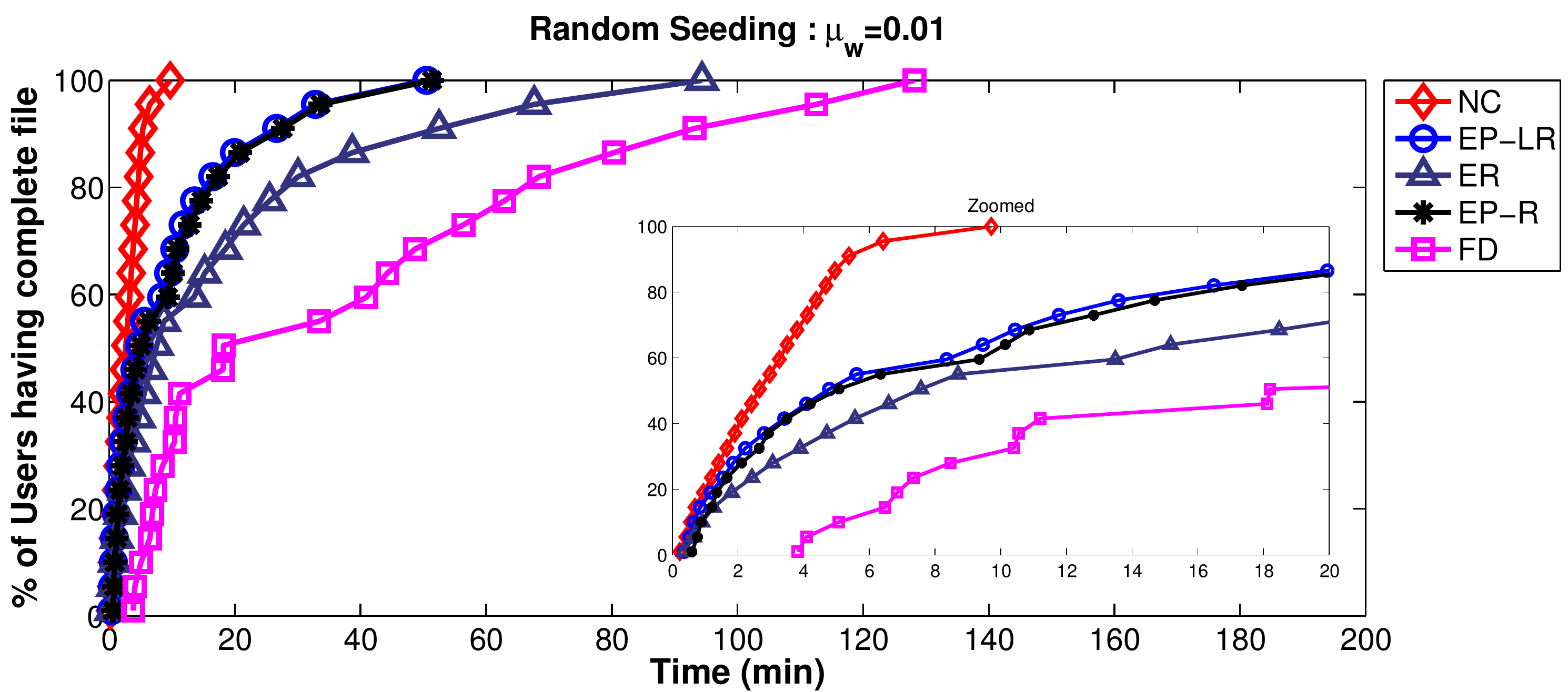}}
	\label{fig:Ch1_Exp1_Graphclus}	
\caption[Dissemination Schemes Comparison for $\mu_w=0.01$ and $\mu_t=0.1$ : 80\%, 100\% and Random Seeding]{Comparison of the expected percentage of users that obtain the file over time for different dissemination strategies with $\mu_w=0.01$. The comparison is shown for the following Seeding: (a) 80\% (b) 100\% (c) Random.} 
\label{DSC2}
\end{figure}

In the previous section we compared different content dissemination schemes in a mobile social network with strongly bounded communities i.e. the nodes within the same communities meet more frequently (stronger connections) than with nodes outside communities (weak connections). As discussed in section~\ref{SGG} the parameter $\mu_w$ controls the assignment of link weights within and outside communities for each node. Previously the comparison was performed by choosing a very small value of $\mu_w=0.001$, in this section we are increasing strength of inter community links by choosing $\mu_w=0.01$.\ As we discussed in section~\ref{SGG}, $\mu_w=0.001$ corresponds to the case where average inter-meeting interval between nodes of different communities is 10 times higher than average inter-meeting interval between nodes of same community. For $\mu_w=0.01$ the average inter-meeting interval between nodes belonging to same and different communities are same.\ For brevity we will compare the schemes for 80\%, 100\% and random seeding scenarios as the conclusion for 90\% and 150\% seeding is similar to that of 80\% and 100\% respectively. Figure~\ref{DSC2} shows the latency plots for different seeding strategies. 

Comparing Figure~\ref{DSC} and \ref{DSC2} for respective seeding schemes, it is evident that as the interaction among communities increases all the dissemination schemes start performing better because the time required to disseminate the file within the network decreases. This effect is more prominent for dissemination schemes and seeding strategies in which complete file retrieval is dependent on neighbouring communities. As the nodes meet with other nodes outside their communities more frequently the chances of retrieving missing packets from neighbouring communities increase. Network coding and erasure coding with 100\% seeding are insensitive to this parameter change as the latency is the same because of equal innovative packet generation opportunities irrespective of $\mu_w$. Table~\ref{Table:INITNC} shows that for network coding with 80\%, 90\% and random seeding cases the number of non-innovative transmissions decrease with $\mu_w=0.01$ as compared to $\mu_w=0.001$.  

It is however important to note that changing $\mu_w$ does not change the relative performance of dissemination schemes i.e. network coding still performs best in all seeding strategies, followed by epidemic routing - local rarest then random, erasure coding and flooding. Different seeding schemes also yield the same result i.e. seeding 100\% packets within each community is better than seeding 80\% packet or random seeding without considering communities.   
 
\begin{table}[H]
	\centering
    	\begin{tabular}{|c|c|c|c|}
	\hline
					   	\multicolumn{4}{|c|}{Network Coding}    \\    \cline{1-4} 
	$\mu_w$ &Seeding 	&	Innovative Transmissions &	Non-Innovative Transmissions	   \\ \hline

	     &150\%	    &	14320	    &	13		   \\
	     & 100\%	&	14880		&	218		  \\
 0.001   &90\%		&	14992		&	20271		 \\ 
	     &80\%		&	15104		&	48488		 \\
	     &Random	&	14880		&	31725		 	\\ \hline
    
         &150\%	    &	14320	    &	12		   \\
	     &100\%		&	14880		&	105		  \\
 0.01    &90\%		&	14992		&	692		 \\ 
	     &80\%		&	15104		&	1970		 \\
	     &Random	&	14880		&	2352		 	\\ \hline
    
    \end{tabular}
    \caption[Innovative and Non-Innovative Transmissions for Network Coding for different $\mu_w$]{Table compares Innovative and Non-Innovative Transmissions in Network Coding for different $\mu_w$.} 
        \label{Table:INITNC}
\end{table}

\subsection{Effect of varying the size of the network}
In this section we will compare the performance of content dissemination schemes by varying the size of the network. So far we have considered a network of 200 nodes. Now we check different content dissemination schemes on a network of 400 nodes. The graph of this network has the same parameters as the graph with 200 nodes generated in Section~\ref{DSC}. The two graphs are shown in figure 4.10. The number of communities are 14 in both graphs and the network with 400 nodes comprises of communities with more variable sizes as compared to 200 nodes. For comparison we will consider 80\%, 100\% and random seeding strategies. Figure~\ref{fig:CEGclus} contains the latency plots.

\begin{figure}[H]
    \centering
	\subfigure[]{
	\includegraphics[angle=270,scale=0.45,  clip,type=pdf,ext=.pdf,read=.pdf]{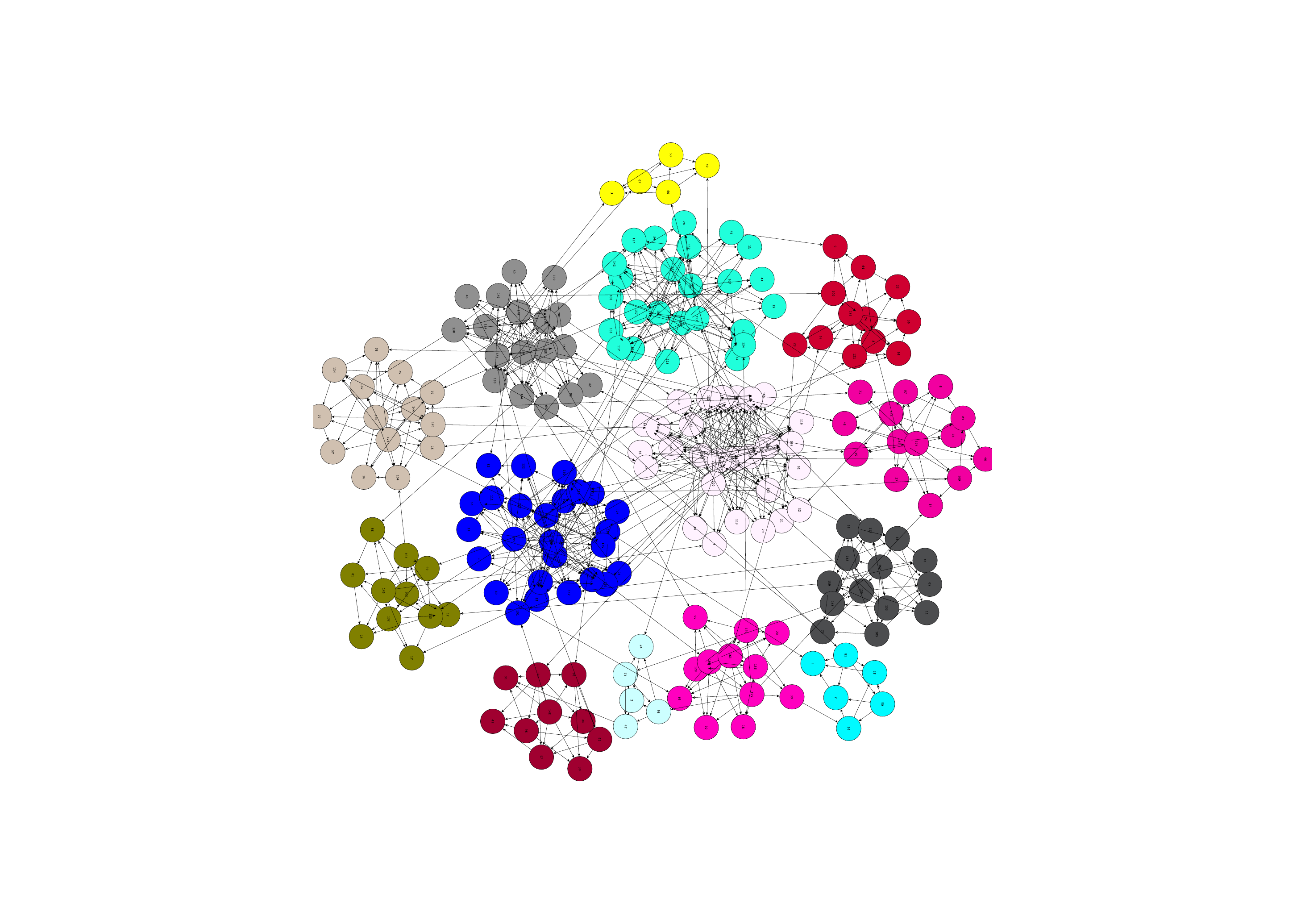}}
	\label{fig:Ch1_Exp1_Graph}
	
\subfigure[]{	
	\includegraphics[angle=270,scale=0.45,  clip, type=pdf,ext=.pdf,read=.pdf]{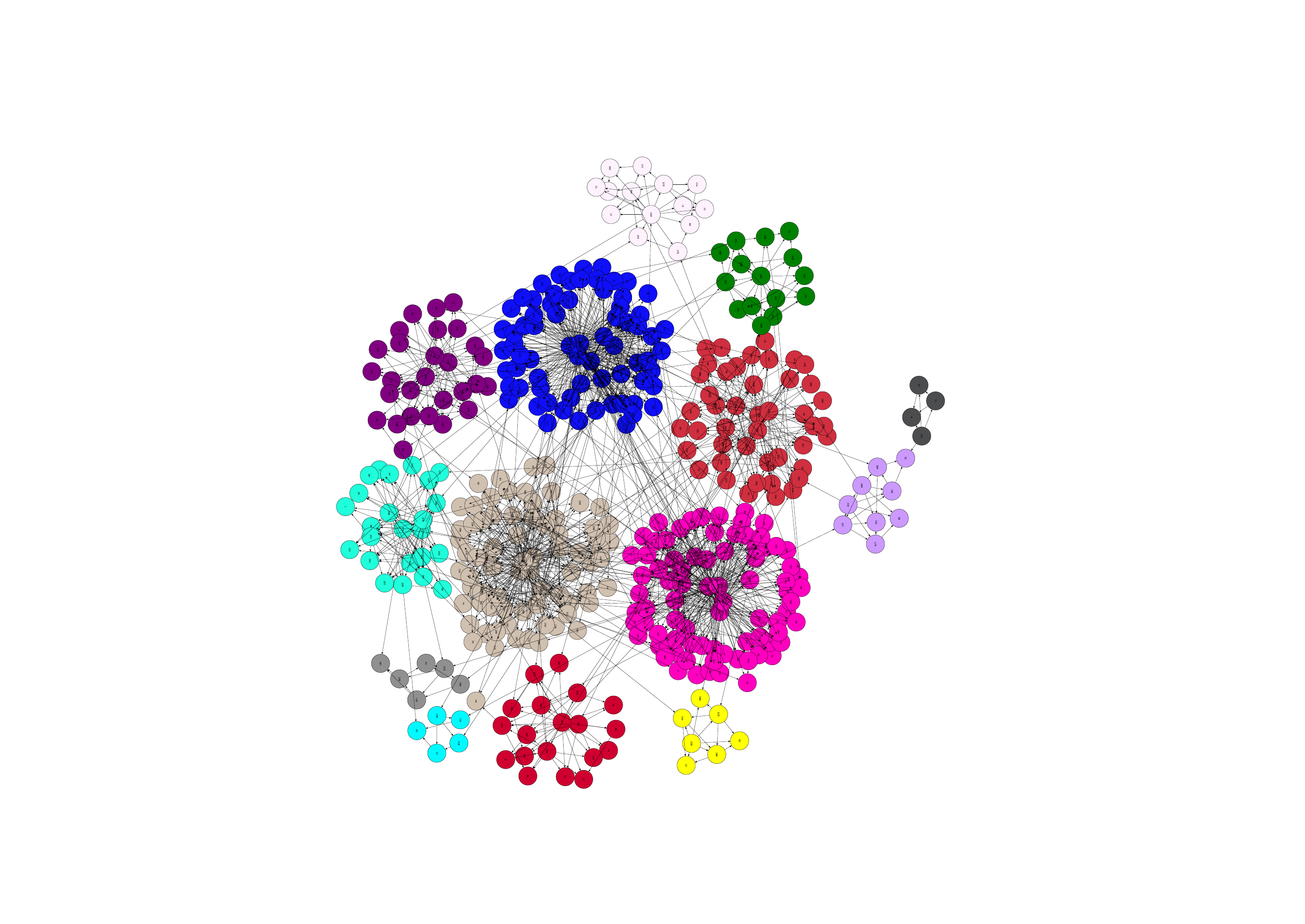}}
	\label{fig:Ch1_Exp1_Graphclus}
	\caption[Graph for (a) 200 users (default parameters) and (b) 400 users]{Graph for (a) 200 users (default parameters) and (b) 400 users. Communities are distinguished by different colors and users meet if they have an edge between them.}
\end{figure}	

\begin{figure}[H]
    \centering
	\subfigure[]{
	\includegraphics[angle=270,scale=0.35,  clip,type=pdf,ext=.pdf,read=.pdf]{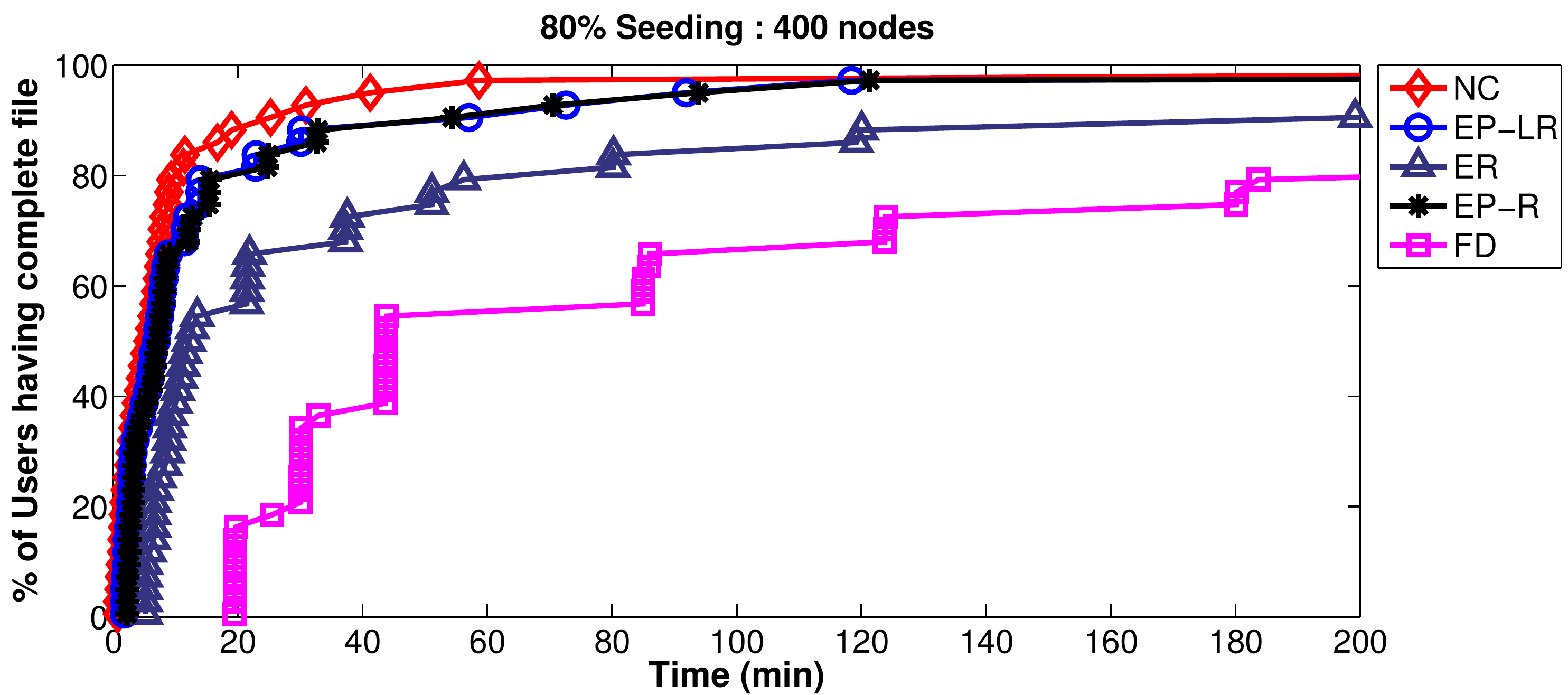}}
	\label{fig:Ch1_Exp1_Graph}
	
\subfigure[]{
	\includegraphics[angle=270,scale=0.35,  clip, type=pdf,ext=.pdf,read=.pdf]{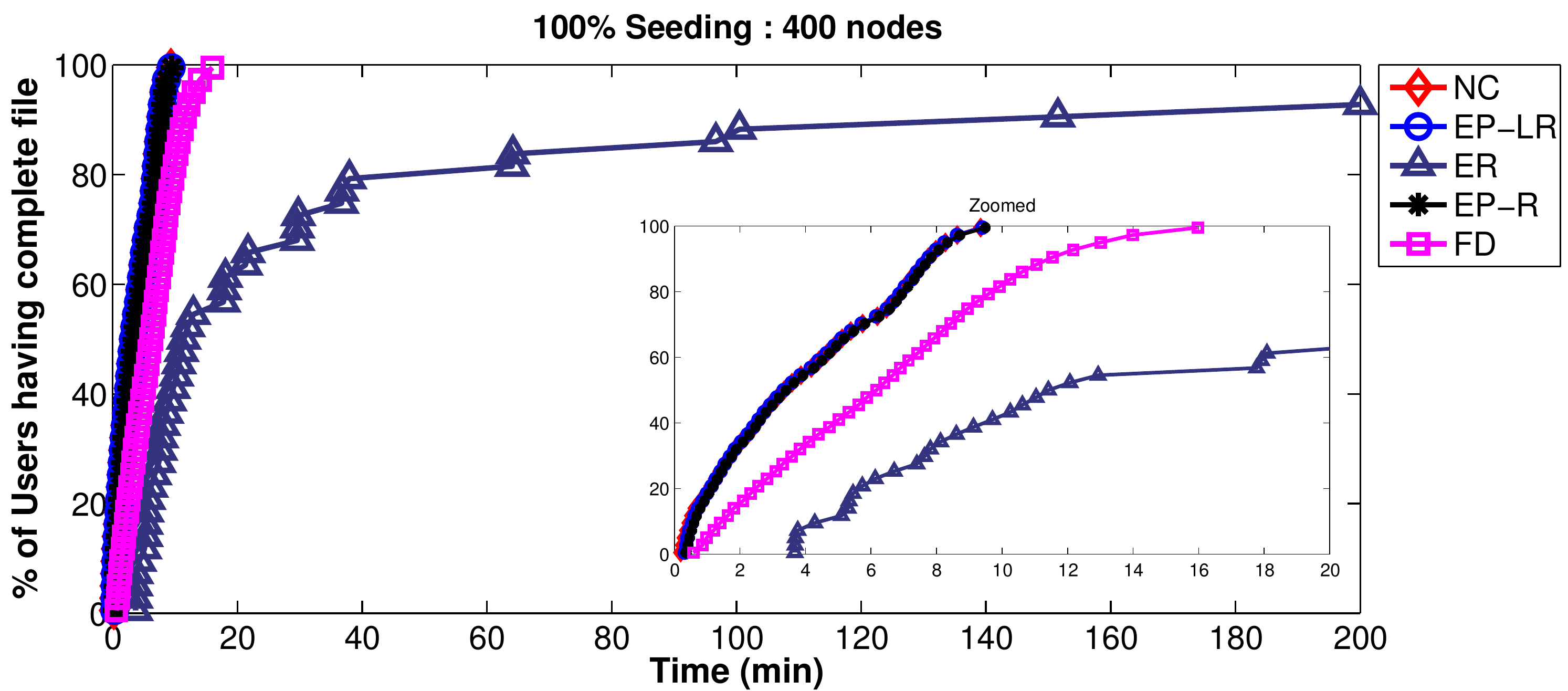}}
	\label{fig:Ch1_Exp1_Graphclus}
	
\subfigure[]{
	\includegraphics[angle=270,scale=0.35,  clip, type=pdf,ext=.pdf,read=.pdf]{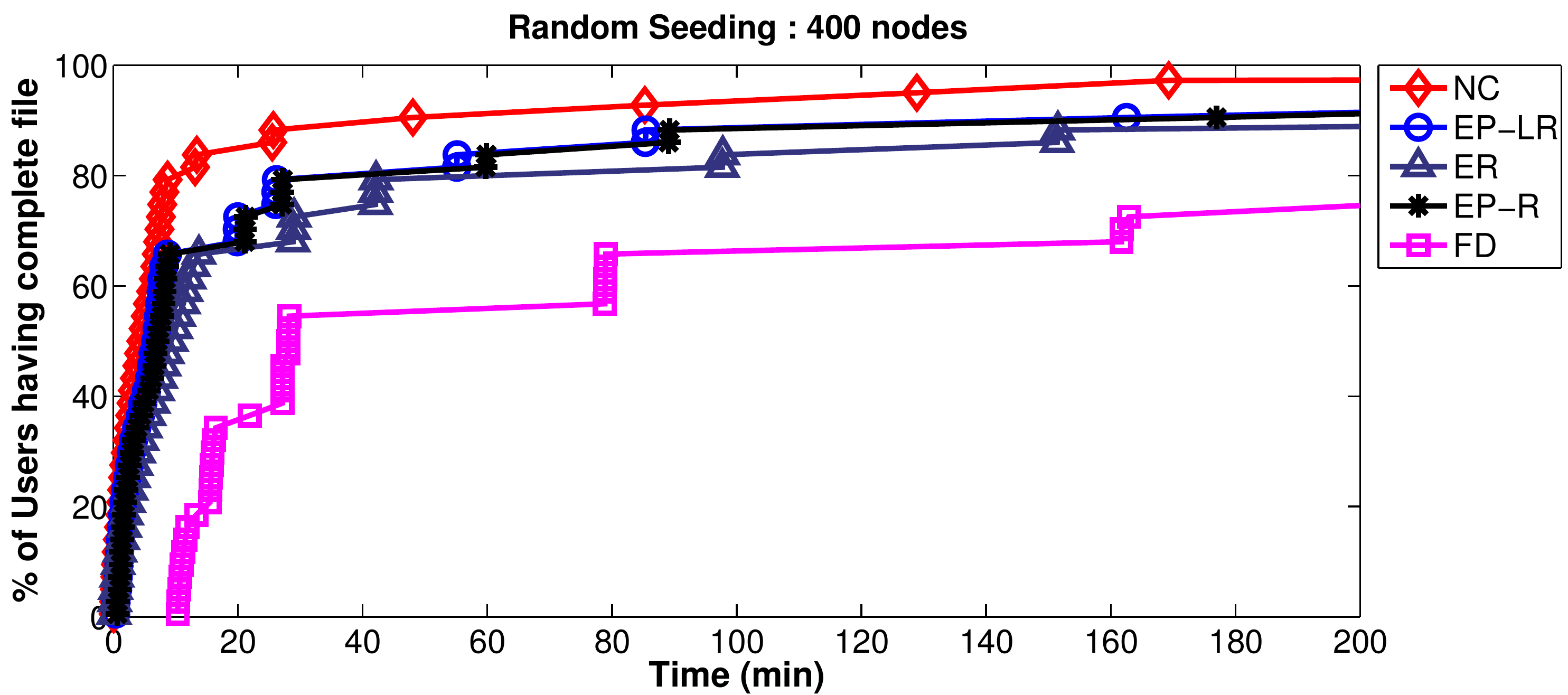}}
\caption[Dissemination Schemes Comparison for different seeding in 400 nodes network]{Comparison of the expected percentage of users that obtain the file over time for different Dissemination strategies in 400 nodes Network. The comparison is shown for the following Seeding: (a) 80\% (b) 100\% (c) Random.}	
	\label{fig:CEGclus}

\end{figure}

Comparing Figure~\ref{DSC} and \ref{fig:CEGclus}, it is evident that as the number of nodes increase to 400 in the network, all the schemes take more time to disseminate the complete file as compared to 200 nodes network. The community size increases with 400 nodes, and the nodes in the larger communities take more time to receive the necessary packets to retrieve the complete file due to an increase in the hop distance for users at the periphery of their communities. Increasing the number of nodes also increases the number of transmissions as expected. The performance comparison of dissemination schemes and seeding strategies remains the same. Network Coding still performs best, followed by epidemic routing, erasure coding and flooding. For 100\% seeding flooding performs better than erasure coding because unlike erasure coding it does not depend on packet retrieval from neighbouring communities. Seeding the complete file to each community is beneficial as compared to seeding fewer than the required file packet count within each community or seeding randomly without considering communities. Table~\ref{Table:MFTE} summarizes median finish times along with standard deviation for 200 and 400 nodes networks.

\begin{table}[H]
	\centering
    	\begin{tabular}{|c|c|c|c|c|c|c|}
	\hline
					   	\multicolumn{7}{|c|}{Finish Times}    \\    \cline{1-7} 
Nodes & Seeding 	&	Network &	Epidemic	&	Epidemic     &	Erasure  & 	Flooding   \\ 
		
&	\%           &  Coding 	&	(LR)	& (Rnd)	&	Coding 
		&  	 	   \\ \hline
	& 150\%	    &	5.8(0.4)	    &	6.00(0.43)		&	6.1(0.4)		&	149.2(65.7)	    & 	12.85(0.8)   \\
	& 100\%		&	6.1(0.3)		&	6.1(0.3)		&	6.2(0.3)		&	494.1(118)		&	12.83(0.8)  \\
200	& 90\%		&	26.9(20.9)		&	76.5(28.5)		&	83.5(21.7)		&	463.3(130.8)		&	988(164.6)  \\ 
	& 80\%		&	42(31.1)		&	149.5(30.2)		&	161.9(20.1)		&	508.4(90.8)		&	1001.3(115.7)  \\
	& Random		&	84(36)		&	502.6(70.5)		&	483.4(85.7)		&	782.7(173.6)		&	1061.1(86.4)  	\\ \hline

& 150\%	    &	9.9(0.5)	    &	9.8(0.3)		&	9.9(0.3)		&	480.89(620.9)	    & 	17.22(0.91)   \\
	& 100\%		&	9.7(0.4)		&	10(0.4)		&	10.1(0.4)		&	3604(1403)		&	17.3(1)  \\
400	& 90\%		&	140(85)		&	511(176.7)		&	445(185.5)		&	3394.2(2237)		&	3608.7(657.1)  \\ 
	& 80\%		&	308(284.8)		&	946.1(301.8)		&	981.3(275.1)		&	3995.1(1344.2)		&	4196.4(751.4)  \\
	& Random		&	1331(198.2)		&	4224.2(489)		&	4348.4(456.4)		&	6707(1079.7)		&	4809(624.2)  	\\ \hline

    \end{tabular}
    \caption[Median and Standard Deviation of Finish Times comparison for different network sizes]{Table shows the comparison of median finish times (time for all users in the network to obtain the complete file) for all seeding across networks with different routing strategies and network size. The standard deviation is shown in brackets.}
    \label{Table:MFTE}
\end{table}
\newpage

\subsection{Effect of number of clusters in network}
\begin{figure}[H]
    \centering
\subfigure[]{
	\includegraphics[angle=270,scale=0.48,  clip,type=pdf,ext=.pdf,read=.pdf]{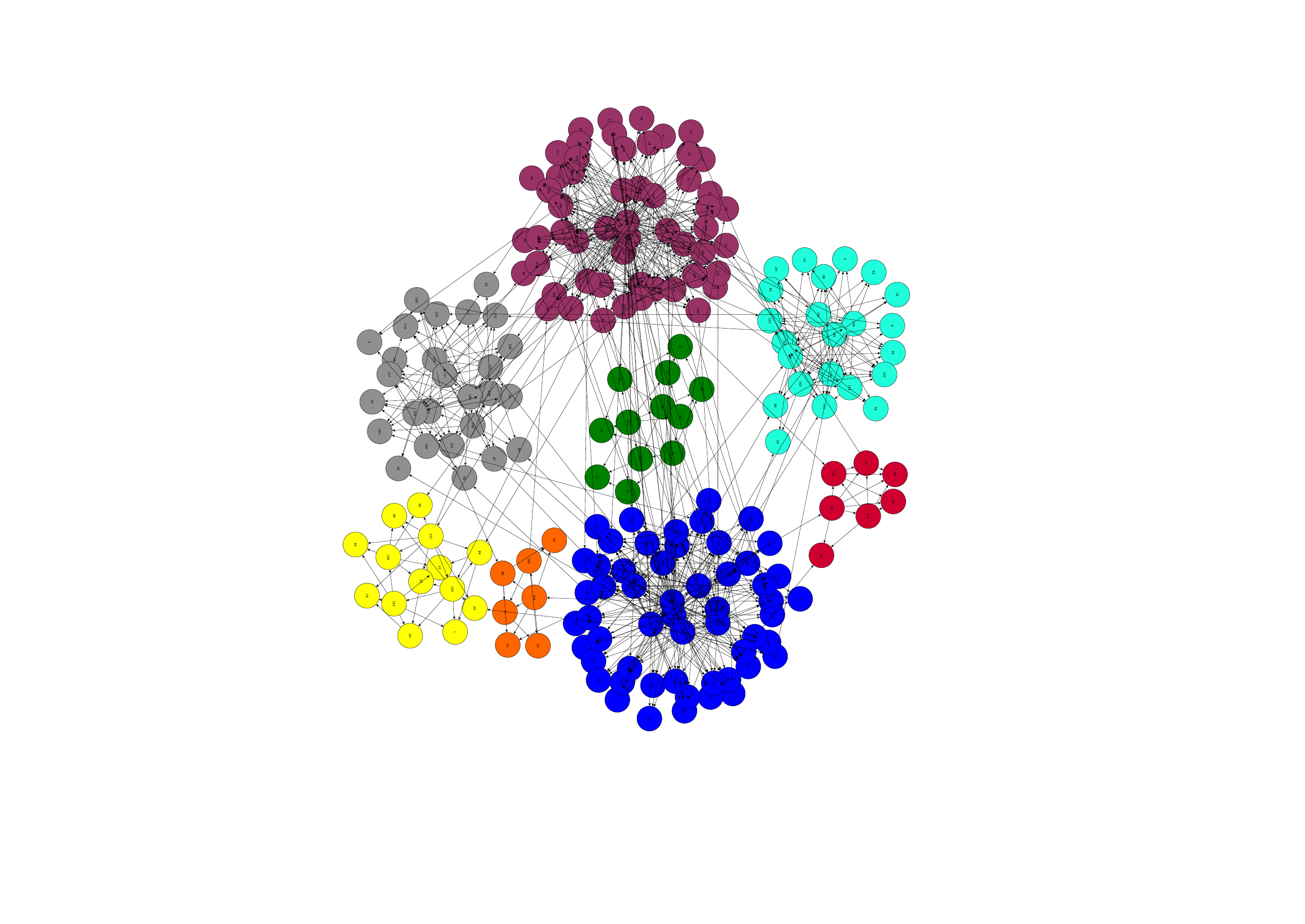}}
	\label{fig:CEG8}
	
\subfigure[]{
	\includegraphics[angle=270,scale=0.48,  clip, type=pdf,ext=.pdf,read=.pdf]{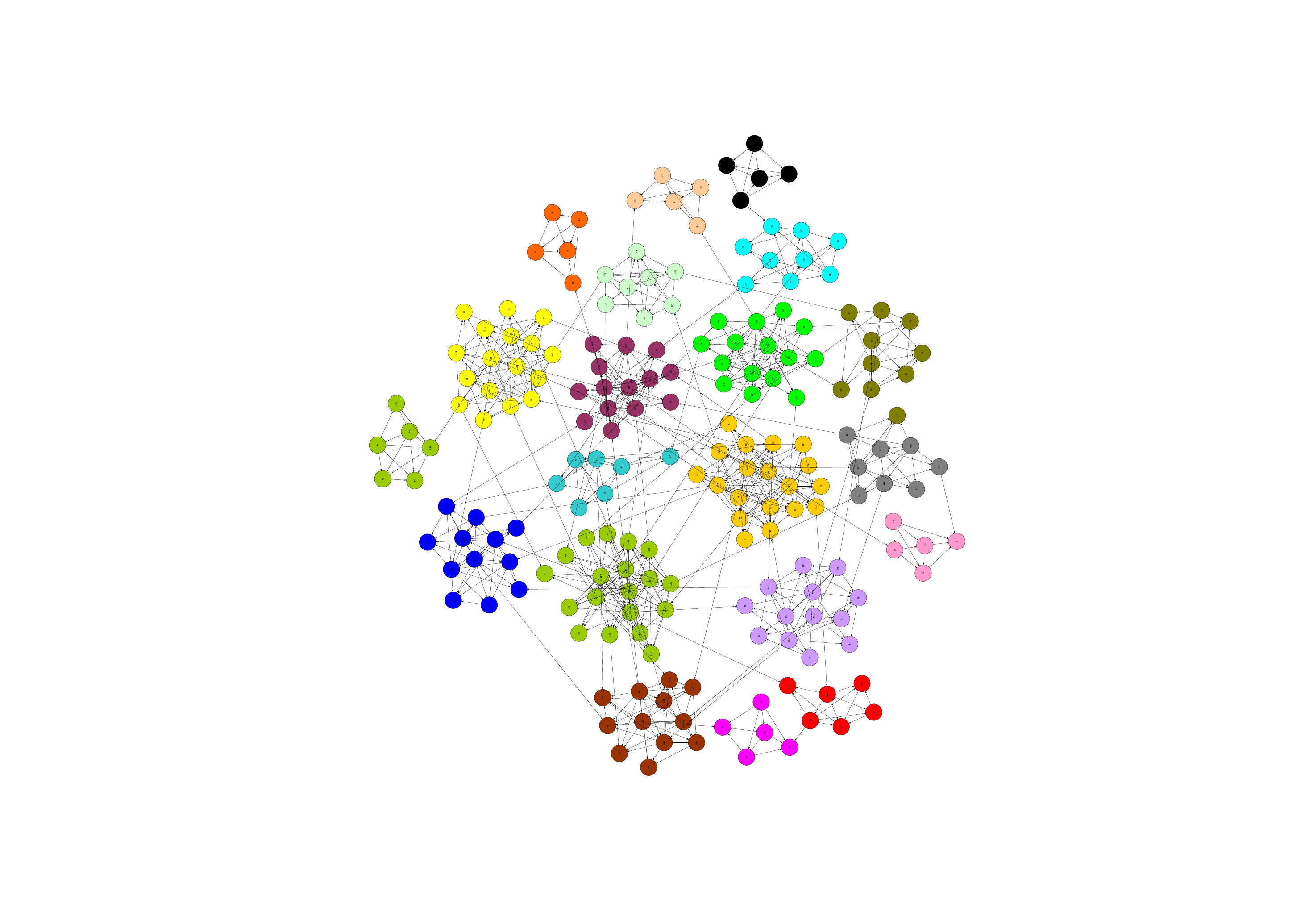}}
	\label{fig:CEG20}
	\caption[Graph having 8 communities in (a) and 20 communities in (b) in 200 nodes network]{Graph with (a) 8 communities  (b) 20 communities in 200 nodes network.}
	\label{fig:CEG8_20}
\end{figure}

In this section we compare the performance of dissemination schemes by varying the number of communities in the network. So far we have considered 14 communities, now we will vary the number of communities to 8 and 20 for the 200 node network and compare the dissemination schemes for these three community counts. The rest of the graph parameters remain the same. The graph with 14 communities is already presented in Figure~\ref{fig:CH1_Exp1_individual_nodes_150percent0001_a} while Figure~\ref{fig:CEG8_20} represents the graphs with 8 and 20 communities. 

\begin{figure}[H]
    \centering
	\subfigure[]{
	\includegraphics[angle=270,scale=0.35,  clip,type=pdf,ext=.pdf,read=.pdf]{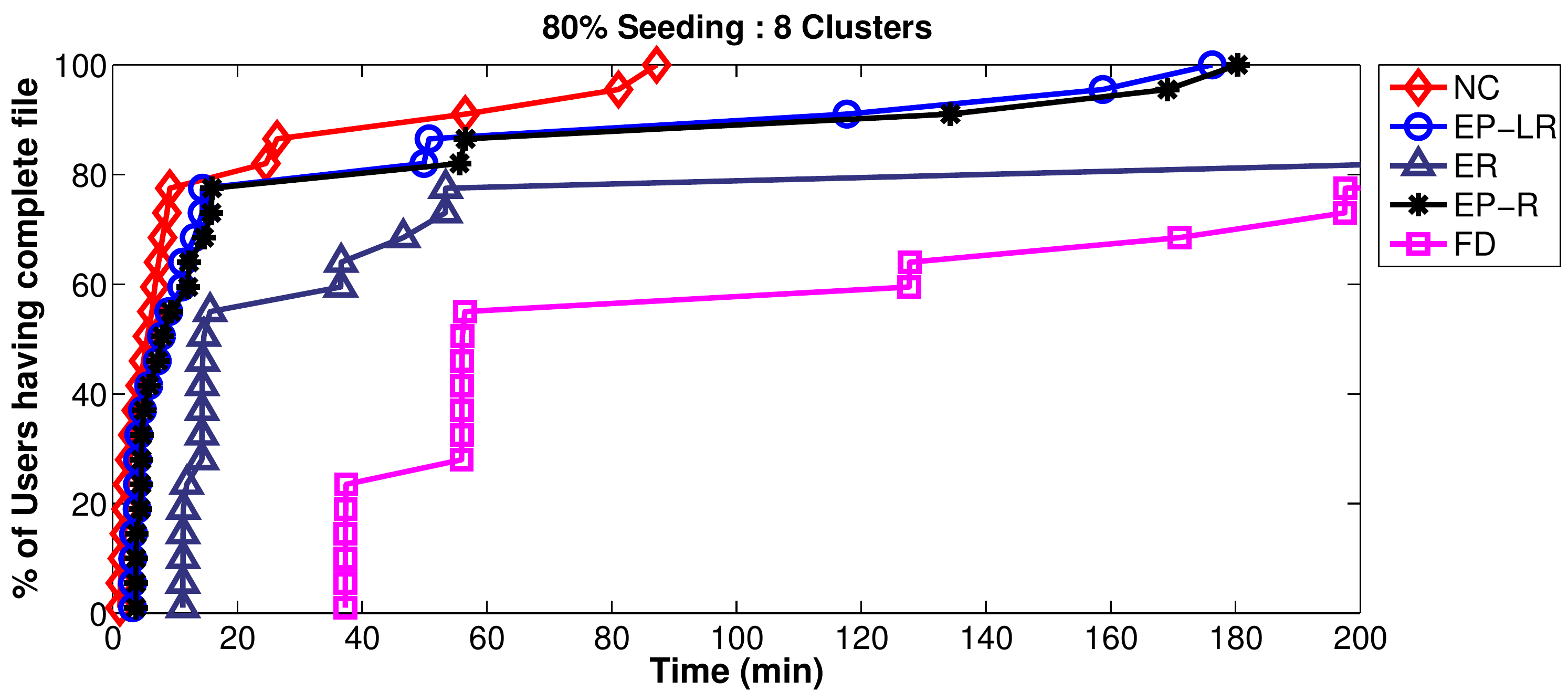}}
	\label{fig:Ch1_Exp1_Graph}
	
\subfigure[]{
	\includegraphics[angle=270,scale=0.35,  clip, type=pdf,ext=.pdf,read=.pdf]{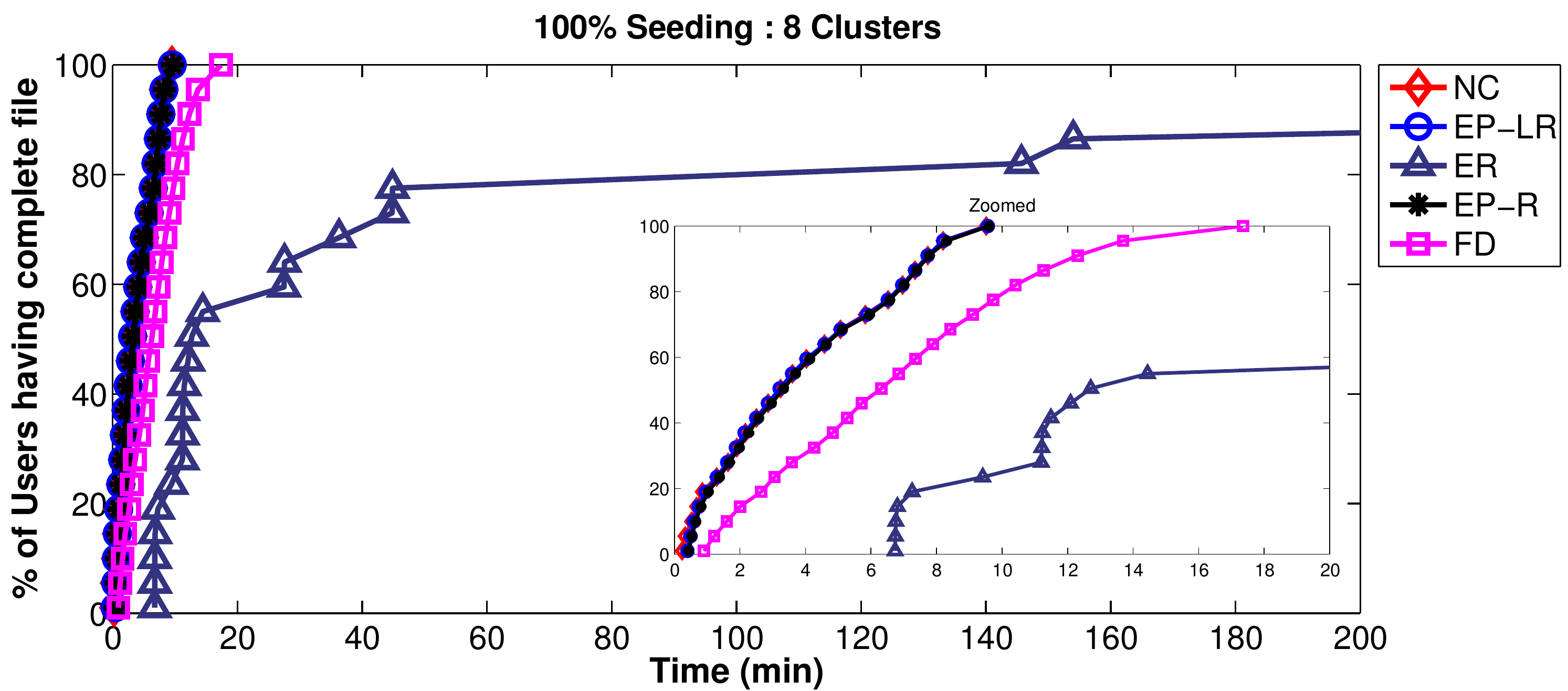}}
	\label{fig:Ch1_Exp1_Graphclus}
	
\subfigure[]{	
	\includegraphics[angle=270,scale=0.35,  clip, type=pdf,ext=.pdf,read=.pdf]{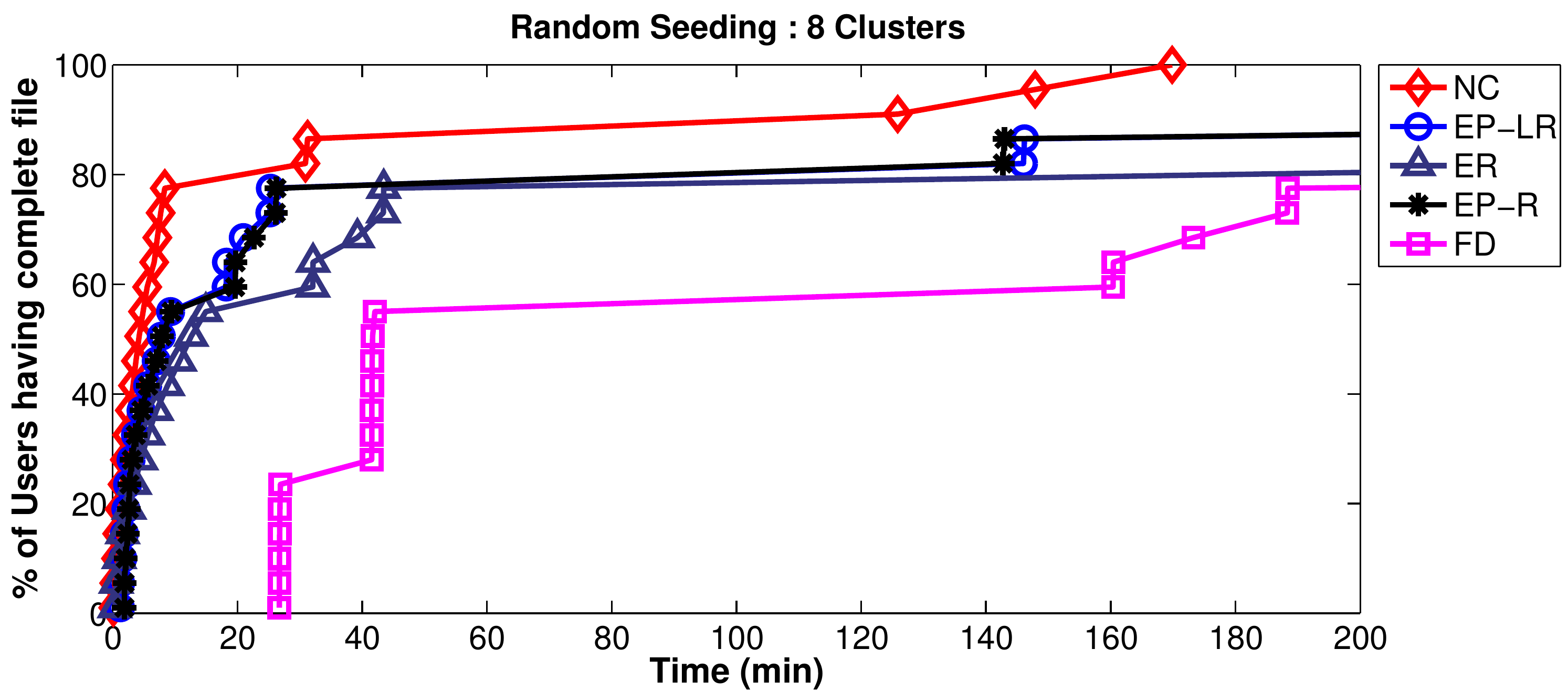}}
\caption[Dissemination Schemes Comparison for different seeding in 8 Clusters 200 nodes network]{Comparison of the expected percentage of users that obtain the file over time for different Dissemination strategies in 8 Clusters 200 nodes Network. The comparison is shown for the following Seeding: (a) 80\% (b) 100\% (c) Random.}	
	\label{fig:CEG1}
	\end{figure}

\begin{figure}[H]
    \centering
	\subfigure[]{
	\includegraphics[angle=270,scale=0.35,  clip,type=pdf,ext=.pdf,read=.pdf]{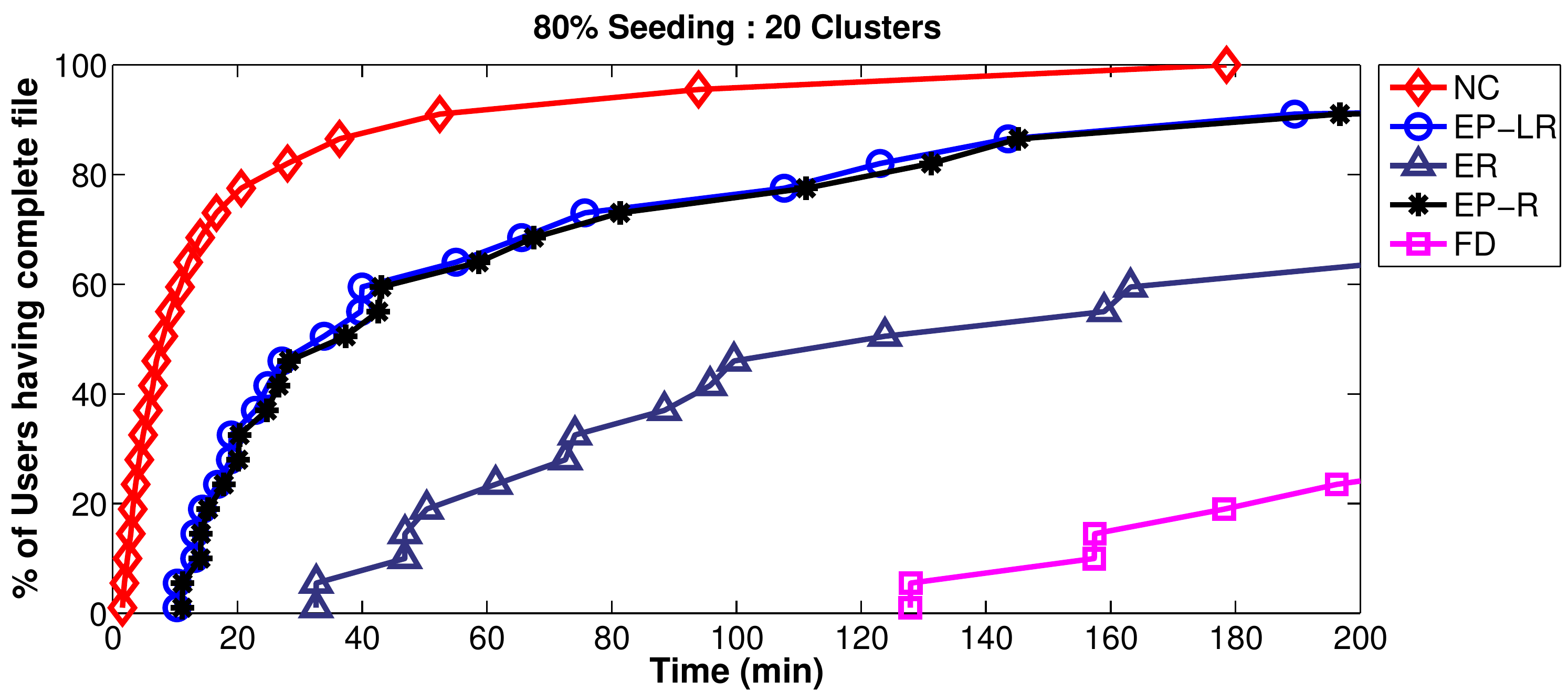}}
	\label{fig:Ch1_Exp1_Graph}
	
\subfigure[]{
	\includegraphics[angle=270,scale=0.35,  clip, type=pdf,ext=.pdf,read=.pdf]{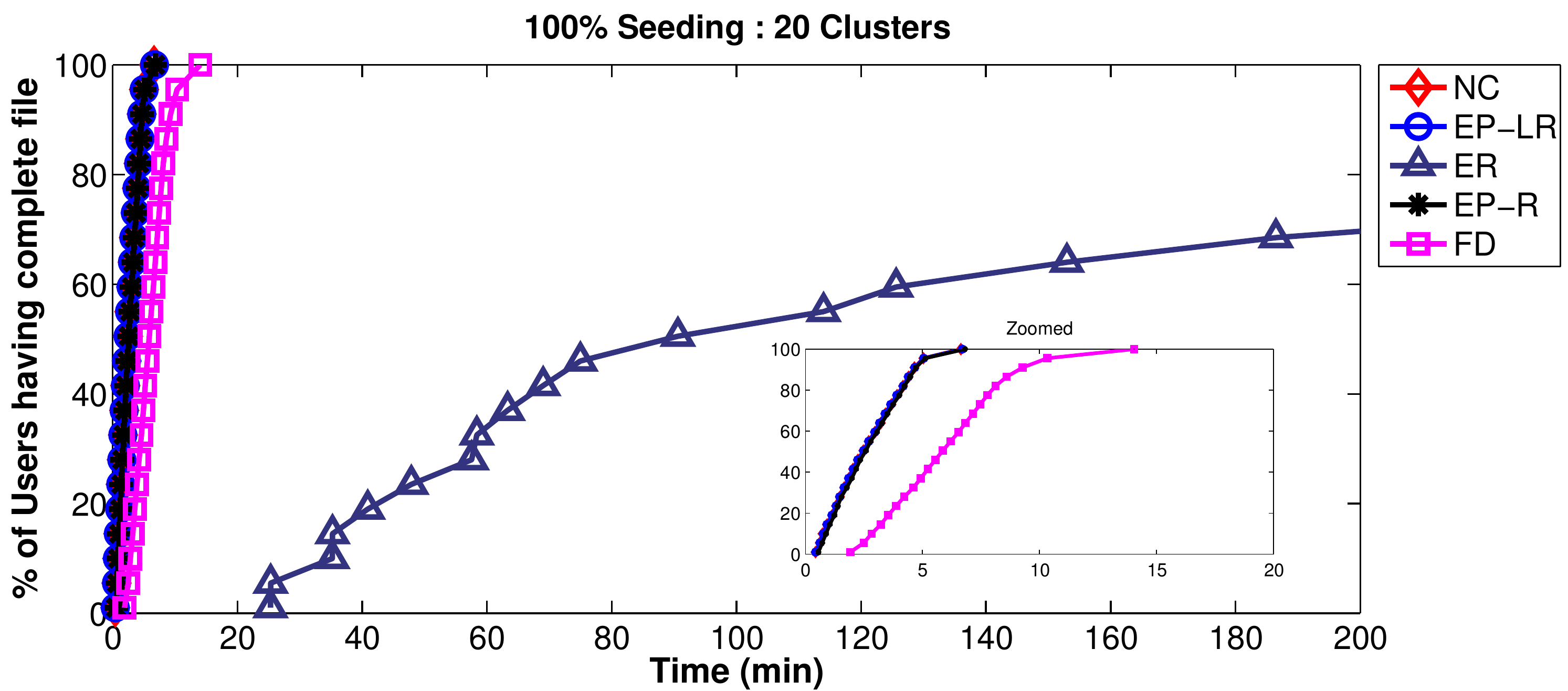}}
	\label{fig:Ch1_Exp1_Graphclus}
	
\subfigure[]{
	\includegraphics[angle=270,scale=0.35,  clip, type=pdf,ext=.pdf,read=.pdf]{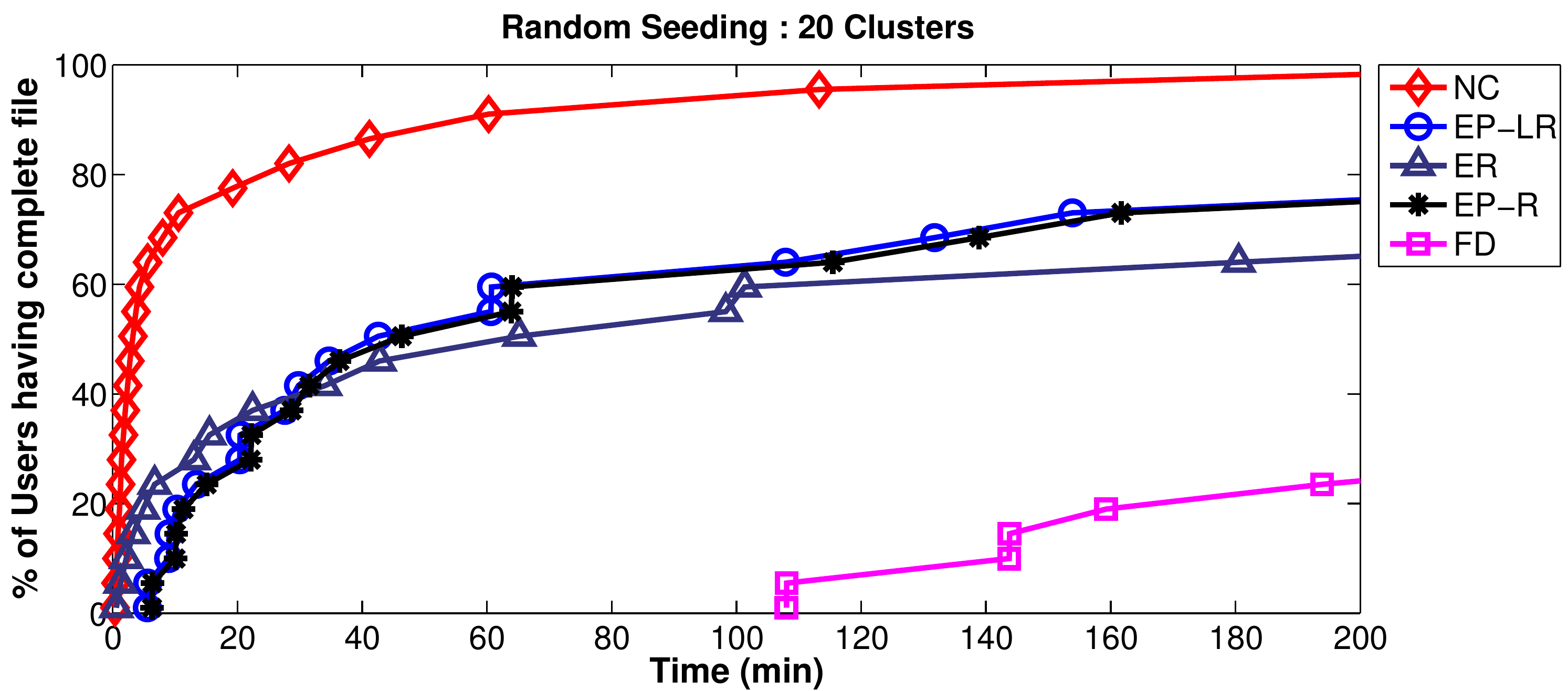}}
	
\caption[Dissemination Schemes Comparison for different seeding in 20 Clusters 200 nodes network]{Comparison of the expected percentage of users that obtain the file over time for different Dissemination strategies in 20 Clusters 200 nodes Network. The comparison is shown for the following Seeding: (a) 80\% (b) 100\% (c) Random.}	
	\label{fig:CEG2}	
	
\end{figure}

Comparing Figures~\ref{DSC}, \ref{fig:CEG1} and \ref{fig:CEG2} it is evident that for 100\% seeding the performance of network coding, epidemic routing - Local Rarest and Random is almost same for 14 and 20 clusters, however with 8 clusters all the mentioned schemes took more time. The reason is that for a fixed sized network as the number of communities is reduced the number of users within each community increases and as a result it takes more time to distribute the content due to an increase in the hop distance for users at the periphery of their communities. Also with 8 communities the copies of the content within the network are less because we are distributing content per community in all seeding strategies. The same behaviour is also observed with 150\% seeding strategy. 

For the other dissemination schemes and seeding strategies, it is observed that the performance of dissemination schemes with 14 clusters is better than that of 8 and 20 clusters. The reason can be understood by visually inspecting the respective graphs of 8,\ 14 and 20 clusters. In the graph of 20 clusters it is evident that most of the small size communities are linked to only one community by single link which is causing longer delay because the missing packets must be brought into the community through this link alone. With 8 communities there are more inter-community links but the communities are larger and fewer copies of the content are seeded to the network.

With 14 clusters the effects of large community sizes and fewer inter community links are balanced i.e the communities are of reasonable size and there exist sufficient links to neighbouring communities.             

Again it is evident that seeding a single copy of content within each community is enough regardless of the number of communities. Even the performance of 80\% seeding is better than randomly seeding within the network which further motivates community based seeding. 

Table 4.7 summarizes the mean finish time for 8, 14 and 20 communities in network of 200 nodes along with standard deviation in brackets.  

\begin{table}[H]
	\centering
    	\begin{tabular}{|c|c|c|c|c|c|c|}
	\hline
					   	\multicolumn{7}{|c|}{Finish Times}    \\    \cline{1-7} 
Comm & Seeding 	&	Network &	Epidemic	&	Epidemic     &	Erasure  & 	Flooding   \\ 
		
&	\%           &  Coding 	&	(LR)	& (Rnd)	&	Coding 
		&  	 	   \\ \hline
	& 150\%	    &	9.38(0.35)		&	9.42(0.43)		&	9.56(0.41)		&	242.6(201.3)		&	17.26(0.94)  \\
    & 100\%		&	9.59(0.34)	    &	9.51(0.37)		&	9.51(0.36)		&	600.3(158)	    & 	17.26(0.87)   \\
8	& 90\%		&	41(12.73)		&	96.9(20.28)		&	85.9(23.43)		&	476.7(112.1)		&	1216.1(155)  \\ 
	& 80\%		&	79.8(27.84)		&	170.2(38.52)		&	178.5(20.89)		&	731.7(186.6)		&	1234.4(183.09)  \\
	& Random		&	172.7(32.9)		&	526.5(54.1)		&	547.3(65.4)		&	970(208.1)		&	1416.2(112.18)  	\\ \hline

	& 150\%	    &	5.84(0.44)	    &	6.00(0.43)		&	6.07(0.41)		&	149.2(65.7)	    & 	12.85(0.89)   \\
	& 100\%		&	6.19(0.39)		&	6.05(0.38)		&	6.18(0.37)		&	494.1(118.0)		&	12.83(0.85)  \\
14	& 90\%		&	26.92(20.98)		&	76.55(28.55)		&	83.50(21.77)		&	463.3(130.8)		&	988.03(164.68)  \\ 
	& 80\%		&	42(31.02)		&	149.5(30.24)		&	161.9(20.12)		&	508.4(90.8)		&	1001.30(115.74)  \\
	& Random		&	84(36.03)		&	502.6(70.50)		&	483.4(85.73)		&	782.7(173.6)		&	1061.1(86.45)  	\\ \hline

	& 150\%	    &	6.2(0.58)	    &	6.15(0.61)		&	6.2(0.62)		&	689.4(428.2)	    & 	13.7(0.87)   \\
	& 100\%		&	6.5(0.6)		&	6.7(0.6)		&	6.7(0.7)		&	2502.8(1065.3)		&	14.2(1.3)  \\
20	& 90\%		&	85.7(51.56)		&	350(205.88)		&	455(154.25)		&	2632.6(874.6)		&	2944.3(455.64)  \\ 
	& 80\%		&	151.3(86)		&	682.2(277.4)		&	689.4(247.9)		&	3012.8(1074.2)		&	2801.8(449)  \\
	& Random		&	231.9(126.9)		&	2192.8(478.1)		&	2134.4(409.7)		&	3979.3(1034.9)		&	2897.4(537.8)  	\\ \hline
\end{tabular}

\caption[Median and Standard Deviation of Finish Times comparison for different Community Counts]{Table shows the comparison of median finish times (time for all users in the network to obtain the complete file) for all seeding across networks with different dissemination strategies and number of communities. The standard deviation is shown in brackets.}
\label{Table:median_finish_time_exp1}

\end{table}

\typeout{}
\chapter{Efficient seeding inside communities}\label{Ch4}

In the previous section, we observed that seeding each community with the complete file is the best strategy to ensure small delays in file retrieval while reducing the server bandwidth consumption. We further observed that \textit{network coding} provides the best results in terms of latency in retrieving the file for users. In this section we will investigate effective intra-community seeding strategies to improve file dissemination using \textit{network coding} inside a community. An effective intra-community seeding strategy should reduce the expected delay faced by members of the community in obtaining the file while also minimizing the number of \textit{non-innovative} transmissions that occur during opportunistic contacts between users.

\section{Node Centrality}

Mobility of users is expected to play an important role in the way packets are propagated in a network. A user who frequently encounters other users or meets with a larger subset of users is expected to play an important role in spreading network-coded packets within a community. Evaluating the importance of a node in a network has been widely studied in graph theory and network analysis \cite{123,124,125}. Centrality of a node is a metric used to measure the relative importance of a node in a graph. In the context of a social network with users represented as vertices and edge weights corresponding to the frequency of contacts between pairs of users, we can utilize different centrality measures to identify important users in the network. These centrality values can then be used to select users for initial seeding of network-coded packets from the service provider. We consider the following centrality measures for our analysis: (i) degree; (ii) betweenness; and (iii) closeness.
\\ \\
\noindent \textbf{Degree} centrality measures the number of direct neighbours of a node. We consider our graph to be static. This implies that a user can only meet another user if they share an edge between them in the graph. A higher centrality value is assigned to a node with more neighbors or edges connected to it. For a graph $G: = (V,E)$ with $n$ vertices, the degree centrality $C_{d}(v)$ for vertex $v$ is given by:

$$ C_{d}(v) = \frac{degree(v)}{n-1}$$ \\
\noindent \textbf{Betweenness} centrality assigns a higher value to nodes through which a larger number of shortest path connections are formed between nonadjacent nodes. Such nodes are expected to govern the rate at which information flows between nonadjacent nodes in the network. For a graph $G:= (V,E)$ with $n$ vertices, the betweenness $C_{b}(v)$ for vertex $v$ is determined as follows:
\begin{enumerate}
 \item Compute the shortest paths between each pair of vertices in the graph.
 \item For each vertex $v$, determine the fraction of all shortest paths between a specific pair of vertices (i,j) that pass through $v$, where  $i \neq v \neq j$.
 \item Sum this over all pairs of vertices (i,j).
 \end{enumerate}
In mathematical terms, this can be written as:

$$ C_{b}(v) = \sum_{i\neq v\neq j\in V} \frac{\sigma_{ij}(v)}{\sigma_{ij}}$$

\noindent where $\sigma_{ij}$ is the number of shortest paths from $i$ to $j$ and $\sigma_{ij}(v)$ is the number of shortest paths from $i$ to $j$ which pass through the vertex $v$. \\ \\
\noindent \textbf{Closeness} centrality ranks nodes in terms of their average shortest paths to all other users in the network. A higher centrality value establishes the importance of a node as being well connected to all other nodes in the network. It gives a measure of how long it would take data to spread from a user to all other users in the network. Therefore, we anticipate that it could be significant in our context. For a connected graph, the closeness centrality for a vertex $v$ is written as:

$$C_{c}(v) = \frac{1}{\sum_{j \in V}d_{G}(v,j)}$$
\noindent where $d_{G}(v,j)$ is the shortest path distance from node $v$ to $j$.

\section{Experimental Setup}\label{compare}

In this section the performances of different centrality-based seeding strategies are compared in terms of the expected percentage of users in the network who obtain the file at any given time and the number of transmissions required. The analysis is performed on the graph with the default parameters mentioned in Section~\ref{DSCsec}. The complete file is seeded in each community in the network. This seeding strategy was shown to provide the best tradeoff in terms of server bandwidth consumption and the latency users face in obtaining the file in Chapter~\ref{Ch3}. The current chapter focuses on extracting additional performance benefits by optimizing the distribution of file packets within each community. 

\textit{Betweenness, degree} and \textit{closeness} centrality values are used to identify important users in each community. A practical way of evaluating the centralities could be by asking the users to keep track of their meetings and relaying them back to a central server. The meetings log can be used to estimate the average rate of meetings between pairs of users in each community. The central server could then be used to evaluate the centrality measures based on the average rate of meetings between users in a community. After determining the centrality value for each user, file dissemination process is simulated for all centrality-based seeding strategies. The default case of $Random$ seeding is also included from Chapter~\ref{Ch3}. 

Experiments are performed for two seeding schemes based on the centrality values of the users in their respective communities. In the first scheme, the number of network coded packets seeded to each user is proportional to the centrality value of a user within its community. We call this scheme \textbf{S1}. In the second scheme \textbf{S2}, only one user is seeded with the complete file in each community. The user selected for seeding is the \textit{most central user} in its community. The normalized degree, betweenness and closeness centrality values of each user are summed and the user which has the highest aggregate value is selected as the \textit{most central user}. The normalized centrality values for each user within respective communities is obtained using the \textbf{networkx} tool for python. The process of allocating network coded packets to users under seeding scheme \textbf{S1} is performed using the following operation:

\begin{equation} 
P_{k}^{S1}(i) = \frac{C_{k}(i)}{\sum_{i = 1}^{n}C_{k}(i)} \cdot N
\label{S1}
\end{equation}

\noindent Here,  $k$ is the centrality measure (degree(d), betweenness(b) or closeness(c)), $P_{k}^{S1}(i)$ is the number of packets assigned to user $i$ in scheme \textbf{S1}, $C_{k}(i)$ is the centrality value of user $i$ and $N$ is the file length. The value of $P_{k}^{S1}(i)$ is rounded-off to the nearest integer value.

\section{Simulation and Results}\label{SimResch4}

Experiments are run individually for each centrality-based seeding strategy mentioned in Table~\ref{Table:Exp_default}. After the initial seeding is completed, users are allowed to obtain packets from each other opportunistically until all users obtain the complete file. Nodes in the community announce completion once they are able to decode the file. This allows us to keep track of the file completion time of individual users in the community. Users that obtain the file remain in the network and aid other users in obtaining the file. The dissemination process is performed for 50 Monte Carlo simulations for each seeding strategy and the results presented are the averages of these runs. The expected percentage of users in the network having the file at different times is shown in Table~\ref{Table:Exp_default}.

 \begin{table}[htbp]
	\centering
    \begin{tabular}{|l|l|c|c|c|c|c|c|}
     \hline
	Seeding 			&	Seeding strategy			&   	\multicolumn{6}{|c|}{Expected percentage of users having the file} \\ 	\cline{3-8}
	scheme			&						&	1 min	&	2 min	&	4 min	&	6 min	&	8 min	&	9min\\ \hline
	S1				&	Degree centrality		&	1(0.9)	&	29(1.8)	&	50(3.6)	&	71(3.1)	&	85(3.5)	&	95(1.6)\\
	S1				&	Betweenness centrality	&	3(1.8)	&	33(1.5)	&	50(2.7)	&	70(1.6)	&	86(3.3)	&	96(1.9)\\
	S1				&	Closeness Centrality		&	2(1.3)	&	31(1.7)	&	53(2.2)	&	72(3.4)	&	85(3.5)	&	93(2.1)\\ 
	S1				&	Random 				&	2(1.6)	&	29(2.0)	&	53(2.7)	&	68(2.9)	&	84(4.2)	&	95(1.7)\\
	S2				&	Most central user		&	1(1.5)	&	28(1.4)	&	52(1.8)	&	69(1.5)	&	82(3.3)	&	94(2.9)\\ \hline
    \end{tabular}
    \caption[Intra-community seeding comparison for network in Section~\ref{compare}]{Expected percentage of users in the network with default parameters which have the file at different times. The results are the averages for 50 Monte Carlo simulation. The values in the brackets represent the standard deviation.}
    \label{Table:Exp_default}
\end{table}

From Table~\ref{Table:Exp_default}, it is hard to identify any one particular seeding strategy as optimal since all values are closely matched. This suggests that the initial distribution of packets seeded in each community has little influence on the rate at which users acquire the network-coded packets of the file. The presence of hubs in the communities~\cite{34,116} enables the movement of packets from one user to another within the community with few hop counts. The rate at which a user acquires packets of the file then depends only on the frequency with which the user meets other users.
 
 The number of non-innovative transmissions is listed in Table~\ref{Table:median_finish_time_default}. The seeding scheme \textbf{S2} which involves seeding the \textit{most central user} with the complete file performs the best in terms of the number of \textit{non-innovative} packets transmitted during the file collection process. The second ranked seeding strategy is based on \textit{Degree centrality} but leads to 3 times more \textit{non-innovative} packets. 
 
  \begin{table}[htbp]
	\centering
    	\begin{tabular}{|c|c|c|c|c|}
	\hline
	MCU(\textbf{S2})		&	DC(\textbf{S1})		&	BC(\textbf{S1})		&	CC(\textbf{S1})	& 	$Random$ \\ \hline
	210(10)				&	700(18)			&	980(15)			&	672(13)		&	812(55) \\
	\hline
    \end{tabular}
    \caption[Non-innovative packet exchanges for Intra-community seeding schemes]{Median number of non-innovative transmissions for each intra-community seeding strategy. S2 represents Seeding Scheme 2 and S1 is Seeding Scheme 1. The values in brackets are the standard deviations. MCU: Most Central User, DC: Degree Centrality, BC: Betweenness Centrality, CC: Closeness Centrality. }
    \label{Table:median_finish_time_default}
\end{table}
 
\begin{figure}[tbp]
	\centering
	\subfigure[]{
	\includegraphics[scale=0.25, angle=-90, type=pdf,ext=.pdf,read=.pdf]{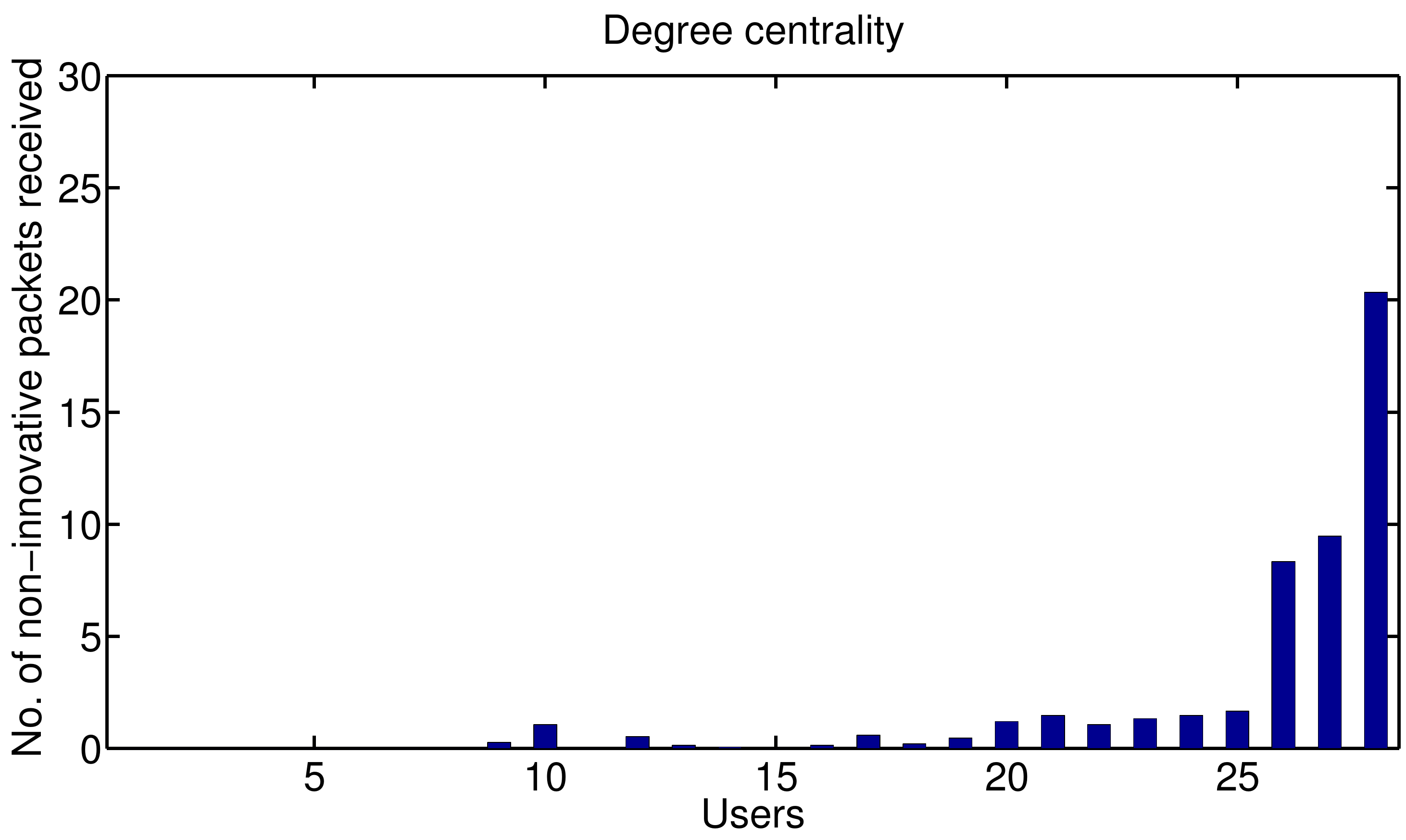}
	\label{fig:28_users_non_inn_DC}
	}
	\subfigure[]{
	\includegraphics[scale=0.25, angle=-90, type=pdf,ext=.pdf,read=.pdf]{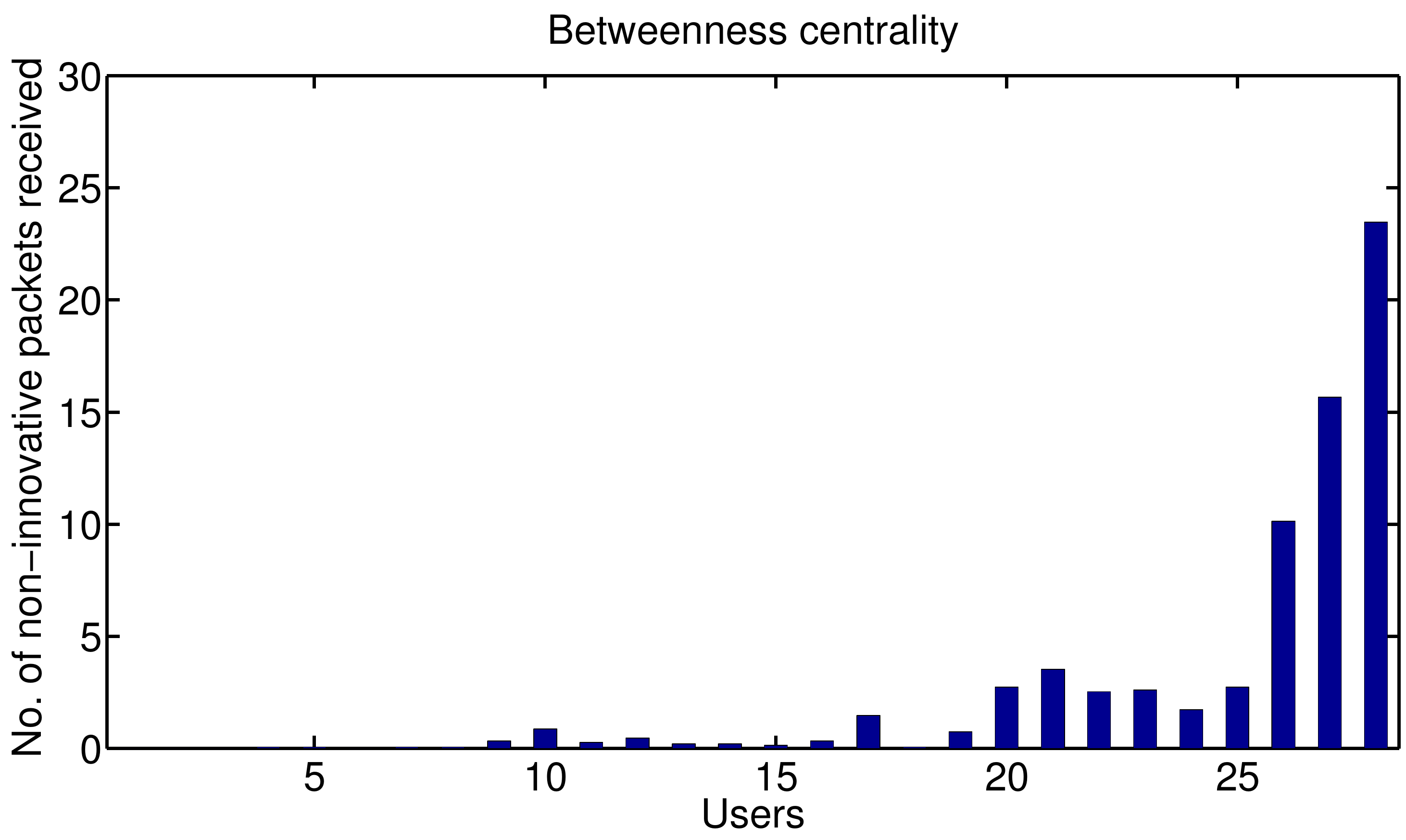}
	\label{fig:28_users_non_inn_BC}
	}	
	\subfigure[]{
	\includegraphics[scale=0.25, angle=-90, type=pdf,ext=.pdf,read=.pdf]{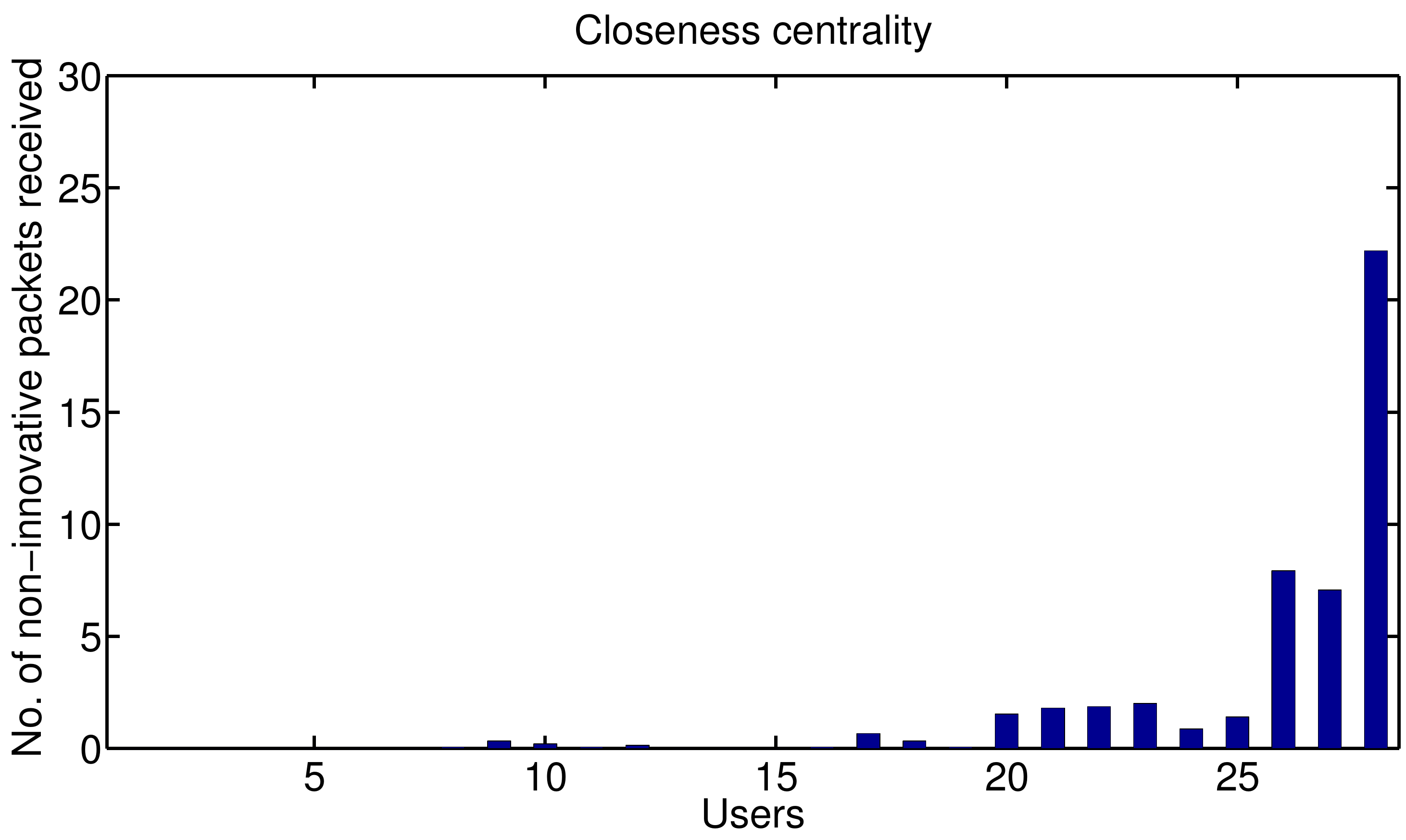}
	\label{fig:28_users_non_inn_CC}
	}
	\subfigure[]{
	\includegraphics[scale=0.25, angle=-90, type=pdf,ext=.pdf,read=.pdf]{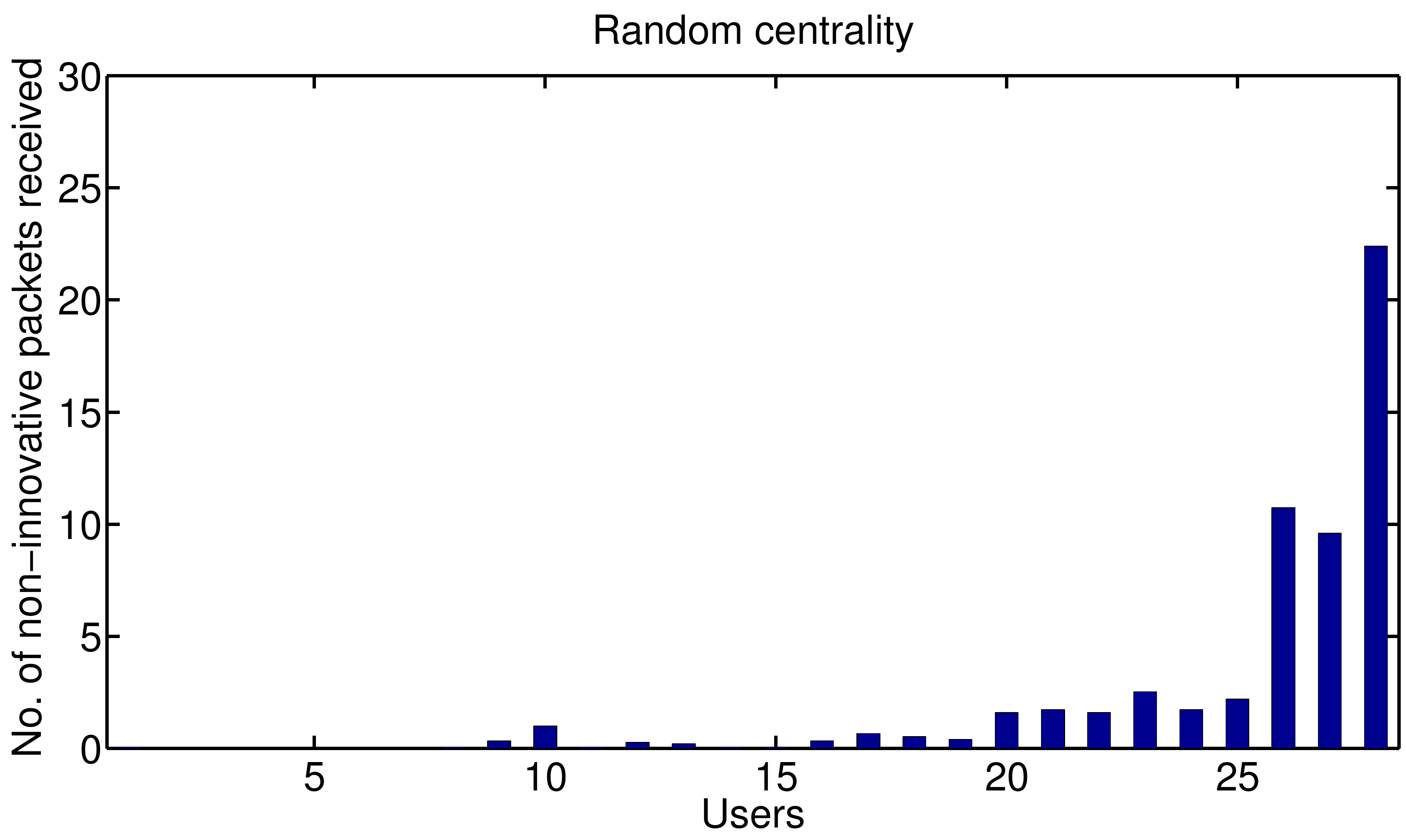}
	\label{fig:28_users_non_inn_R}
	}
	\subfigure[]{
	\includegraphics[scale=0.25, angle=-90, type=pdf,ext=.pdf,read=.pdf]{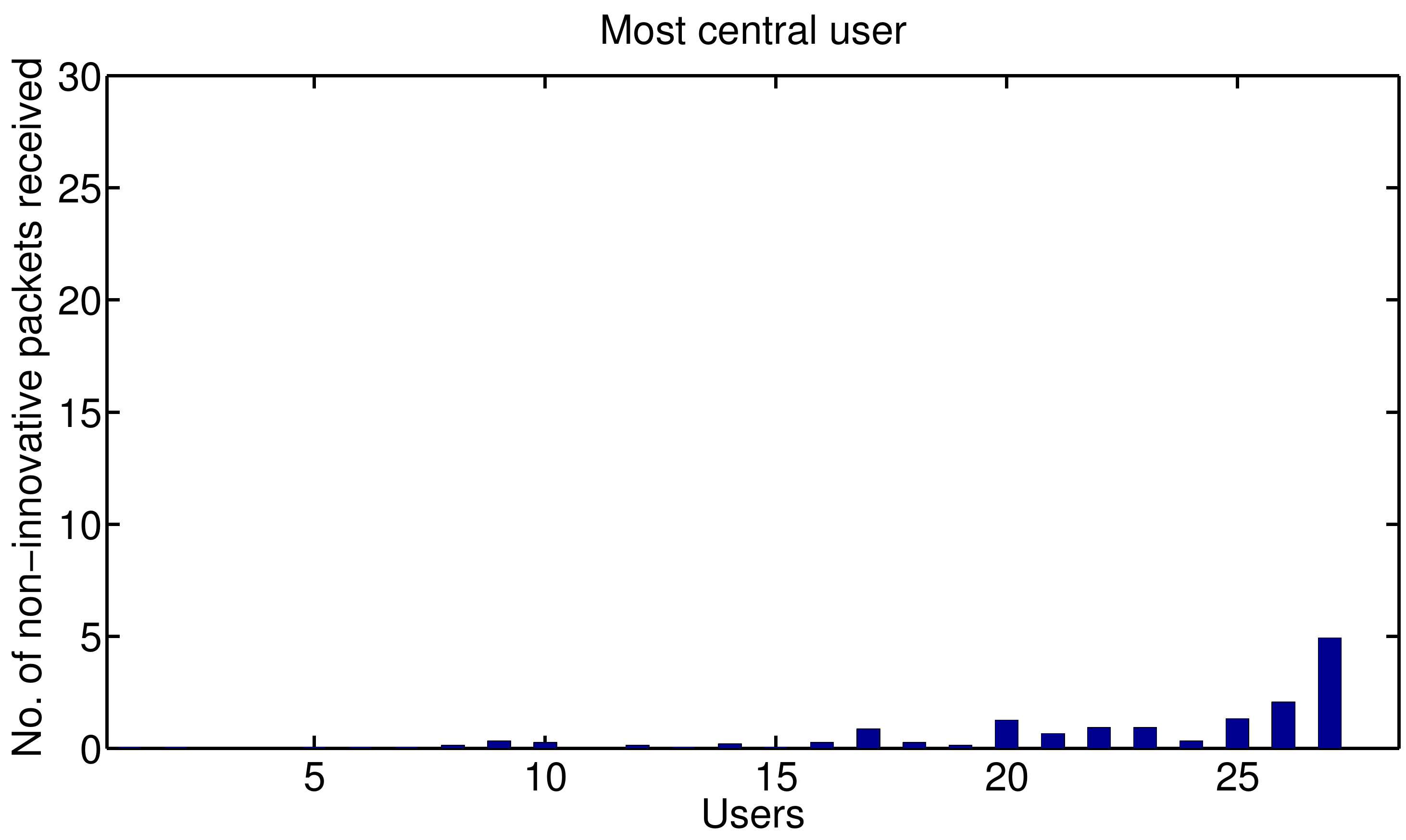}
	\label{fig:28_users_non_inn_MC}
	}
\caption[Box plot of non-innovative packet exchanges]{Distribution of average number of non-innovative packets received by each user in the community for seeding strategy \subref{fig:28_users_non_inn_DC} degree centrality, \subref{fig:28_users_non_inn_BC} betweenness centrality, \subref{fig:28_users_non_inn_CC} closeness centrality, \subref{fig:28_users_non_inn_R} \textit{Random} and \subref{fig:28_users_non_inn_MC} Most central user. Users are arranged in increasing order of centrality. The total number of users in the community are $28$ and the $28^{th}$ user is also the \textit{most central user} in the community.}
\label{fig:Non_inn_dist_28}
\end{figure}

\begin{figure}[tbp]
	\centering
	\subfigure[]{
	\includegraphics[scale=0.25, angle=-90, type=pdf,ext=.pdf,read=.pdf]{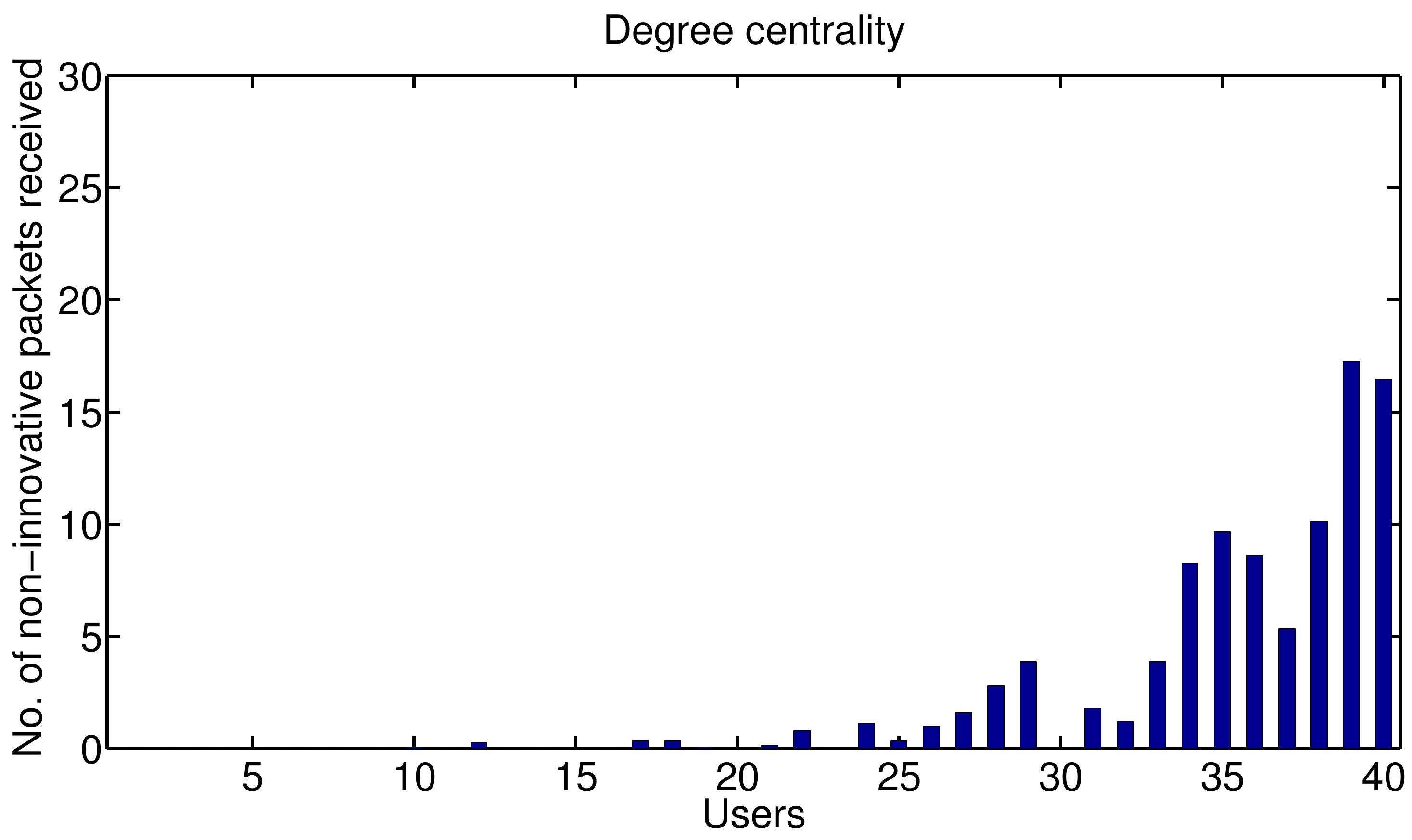}
	\label{fig:40_users_non_inn_DC}
	}
	\subfigure[]{
	\includegraphics[scale=0.25, angle=-90, type=pdf,ext=.pdf,read=.pdf]{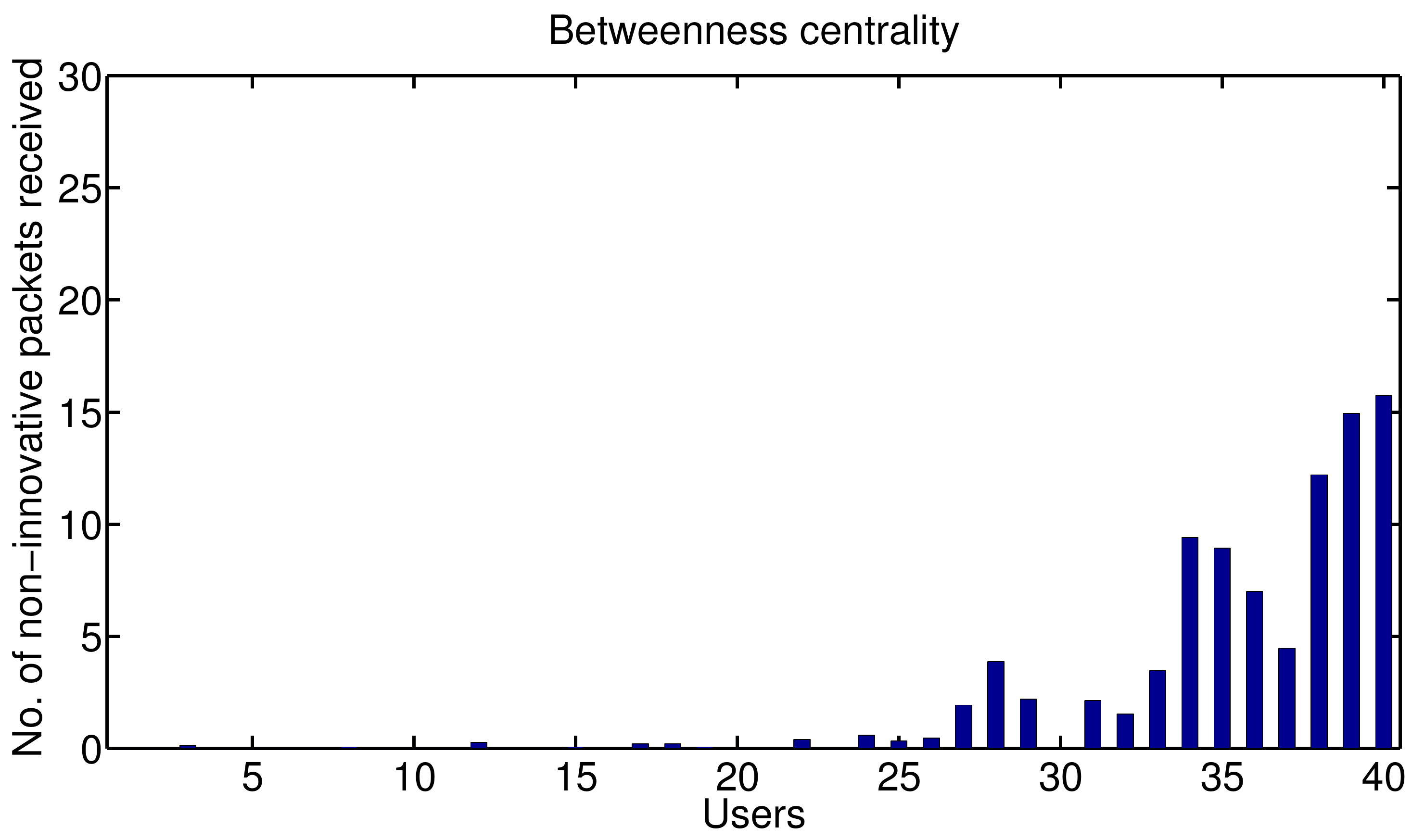}
	\label{fig:40_users_non_inn_BC}
	}	
	\subfigure[]{
	\includegraphics[scale=0.25, angle=-90, type=pdf,ext=.pdf,read=.pdf]{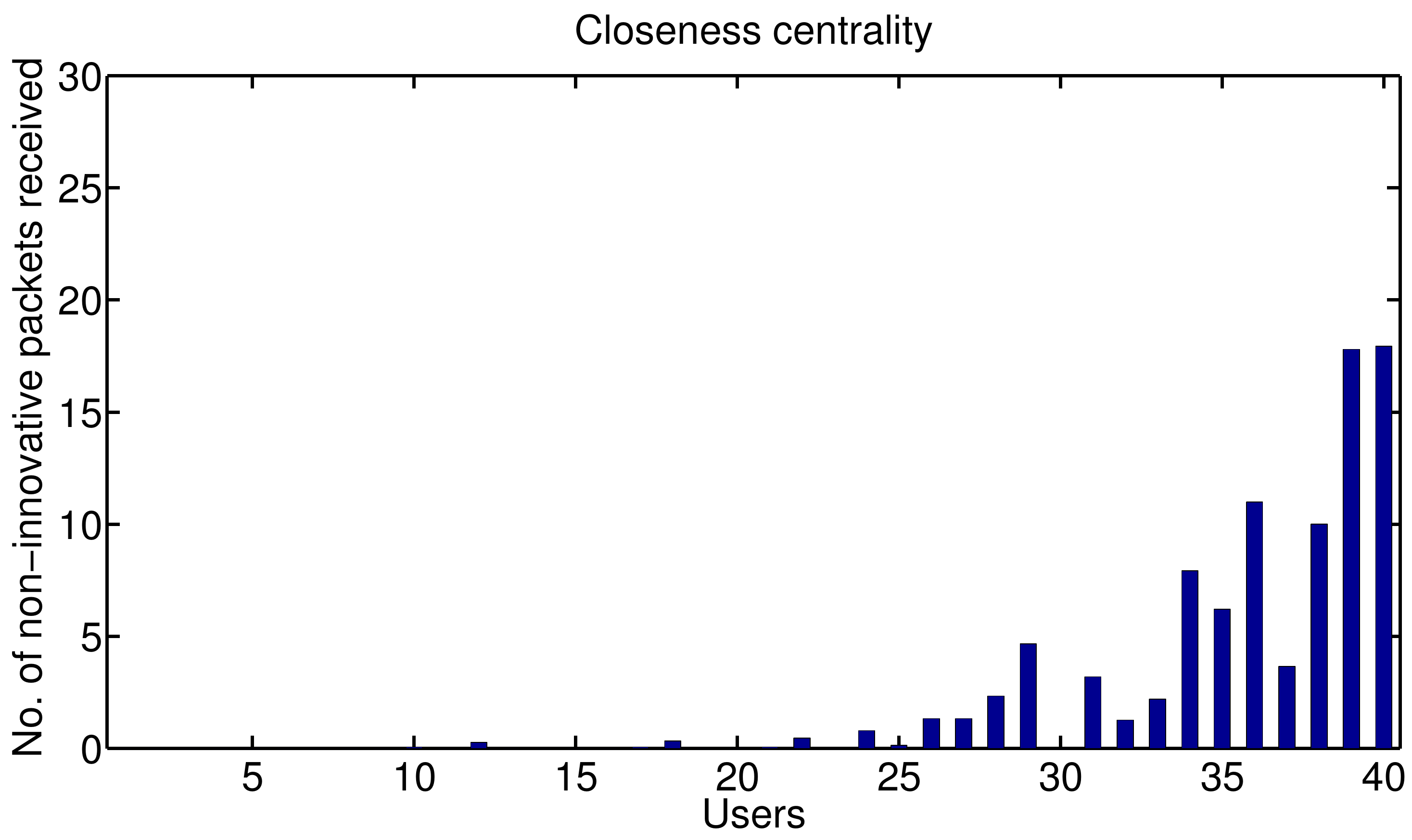}
	\label{fig:40_users_non_inn_CC}
	}
	\subfigure[]{
	\includegraphics[scale=0.25, angle=-90, type=pdf,ext=.pdf,read=.pdf]{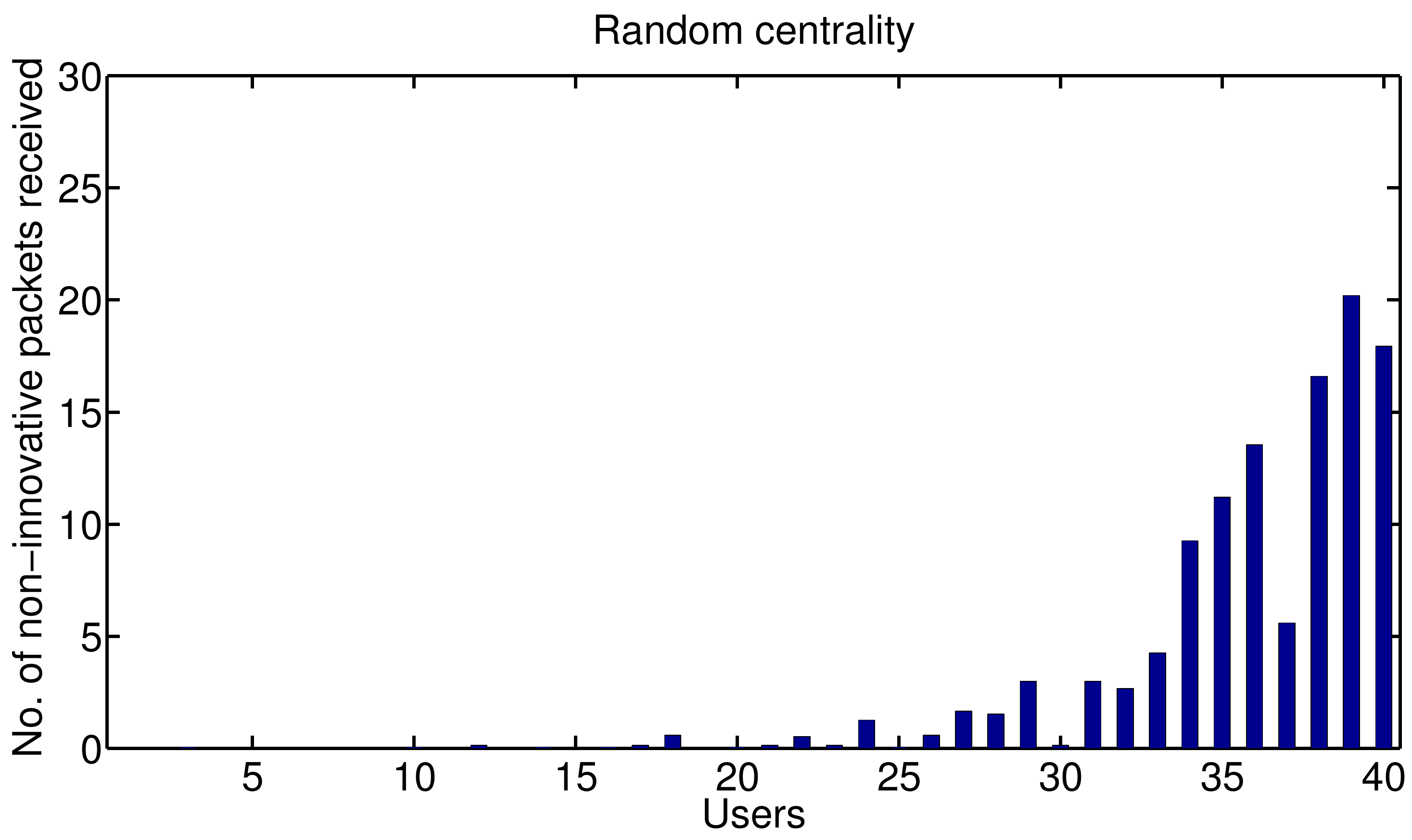}
	\label{fig:40_users_non_inn_R}
	}
	\subfigure[]{
	\includegraphics[scale=0.25, angle=-90, type=pdf,ext=.pdf,read=.pdf]{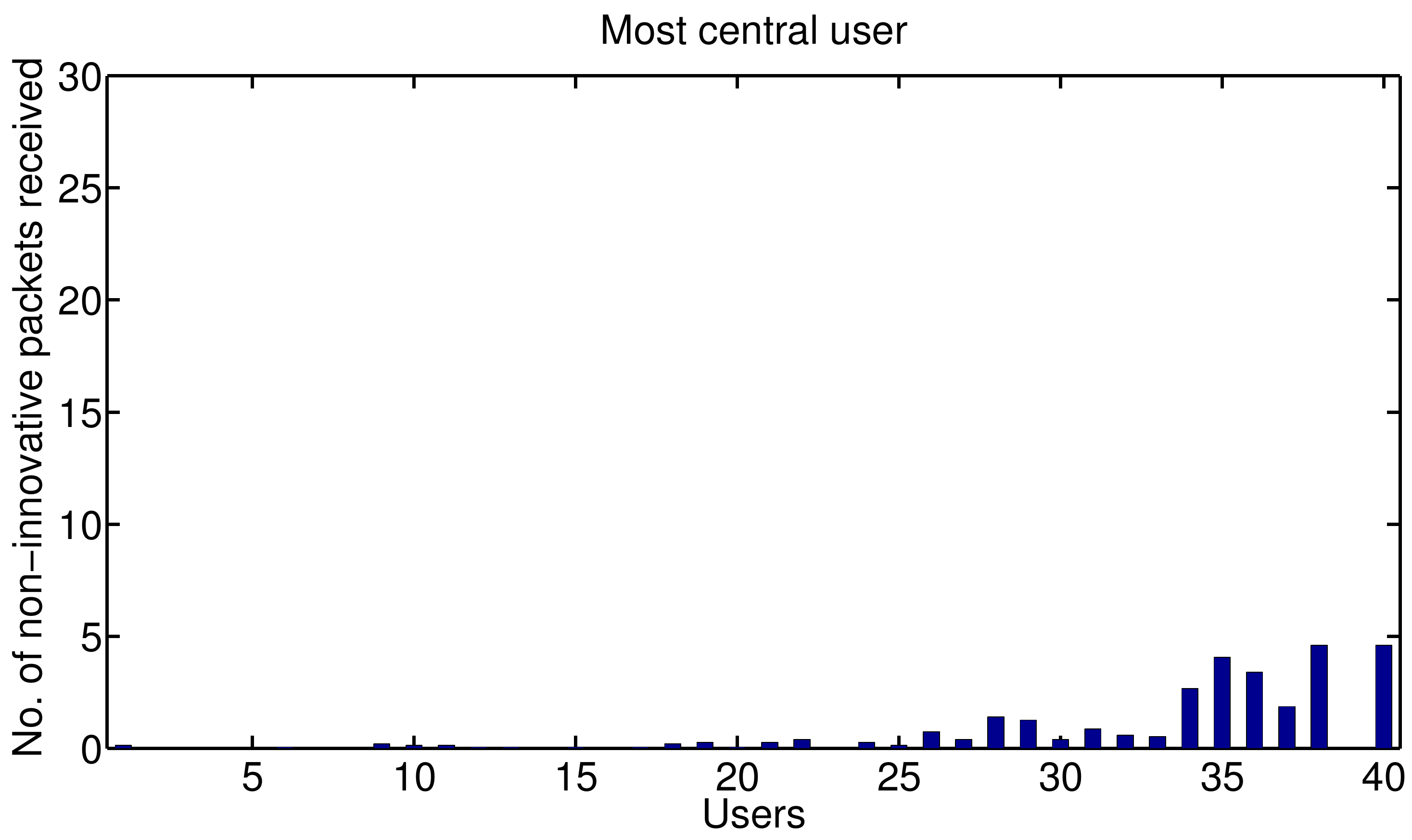}
	\label{fig:40_users_non_inn_MC}
	}
\caption[Non-innovative packets received by subscribers]{Distribution of average number of non-innovative packets received by each user in the community for seeding strategy \subref{fig:40_users_non_inn_DC} degree centrality, \subref{fig:40_users_non_inn_BC} betweenness centrality, \subref{fig:40_users_non_inn_CC} closeness centrality, \subref{fig:40_users_non_inn_R} \textit{Random} and \subref{fig:40_users_non_inn_MC} Most central user. Users are arranged in increasing order of centrality. The total number of users in the community are $40$ and the $39^{th}$ user is the \textit{most central user} in the community.}
\label{fig:Non_inn_dist_40}
\end{figure}

  Users which act as \textit{hubs} in social networks meet many other users and their encounters are more frequent. This allows them to obtain packets from many different users. Therefore, any user which encounters these \textit{hubs} can obtain an \textit{innovative} packet with high probability. On the other hand, if these \textit{hubs} are in search of innovative packets themselves, the task is more challenging especially if there is no book-keeping of which packets were obtained from which users. Figures~\ref{fig:Non_inn_dist_28} and \ref{fig:Non_inn_dist_40} show the average number of \textit{non-innovative} packets received by users in two different communities of our network. In both figures, the users are arranged in increasing order of centrality values (user 10 will always have a higher centrality than user 9). It is seen from the figures that the users which have higher centrality values also obtain the most number of non-innovative transmissions. In Figure~\ref{fig:Non_inn_dist_40}, the \textit{most central user} receives the most number of non-innovative transmission for all seeding strategies except the MCU-based seeding strategy.
  
\section{Robustness}\label{Robust}
In this section we evaluate the robustness of our community-based seeding strategies --- 100$\%$ and \textit{most central user} --- in the presence of node failure or node departure. Robustness against node failure is measured by determining the reduction in performance of the seeding strategies when some nodes are randomly removed from the network. Three scenarios are considered for simulation: no user fails, some randomly selected user fails every minute and a randomly selected user fails after every 30 seconds. The failing user is chosen randomly from the network and could belong to any community. The results are derived from 100 Monte Carlo simulations for each scenario.

\begin{figure}[tbp]
	\centering
	\subfigure[]{ex
	\includegraphics[scale=0.4, angle=-90, type=pdf,ext=.pdf,read=.pdf]{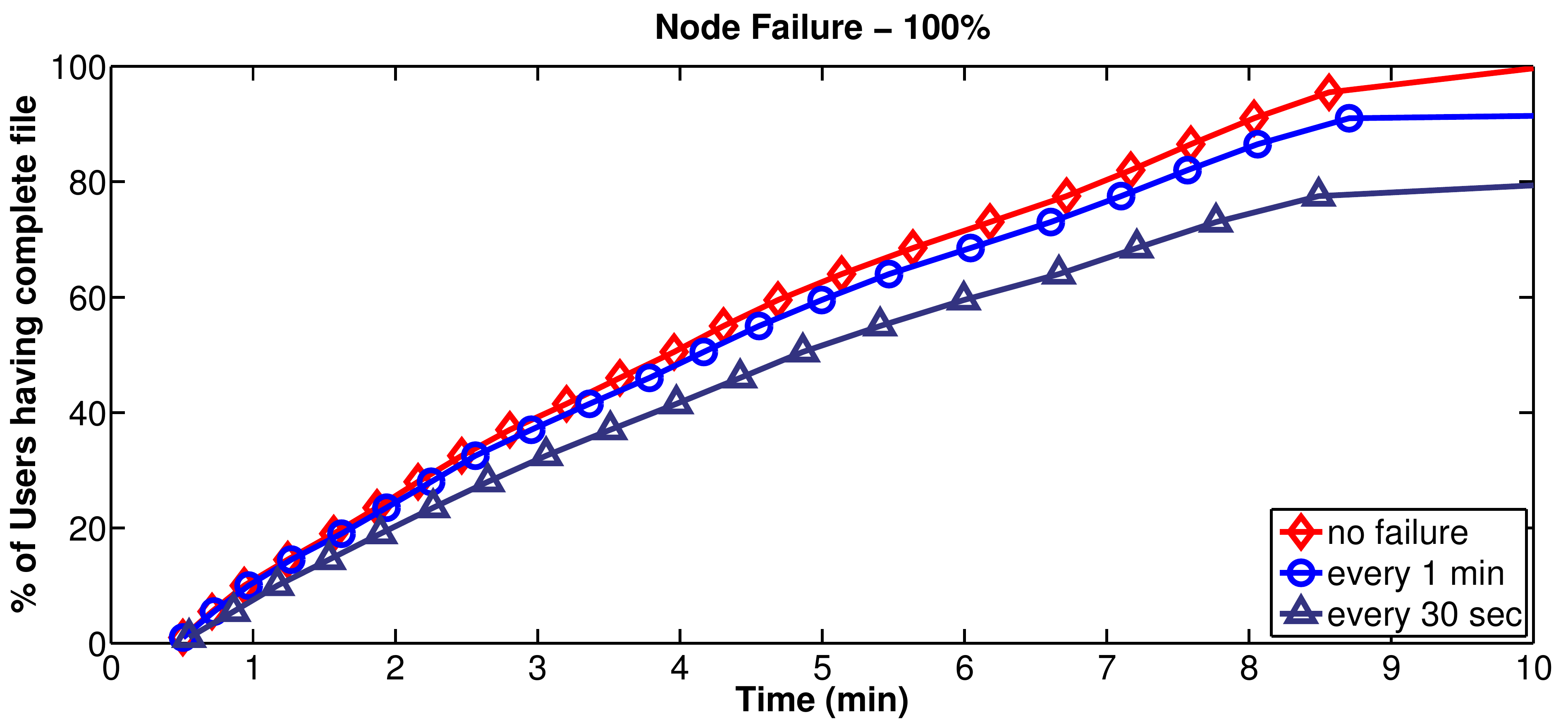}
	\label{fig:node_failure_users_NC}
	}
	\subfigure[]{
	\includegraphics[scale=0.4, angle=-90, type=pdf,ext=.pdf,read=.pdf]{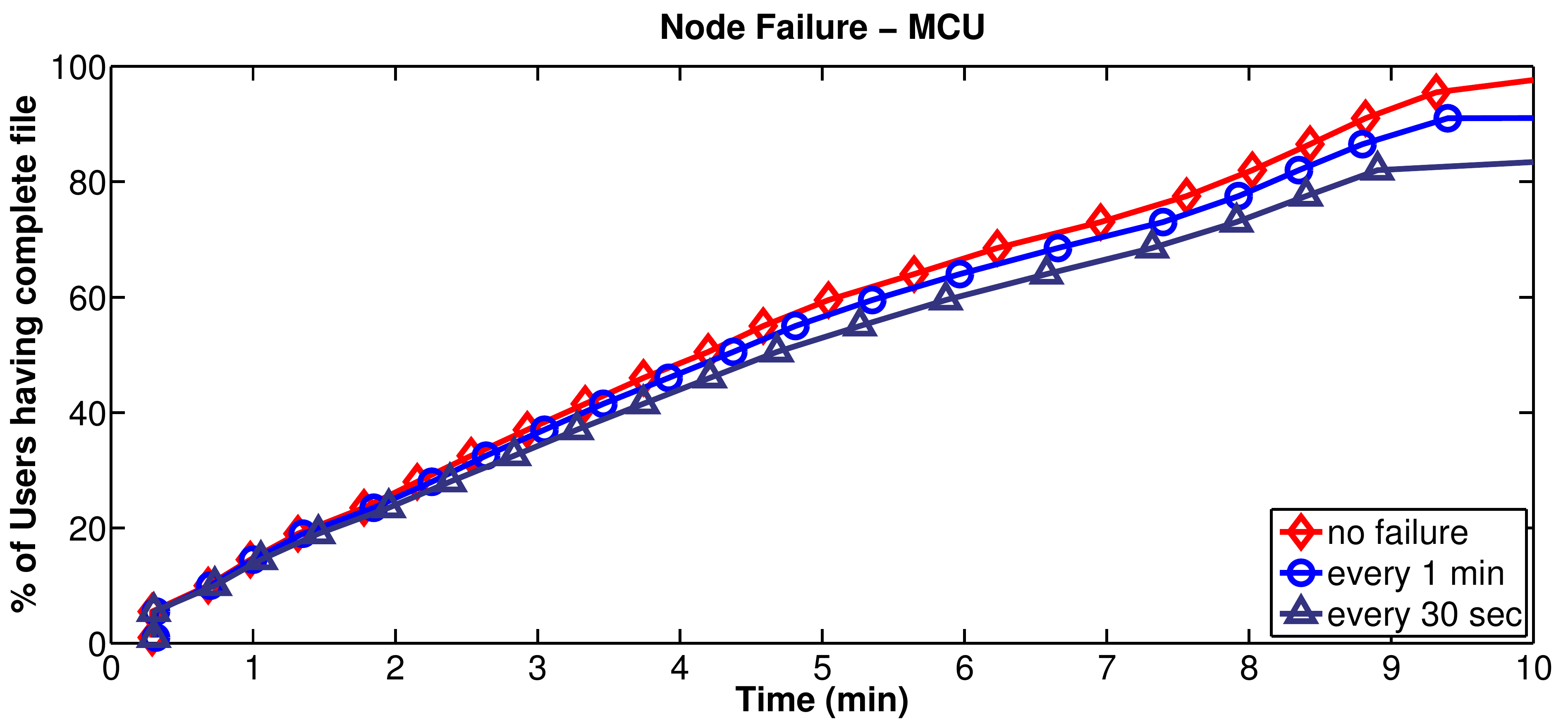}
	\label{fig:node_failure_users_box_NC}
	}
\caption[Percentage of users having complete file]{The plots shows the percentage of users which obtain the complete file with time for \subref{fig:node_failure_users_NC}100$\%$  and \subref{fig:node_failure_users_box_NC}\textit{most central user} based seeding strategy in the presence of node failure. Three scenarios are considered. Two scenarios involve nodes failing randomly after every 1 minute and 30 seconds. In the third scenario there is no failure. The results calculated are based on 100 Monte Carlo simulations for each scenario.}
\label{fig:node_failure_users_NC_FIG}
\end{figure}

\begin{figure}[tbp]
	\centering
	\subfigure[]{
	\includegraphics[scale=0.4, angle=-90, type=pdf,ext=.pdf,read=.pdf]{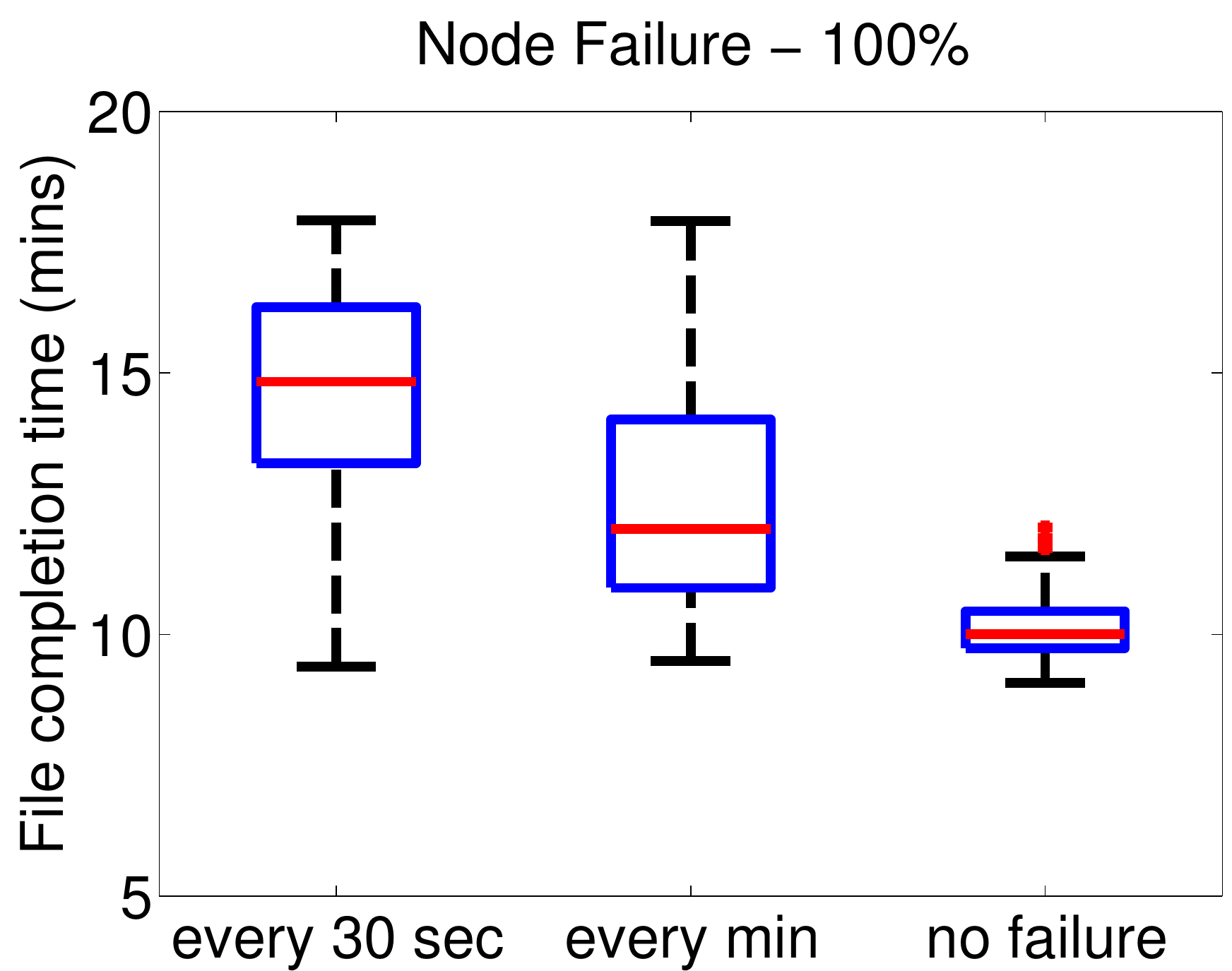}
	\label{fig:node_failure_users}
	}
	\subfigure[]{
	\includegraphics[scale=0.4, angle=-90, type=pdf,ext=.pdf,read=.pdf]{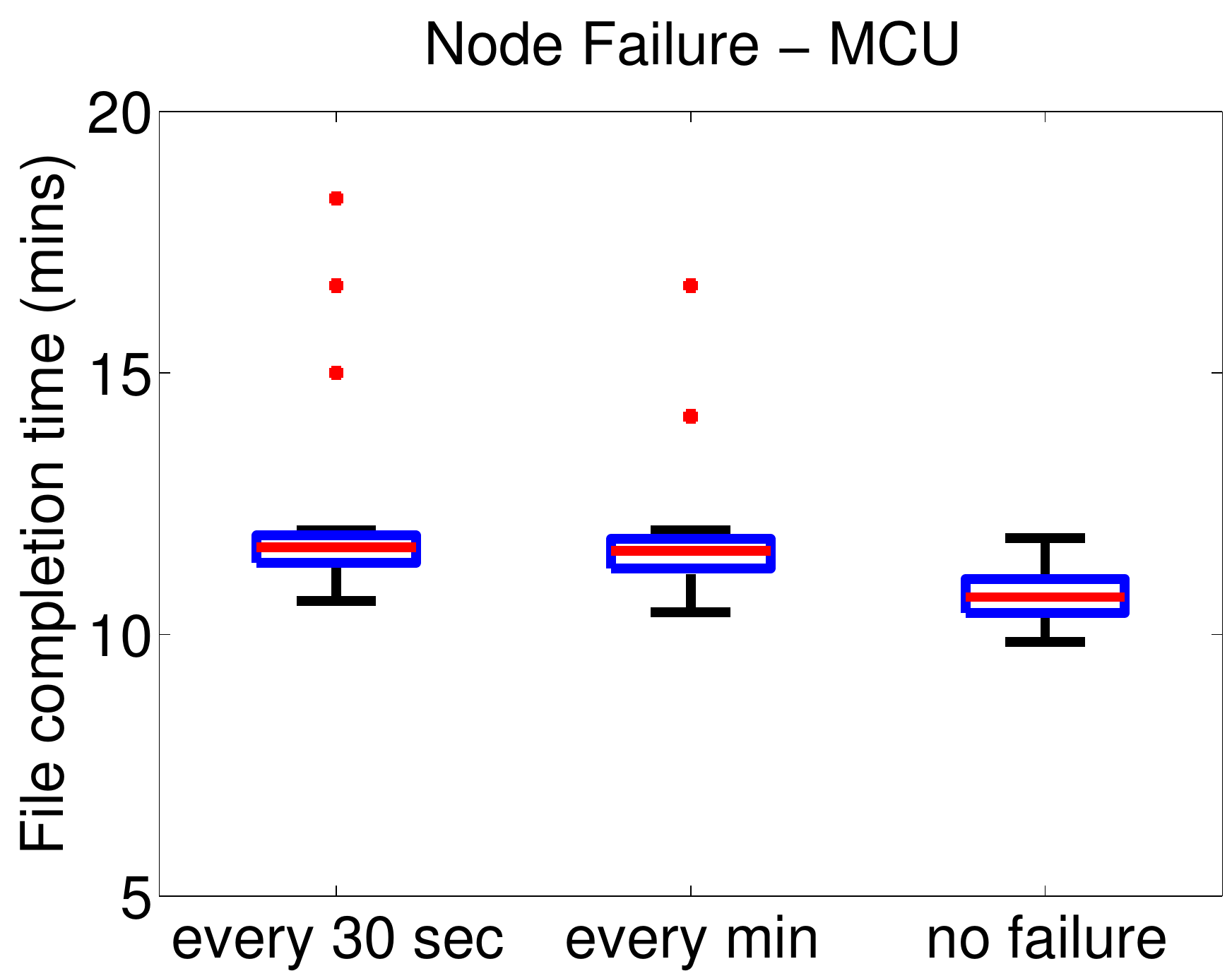}
	\label{fig:node_failure_users_box}
	}
\caption[Node failure in Most Central User (MCU) based seeding]{The plots shows the distribution of finish times for \subref{fig:node_failure_users}100$\%$ and \subref{fig:node_failure_users_box}\textit{most central user} based seeding strategy in the presence of node failure. Three scenarios are considered. Two scenarios involve nodes failing randomly after every 1 minute and 30 seconds. In the third scenario there is no failure. The results calculated are based on 100 Monte Carlo simulations for each scenario.}
\label{fig:node_failure_users_MCU_FIG1}
\end{figure}

The results for 100$\%$ and \textit{most central user} based community seeding in the presence of node failure is shown in Figure~\ref{fig:node_failure_users_NC_FIG} and Figure~\ref{fig:node_failure_users_MCU_FIG1}. Figure~\ref{fig:node_failure_users_NC_FIG} shows the expected percentage of users which obtain the complete file at any given time. The curves obtained, by varying the frequency of node failure, are less separated in Figure~\ref{fig:node_failure_users_box_NC} compared to Figure~\ref{fig:node_failure_users_NC}. These results imply that seeding based on \textit{most central user} is more robust to node failures compared to 100$\%$ community seeding. During the initial phase of the 100$\%$ seeding strategy, the server randomly selects users from each community and provides them with file packets. The total number of packets seeded to users lying in the same community equals the number of packets in the original file. This allows users to obtain the required file packets through frequent meetings with other users in their community. Meetings between users in different communities are less frequent. If a node fails, prior to transferring all of its packets obtained from the server onto its neighbouring nodes, some packets are lost. The users belonging to the community can only reconstruct the file when these lost packets are acquired through interactions with users lying in adjacent communities. The inter-community meetings are less frequent and cause a longer delay for the users residing in the former community to obtain the necessary packets to reconstruct the file. 

During each Monte Carlo simulation all active users, in both seeding strategies, are able to obtain the complete file. The distribution of finish times of the experiments is shown in Figure~\ref{fig:node_failure_users_MCU_FIG1}. From  Figure~\ref{fig:node_failure_users}, it is observed that the variance in finish times and the median finish times increase in the presence of node failures for the 100$\%$ seeding strategy. There are two important factors that effect this change. Firstly, the departing user can cause loss of some initially seeded, and non-replicated packets, which has been discussed in the previous paragraph. The departing user could also be a well-connected user and hence its departure will directly affect the rate at which its neighbouring users obtain innovative file packets. The same trend is not observed for \textit{most central user} based seeding. The variance in the finish times is considerably less compared to 100$\%$ seeding strategy and is similar to the case of \textbf{no failure}. Centralizing all information from the server to a single node removes the possibility of suffering loss of necessary packets within each community except for the case when the \textit{most central user} fails. The median finish times, however, increase with an increase in the frequency of node failure. This increase is because any departing user reduces the rate at which its neighbouring users obtain file packets. 

\begin{figure}[tbp]
	\centering
	\includegraphics[scale=0.4, angle=-90, type=pdf,ext=.pdf,read=.pdf]{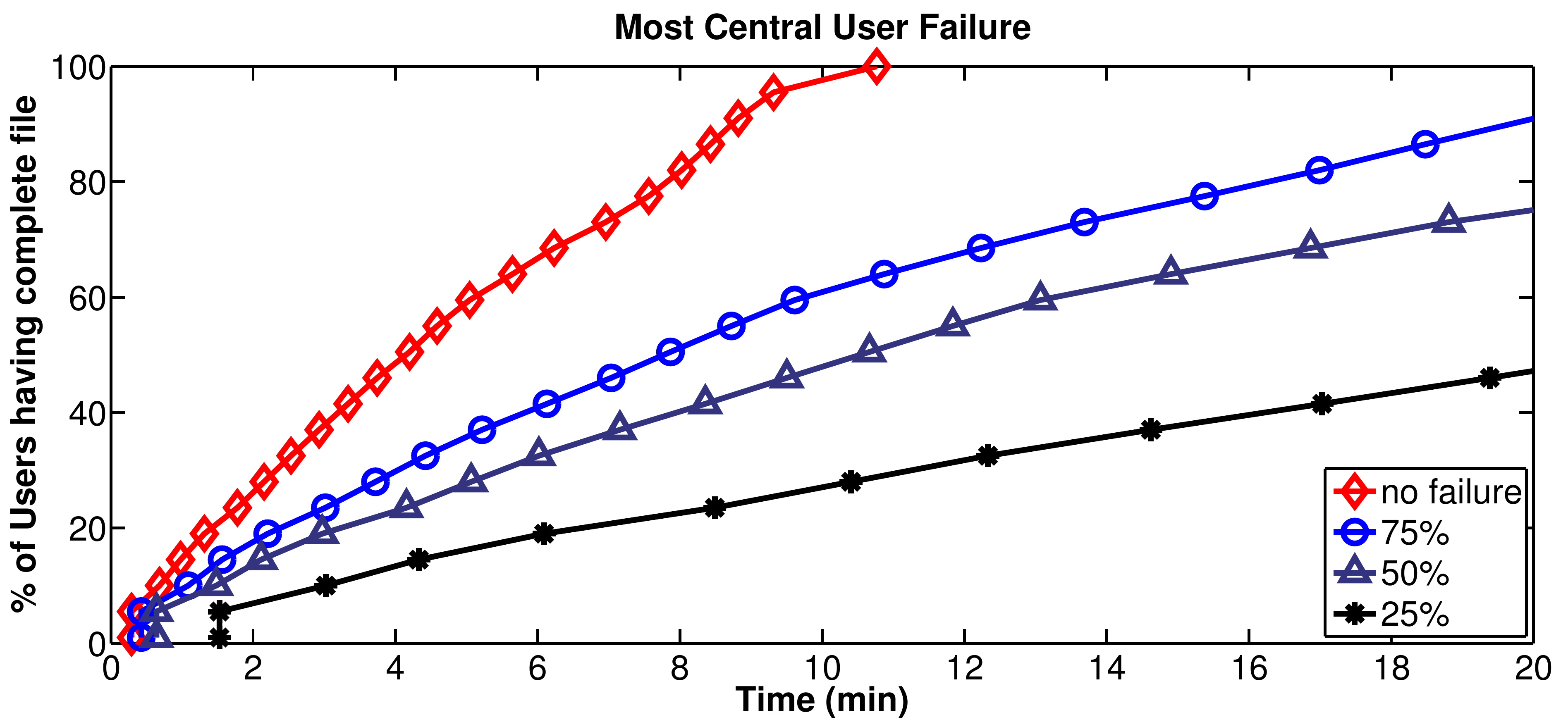}
	
\caption[Node failure in Most Central User (MCU) based seeding]{The plots shows the effect of failure of the most central user's in the network at different instances. The lines represent the percentage of users that obtain the file at any given time when there is no failure(red), most central user fails after distributing 75$\%$ of its packets(blue), 50$\%$ packets(purple) and 25$\%$ packets(black). The results calculated are based on 100 Monte Carlo simulations for each scenario.}
\label{fig:MCU_failure}
\end{figure}

The outliers in Figure~\ref{fig:node_failure_users_box} occur if the \textit{most central user/s} fails. The community which suffers the failure of its \textit{most central user} causes an increase in the finish time because the respective community has to rely entirely on inter-community meetings to obtain the necessary packets to reconstruct the file. Three additional experiments are performed to evaluate the effect of failure of the \textit{most central user}. The \textit{most central user} in each community fails after spreading 25$\%$, 50$\%$ and 75$\%$ of its initially seeded packets. Figure~\ref{fig:MCU_failure} shows that the percentage of users which obtain the file at any given time reduces significantly with the failure of the \textit{most central user}. This necessitates extra precaution to prevent the \textit{most central user} from failing or a re-assignment of the \textit{most central user} from the server in case of failure.

\section{Per node cost of opportunistic distribution}\label{perNode}
Network coding is effective in opportunistic file distribution schemes as it eases the packets scheduling issue, discussed in Section~\ref{scheduling}. The downside, however, is the possibility of occurrence of \textit{non-innovative} transmissions. A transmission is \textit{non-innovative} if contents of the packet received by a user during a meeting provides no new information which can be used to decode the file. An effective seeding strategy would minimize the number of non-innovative packet transmissions. 

\begin{figure}[htbp]
	\centering
	\subfigure[]{
	\includegraphics[scale=0.32, angle=0, type=pdf,ext=.pdf,read=.pdf]{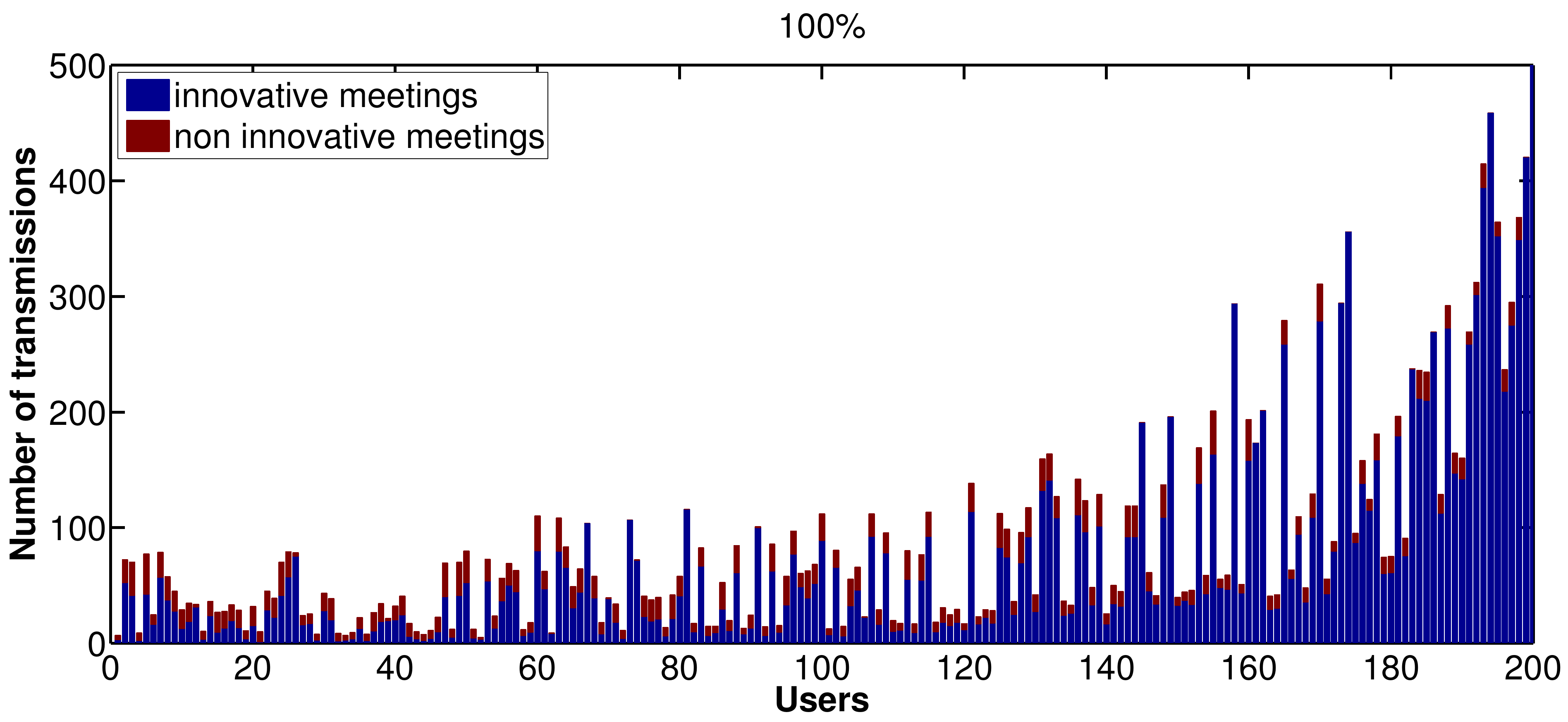}
	\label{fig:perNodeNC100}
	}
	\subfigure[]{
	\includegraphics[scale=0.32, angle=0, type=pdf,ext=.pdf,read=.pdf]{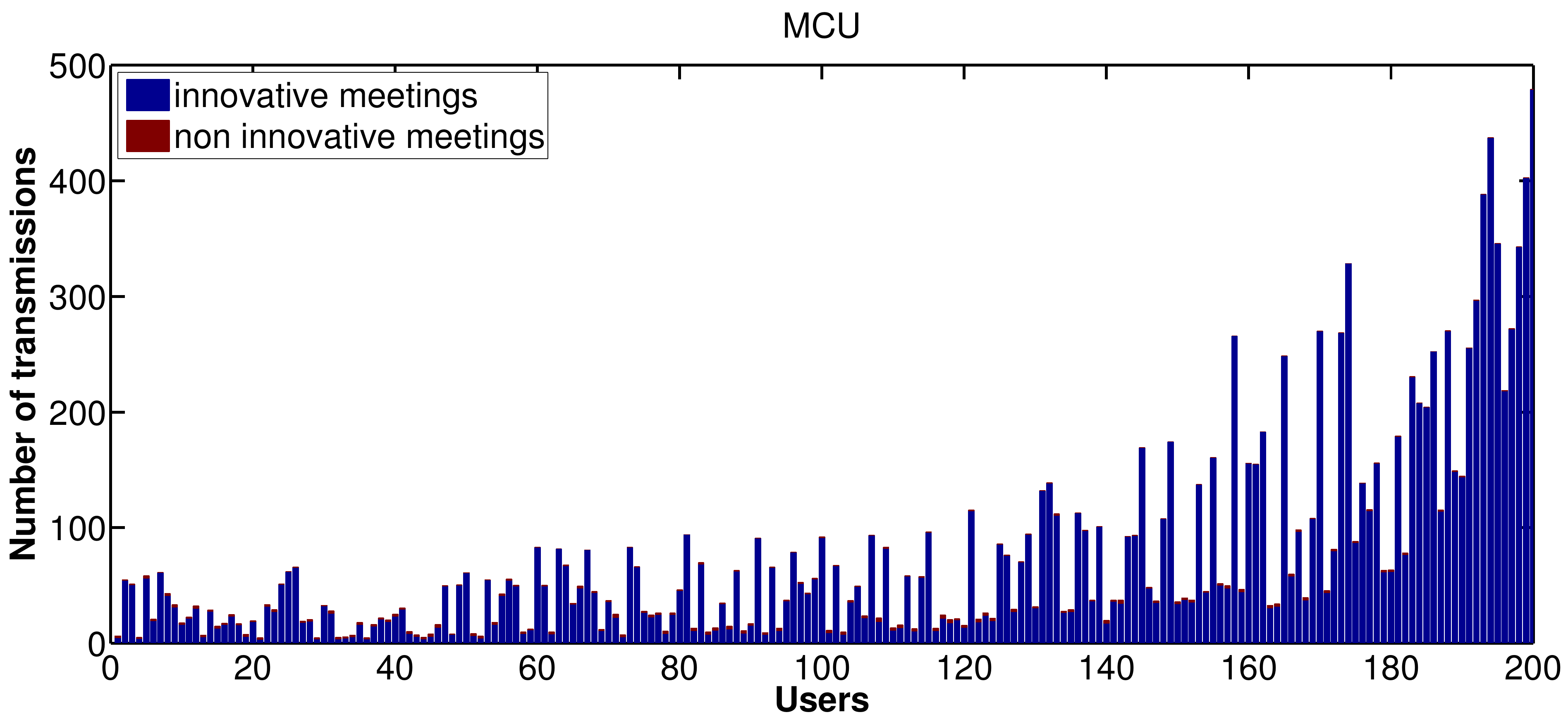}
	\label{fig:perNodeMCU}
	}
\caption[Per node cost of distribution for seeding strategy 100$\%$ and \textit{most central user}]{The plot above represents the per node cost of opportunistic distribution for seeding strategies \subref{fig:perNodeNC100}$100\%$ and \subref{fig:perNodeMCU}\textit{most central user}. The bar plot shows the average number of transmissions that occur at each user during file distribution. The total number of transmissions include \textit{innovative} and \textit{non-innovative} transmissions.}
\label{fig:node_failure_users_MCU_FIG}
\end{figure}

The number of innovative and non-innovative packet transmissions that occur during opportunistic file distribution for seeding schemes 100$\%$ and \textit{most central user} is shown in Figure~\ref{fig:node_failure_users_MCU_FIG}. The number of non-innovative transmissions per node is observed to be less for \textit{most central user} seeding strategy compared to 100$\%$ seeding strategy. It is also observed that the distribution of total number of transmissions occurring at the nodes is skewed and some users are required to make a larger number of transmissions compared to other users. The users which are more mobile interact with more users and are, therefore, required to make a larger number of transmissions. Incentives, such as lower subscription costs for users exhibiting higher mobility, could be used to attract users to participate in opportunistic distribution. 

 \typeout{}
\section{Impact of community detection on file distribution} \label{Ch5}

Our previous results assumed perfect knowledge of communities in the network, but in real networks this information is not always available. In this chapter we will use the GANC (Greedy Agglomerative Normalized Cut) algorithm proposed by Tabatabaei et al.~in \cite{118} to extract community information from networks generated using the LFR graph generation tool. File dissemination is then performed on the communities detected by GANC and the results are compared to the ones obtained on the LFR provided communities. The comparison is important to obtain some insight on the performance of our centrality-based community seeding scheme for real-world traces where community information is often not known a priori. 

The cut associated with some cluster $U$ is the sum of the weights of the edges between nodes in cluster $U$ and nodes in other clusters. Therefore, minimizing the maximum cut can be used to identify communities which have the weakest ties between them. The downfall is that such a technique results in the formation of several small clusters which do not completely capture the characteristics of the underlying structure of the graph. The normalized cut criteria was first put forward by Shi and Malik in~\cite{31} and prevents this shortcoming by normalizing each cut by the total weight of the associated cluster. This helps to penalize the development of small clusters due to their low aggregate weight. Normalized cut minimization is an NP-complete problem \cite{30}. Although spectral methods exist which approximately solve the problem by determining the eigenvectors of the Laplacian graph, these methods have high computational complexity that generally grow rapidly with the number of nodes in the graph.

The advantage of using the GANC clustering algorithm by Tabatabaei et al.~\cite{118} is two-fold. The algorithm is fast, scaling almost linearly with the number of nodes in the graph, and contains a model order selection method to determine the number of communities in the network. 

\subsection[Performance evaluation on communities detected using GANC]{Performance evaluation on communities detected using GANC~\cite{118}}\label{GANCsec}
In this section, we will estimate the performance of our centrality-based community seeding algorithm in situations where community information is not available. Community information is often not available in real-world networks and the results in this section will give an insight into the performance of our scheme for real-world networks. We utilize the GANC community detection algorithm by Tabatabaei et al.~\cite{118} to identify communities in networks generated by the LFR graph generation tool. The file dissemination process is simulated for the communities detected by GANC and the results are compared to the ones obtained for communities provided by the LFR tool. The performance is compared over three different graph mentioned in Table~\ref{Table:Exp2_T1}. LFR graph generation parameters, $\mu_{w}$ and $\mu_{t}$ are varied to obtain the different graphs. Parameter $\mu_{w}$ controls the ratio between the average intra-community contact interval and the average inter-community contact interval. Increasing its value  moves the ratio more towards unity which makes it harder to distinguish communities. Similarly, the parameter $\mu_{t}$ controls the ratio between the number of edges lying within communities and the number of edges between communities. Increasing the parameter value results in an increase in inter-community connections and consequentially community detection becomes more difficult. 

\begin{table}[htbp]
\centering
\small
\begin{tabular}{|c|c|c|c|c|c|}
\hline
Graph	&	$\mu_{w}$	&	 $\mu_{t}$	& 	$R_{T}$	&	$R_{E}$		&	LFR communities	\\
\hline \hline
A		&	$0.001$ 	        &   $0.1$ 			&	$10$ 	&	$8.36$		&	$14$	 		\\
\hline
B		&	$0.001$		&   $0.3$			&	$38$		&	$2.35$		&	$14$			\\
\hline
C		&	$0.01$		&   $0.1$			&	$1$		&	$8.32$		&	$14$			\\
\hline
\end{tabular}
\caption[Properties of graphs used in Section~\ref{GANCsec}]{Properties of graphs used for evaluating performance of centrality-based community seeding scheme on communities detected by GANC. $R_{T}$ is the ratio between the average inter-community inter-contact time and average intra-community inter-contact time. $R_{E}$ denotes the ratio between the average number of intra-community edges and average number of inter-community edges. The actual number of communities in all graphs is $14$  as provided by the LFR graph generation tool.}
\label{Table:Exp2_T1}
\end{table}

For Graph A and B, GANC is able to identify all communities correctly. The results of the performance of file dissemination on both sets of communities would not be interesting as the performance would be the same for both. For Graph C, the value of parameter $\mu_{w}$ is increased from $0.001$, in Graphs A and B, to $0.01$. An increase in the value of parameter $\mu_{w}$ increases the rate at which users in neighbouring communities interact with each other. It is seen from Table~\ref{Table:Exp2_T1} that the ratio between the average contact interval between a pair of users lying in adjacent communities to the contact interval for pairs of users lying in the same community is $1$, which means that pairs of users lying in neighbouring communities meet at the same expected rate as neighbours having the same membership. This makes detection of communities more difficult. GANC is not able to correctly identify all communities in Graph C. The number of communities identified by GANC are $13$ compared to $14$ provided by the LFR tool. On a closer analysis it is observed that GANC identifies $12$ out of the $14$ communities correctly. The $13^{th}$ and $14^{th}$ communities are, however, merged to form one larger community by GANC. 

To determine the loss in performance resulting from the mislabelling of some users by GANC, the file dissemination process is simulated for the communities detected by GANC and the results are compared to the ones obtained for LFR provided communities. The \textit{most central user} based seeding strategies is employed for both simulations. It is already shown in Section~\ref{compare} that the \textit{most central user} based seeding strategy provides the best results in terms of the delay experienced by users in obtaining the file and the number of non-innovative transmissions.

\begin{figure}[htbp]
	\centering
	\includegraphics[angle=-90, scale=0.45, type=pdf,ext=.pdf,read=.pdf]{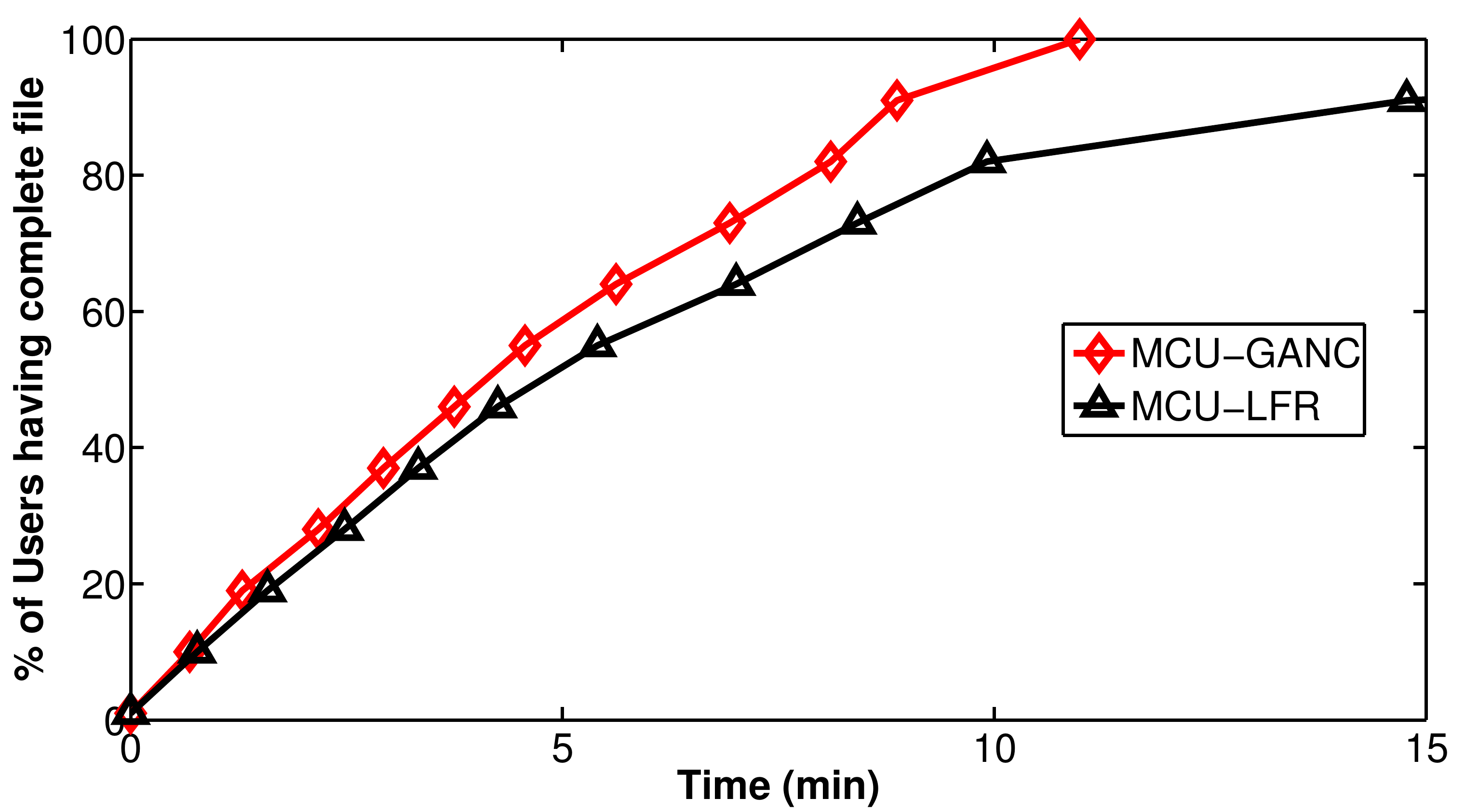}
	\caption[Performance comparison of MCU-GANC and MCU-LFR]{Performance comparison of MCU-LFR and MCU-GANC for Graph C. The curves show the expected percentage of users in the network that obtain the file with time.}
	\label{fig:NA_Curv_muw_ganc_lfr}	
\end{figure}

\begin{figure}[htbp]
	\centering
	\includegraphics[angle=-90, scale=0.5, type=pdf,ext=.pdf,read=.pdf]{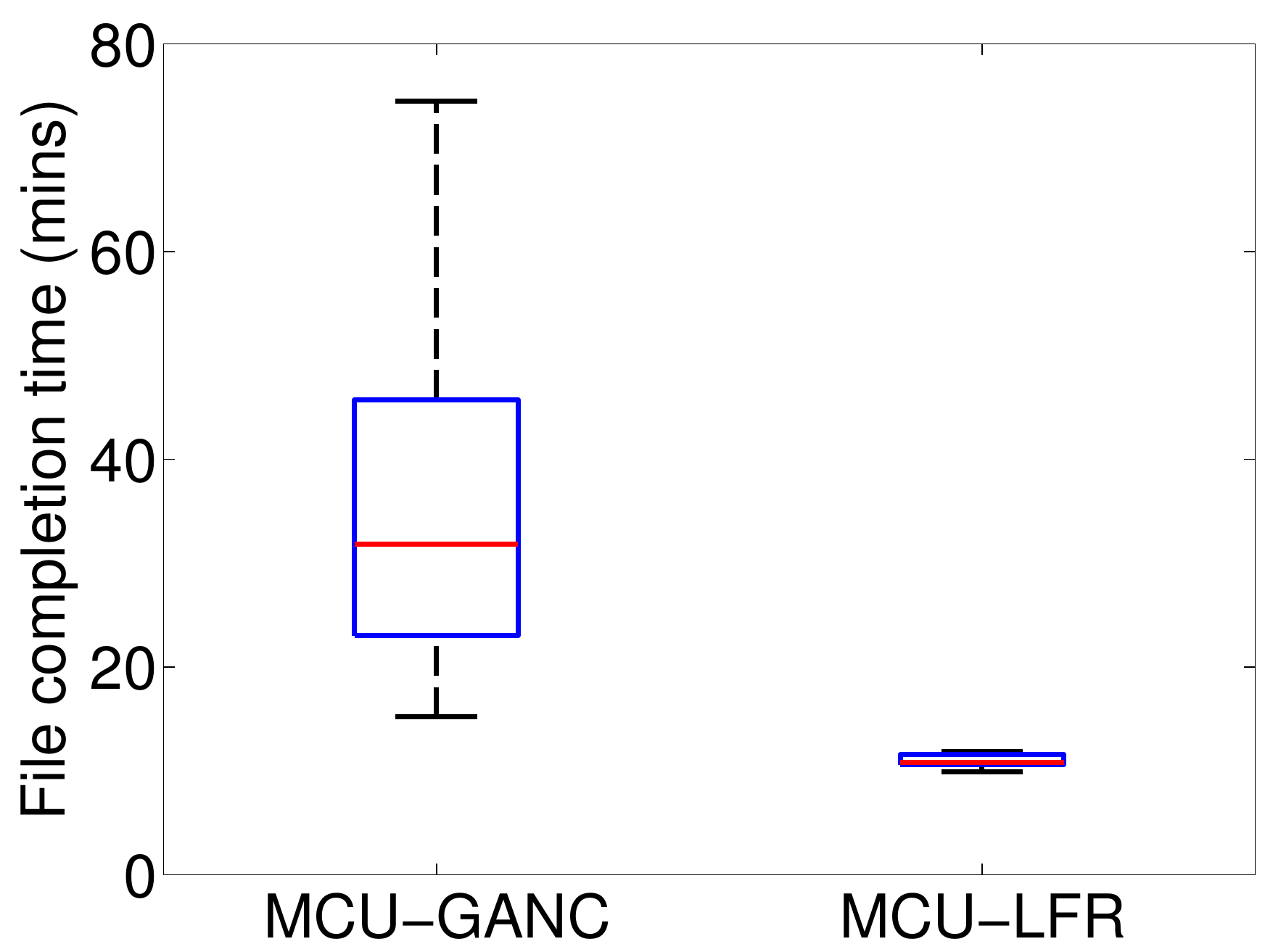}
	\caption[Finish times for MCU-GANC and MCU-LFR]{Distribution of finish time for MCU-GANC and MCU-LFR based seeding strategies on Graph C. The bar represents the median value while the box edges denote the $25^{th}$ and $75^{th}$ percentile values.}
	\label{fig:non_inn_lfr_ganc}	
\end{figure}

Figure~\ref{fig:NA_Curv_muw_ganc_lfr} shows the expected percentage of users that obtain the file at any given time. MCU-GANC denotes the performance on the communities detected via the GANC algorithm while MCU-LFR represents the results for LFR provided communities. The figure shows that the percentage of users that acquire the file at any time is lower for MCU-GANC based seeding. In Figure~\ref{fig:non_inn_lfr_ganc}, the median finish times for MCU-LFR based seeding strategy is 10min compared to 30min for MCU-GANC. It must be pointed out here that the comparison is not fair because the number of packets initially seeded in MCU-GANC is less than the number of initially seeded packets in MCU-LFR. This is because the number of communities detected by GANC is 13 and since the number of packets initially seeded in the network depends on the number of communities in the network (each community is seeded with the file), it is seeded with lesser packets compared to the MCU-LFR which provides 14 communities. If the correct number of communities is provided to the GANC algorithm a priori, it detects the correct membership for all users in the network and the performance is similar to the one observed for MCU-LFR. 
%

\newpage
%
The presence of \textit{hubs} causes swift movement of packets between users within a  community. If each community is seeded with the minimum number of packets required to decode the file, the performance in terms of the expected percentage of users that obtain the file at any time is similar for all centrality-based seeding strategies and the $Random$ seeding strategy. The delay a user faces in acquiring the file is then dependent on the frequency of encounters with other users. The \textit{most central user} based seeding strategy performs the best in terms of the number of non-innovative transmissions that occur during the file dissemination process. An important result is that all other seeding strategies penalize the \textit{most central users} in terms of the number of non-innovative packets it receives. The MCU-based seeding strategy on the other hand ensures that these users do not receive any \textit{non-innovative} packets by providing the file directly to them. This also guarantees the \textit{most central users} of obtaining the file earlier than other users. Such incentives are important to ensure the participation of important users which form \textit{hubs} in their respective communities. 
%
 

%

\renewcommand*{\bibname}{References}
\begin{singlespace}
  \bibliographystyle{IEEEtran}
  \bibliography{main.bib}
\end{singlespace}
\end{document}